\documentclass[12pt,a4paper]{article}

\usepackage{afterpage, amsfonts, amsmath, amssymb, amsthm, bbm, caption, color, float, graphicx, epstopdf, lipsum, longtable, lscape, mathrsfs, multirow, nccmath, nonfloat, pdflscape, rotating, tocloft}

\usepackage[margin=1in]{geometry}

\usepackage[round]{natbib}
\usepackage[doublespacing]{setspace}
\usepackage[table]{xcolor}

\def\be{\begin{equation}}
\def\ee{\end{equation}}
\def\bea{\begin{eqnarray}}
\def\eea{\end{eqnarray}}

\author{}
\title{}

\DeclareMathOperator*{\argmin}{\arg\!\min}

\DeclareMathOperator*{\vect}{\textrm{vec}}

\renewcommand {\arraystretch}{1.20}

\begin{document}
\newcommand\blfootnote[1]{
\begingroup
\renewcommand\thefootnote{}\footnote{#1}
\addtocounter{footnote}{-1}
\endgroup
}

\newtheorem{corollary}{Corollary}
\newtheorem{definition}{Definition}
\newtheorem{lemma}{Lemma}
\newtheorem{proposition}{Proposition}
\newtheorem{remark}{Remark}
\newtheorem{theorem}{Theorem}
\newtheorem{assumption}{Assumption}

\numberwithin{corollary}{section}
\numberwithin{definition}{section}
\numberwithin{equation}{section}
\numberwithin{lemma}{section}
\numberwithin{proposition}{section}
\numberwithin{remark}{section}
\numberwithin{theorem}{section}

\allowdisplaybreaks[4]

\begin{titlepage}

{\small

\begin{center}
{\Large \bf  An Integrated Panel Data Approach \\ to Modelling Economic Growth\blfootnote{
\begin{itemize}
\item[$^{\ast}$] Department of Economics, University of North Texas, Denton, TX 76201, USA. Email: Guohua.Feng@unt.edu

\item[$^\sharp$] Department of Econometrics and Business Statistics, Monash University, Caulfield East, Victoria 3145, Australia. Email: Jiti.Gao@monash.edu

\item[$^{\dag}$] Department of Economics, University of Bath, Bath BA2 7JP, UK. Email: B.Peng2@bath.ac.uk
\end{itemize}
}

} 
\medskip

{\sc Guohua Feng$^{\ast}$, Jiti Gao$^\sharp$ and Bin Peng$^{\dag}$}
\medskip

$^{\ast}$University of North Texas, $^\sharp$Monash University and $^{\dag}$University of Bath

\bigskip\bigskip

\today

\bigskip

\begin{abstract}
Empirical growth analysis has three major problems --- variable selection, parameter heterogeneity and cross-sectional dependence --- which are addressed independently from each other in most studies. The purpose of this study is to propose an integrated framework that extends the conventional linear growth regression model to allow for parameter heterogeneity and cross-sectional error dependence, while simultaneously performing variable selection. We also derive the asymptotic properties of the estimator under both low and high dimensions, and further investigate the finite sample performance of the estimator through Monte Carlo simulations. We apply the framework to a dataset of 89 countries over the period from 1960 to 2014. Our results reveal some cross-country patterns not found in previous studies (e.g., ``middle income trap hypothesis", ``natural resources curse hypothesis", ``religion works via belief, not practice", etc.).

\end{abstract}

\end{center}


\medskip


}

\end{titlepage}

\section{Introduction}\label{Section1}

Following the seminal works of \cite{KormendiMeguire} and \cite{Barro}, a vast amount of studies in the empirical growth literature have attempted to identify salient determinants of economic growth. A main tool used by these studies is ``cross-country growth regressions" --- that is, to regress observed GDP growth on a plethora of possible explanatory variables that could possibly affect growth across countries. Excellent surveys of these studies and their role in the broader context of economic growth theory are provided in \cite{DurlaufQuah}, \cite{Temple} and \cite{DurlaufJohnsonTemple}.

Despite the vast amount of research, the literature has identified a number of problems with conventional growth regressions, among which three deserve particular attention. The first problem is determining what variables to be included in growth regressions. This problem arises because of the nature of growth theories: although a plethora of growth theories have been proposed to identify factors that affect growth, these theories are open-ended in the sense that the validity of one causal theory of growth does not imply the falsity of another (\citealp{BrockDurlauf}). In words of \cite{Durlauf2008}, ``a given body of candidate growth theories defines a space of possible models rather than a single specification". From an empirical perspective, this problem stems from the fact that the number of potential explanatory variables is large (over 140 identified in \citealp{DurlaufJohnsonTemple}) relative to the number of countries with enough data availability, rendering the all-inclusive regression computationally infeasible (\citealp{Martin, DurlaufJohnsonTemple}). In dealing with this problem some studies have resorted to simply ``trying" combinations of variables which could be potentially important determinants of growth and report the results of their preferred specification. However, as noted by \cite{Leamer} and \cite{Martin} such ``data-mining" could lead to spurious inference.

The second problem with conventional growth analysis is that most empirical growth studies assume that the parameters of growth regressions are identical across countries. This assumption complies with the classical Solow model (\citealp{Mankiw}), which assumes that all countries share an identical aggregate Cobb-Douglas production function. However, an increasing number of studies (e.g., \citealp{DurlaufJohnson,Durlauf2001,Salimans}) have suggested that the parameters are heterogeneous across countries. These studies, though using different econometric methods, all suggest that the assumption of a single linear growth model that applies to all countries is inappropriate. For example, \cite{DurlaufJohnson} employ a regression tree analysis to show that a cross-sectional regression using the \cite{Summers} data appears to provide support for several distinct regimes in which aggregate production functions vary among countries according to their level of development, while \cite{Durlauf2001}, employing a varying coefficient growth model, also find strong evidence of parameter heterogeneity across countries.

The third problem is that few studies in the empirical growth literature allow for cross-sectional dependence of individual countries. Panel data econometrics has recently seen an increasing interest in models with unobserved time-varying heterogeneity caused by latent common shocks influencing all units, possibly to a different degree. This type of heterogeneity introduces cross-sectional dependence to individual countries, which, when neglected, can lead to biased estimates and spurious inference (\citealp{Pesaran2006, Bai}). In the context of cross-country growth analysis, the problem of cross-sectional dependence seems particularly salient due to the omnipresence of common global shocks (such as global financial crises and world oil price shocks) that affect all countries through trade and financial linkages (\citealp{Chudik}). \cite{DurlaufQuah} discuss the possibility of cross-sectional dependence in a \cite{Lucas} growth model with human capital spillovers. They find that these spillovers markedly change the dynamics of convergence and the authors call for the modelling of cross-country interactions in empirical convergence analysis.

The three aforementioned problems have received more or less individual attention in the growth literature. For example, \cite{Durlauf2001} address the problem of parameter heterogeneity using a varying coefficient growth model, but do not deal with the problems of variable selection and cross-sectional dependence; both \cite{Martin} and \cite{Benito} select growth determinants using Bayesian averaging, but do not account for parameter heterogeneity and cross-sectional dependence.

The main goal of this study is to propose an integrated framework that is capable of dealing with parameter heterogeneity and cross-sectional dependence, while simultaneously performing variable selection. Specifically, parameter heterogeneity is allowed for by permitting the coefficients to vary across countries according to a country's initial conditions, while cross-sectional dependence is accounted for via a factor structure. We then propose a least absolute shrinkage and selection operator (LASSO) estimator to select growth determinants, establish the associated asymptotic results, and further verify our asymptotic results through extensive simulations, which constitutes another contribution of this paper. We apply this framework to a new dataset of 89 countries over the period 1960-2014. Our findings broadly support the more ``optimistic" conclusion of \cite{Sala-i-Martin1997}, that is, some variables are important regressors for explaining cross-country growth patterns. Moreover, our empirical results also provide support to some important hypotheses in the growth literature, e.g., ``middle income trap hypothesis", ``natural resources curse hypothesis", ``religion works via belief, not practice", etc.

The rest of the paper is organized as follows. Section \ref{Section2} explains how to extend the canonical cross-country growth regression to account for the aforementioned issues. Section \ref{Method} describes a procedure for estimating the extended growth regression model in Section \ref{Section2}, and presents the associated asymptotic properties. Section \ref{Section4} describes the data. The empirical results are presented in Section \ref{Section5}. Section \ref{Section6} concludes. Due to space limitations, preliminary lemmas, proofs of the main theorems and Monte Carlo simulations, together with auxiliary tables and figures, are presented in the supplementary Appendix A of this paper. The proofs of the preliminary lemmas are presented in the supplementary Appendix B of this paper, which can be found at the authors' website (https://papers.ssrn.com/sol3/cf\_dev/AbsByAuth.cfm?per\_id=646779).

\section{A Varying Coefficient Growth Regression Model with Factor Structure and Sparsity}\label{Section2}

A generic representation of the canonical cross-country growth regression is

\begin{eqnarray}\label{b1}
y_{it} =x_{it}'\beta_0 +e_{it},
\end{eqnarray}
where $i=1,2,\ldots,N$ index countries; $t=1,2,\ldots,T$ index time; $y_{it}$ is the rate of economic growth; $x_{it}$ represents a set of observable explanatory variables, including those originally suggested by Solow as well as other growth theories, and $e_{it}$ is an error term. Equation (\ref{b1}) represents the baseline for much of growth econometrics.

However, as discussed in the Introduction, (\ref{b1}) is based on two problematic assumptions. First, it assumes that the parameters (i.e., $\beta_0$) are homogeneous across all countries. Second, it assumes that there is no cross-sectional dependence across countries. 

To relax the two assumptions, in what follows we extend the conventional cross-country growth regression in (\ref{b1}) in two ways. First, in Section \ref{Section21} we allow for parameter heterogeneity by allowing $\beta_0$ to vary across countries according to a country's initial conditions. Second, in Section \ref{Section22} we introduce cross-sectional dependence into the model by means of a factor structure.

\subsection{Parameter Heterogeneity}\label{Section21}

Following \cite{Durlauf2001}, we allow for parameter heterogeneity by generalizing (\ref{b1}) into a varying coefficient model:

\begin{eqnarray}\label{b2}
y_{it} =x_{it}'\beta_0(z_{it}) +e_{it},
\end{eqnarray}
where $z_{it}$ can be interpreted as some measure of ``development" (or initial condition) of a country, and $\beta_0(z)= (\beta_{01}(z), \ldots,\beta_{0p} (z))'$ is a vector of smooth functions that maps the scalar index variable $z_{it}$ into a set of country-specific parameters. 

This generalization in (\ref{b2}) provides a framework within which one can bridge the gap between cross-country regression models and new growth theories. For instance, if one believes that initial GDP per capita causally affects a country's production technology and growth as in \cite{Durlauf2001}, then initial GDP per capita can be introduced as a
``development" index. As pointed out by \cite{DurlaufJohnson}, (\ref{b2}) is compatible both with a model in which economies pass through distinct phases of development towards a unique steady state as well one in which multiple steady states exist.

\subsection{Cross-Sectional Error Dependence}\label{Section22}

Having accounted for parameter heterogeneity, we next introduce the cross-sectional dependence of error terms into (\ref{b2}) using a factor structure:

\begin{eqnarray}\label{b3}
e_{it} = \gamma_{0i}'f_{0t} + \varepsilon_{it},
\end{eqnarray}
where $f_{0t}$ is an $r\times 1$ vector of unobservable common factors, $\gamma _{0i}$ is an $r\times 1$ vector of factor loadings that capture country-specific responses to the common shocks, and $\varepsilon _{it}$ is the idiosyncratic error term. These common factors can be a combination of ``strong" factors, such as world oil price shocks, global financial crises, and recessions in major advanced economies; and ``weak" factors, such as local spillover effects along channels determined by shared culture heritage, geographic proximity, economic or social interaction (\citealp{Chudik2011}). Moreover, the components of the factor structure are allowed to drive both economic growth and explanatory variables, thus partially accounting for potential endogeneity of explanatory variables, which is neglected by the traditional approaches to causal interpretation of cross-country empirical analysis.

\subsection{The Varying Coefficient Growth Regression Model with Factor Structure  and Sparsity}

Substituting (\ref{b3}) into (\ref{b2}) yields the following growth regression model that allows for parameter heterogeneity and cross-sectional dependence

\begin{eqnarray}\label{model1}
y_{it} = x_{it}'\beta_0(z_{it}) + \gamma_{0i}'f_{0t} +\varepsilon_{it}, \ t=1,2,\cdots, T,
\end{eqnarray}
The model in \eqref{model1} extends the local Solow growth model investigated in \cite{Durlauf2001} into a panel data context with interactive fixed effects (or factor structure). From an econometric perspective, \eqref{model1} extends the panel data model with interactive fixed effects in \cite{Bai} into a varying coefficient context. Some closely related studies include, but are not limited to, \cite{DGP2018} on \eqref{model1} with partially observed factor structure, \cite{FGPZ2017} on a special case of \eqref{model1} with discrete $z_{it}$ and $\gamma_{0i}'f_{0t}$ being reduced to fixed-effects $\alpha_i$, \cite{LGY2018} on a time-varying heterogeneous model with $z_{it} = \frac{t}{T}$, and \cite{Malikov} on a binary varying-coefficient panel data setting with endogenous selection and fixed-effects. There are some key differences between this paper and the relevant literature. First, it is worth pointing out that these previous studies introduce parameter heterogeneity and cross-sectional dependence in different manners. Second, none of them consider performing variable selection on their varying coefficient models as we will show below particularly in the high-dimensional setting. Finally and most importantly, both model (\ref{model1}) and its asymptotic theory in Section 3 below are naturally motivated by the relevant empirical literature in economic growth.  

In addition to parameter heterogeneity and cross-sectional dependence, we are also interested in another issue that is prominent in the empirical growth literature --- variable selection. This issue is important because (1) the dimension of $x_{it}$ can be very large; and (2) not all elements of $x_{it}$ drive economic growth. In other words, for those factors not driving economic growth, it is reasonable to assume that their associated coefficients are zero, which is called ``sparsity" in the literature of high dimensional econometrics.

In order to formally introduce the sparsity to the model (\ref{model1}), we assume that there exists an unknown set $\mathcal{A}^\dagger\subseteq \{1,\ldots,p\}$ satisfying that $E|\beta_{0j} (z_{it})|^2 = 0$ if and only if $j\in \mathcal{A}^\dagger$. For notational simplicity, we assume $\mathcal{A}^\dagger=\{p^*+1,\ldots,p\}$ for an unknown integer $p^*$ satisfying $1\le p^*<p$. Further, let $\mathcal{A}^* = \{ 1,\ldots, p^*\}$, $x_{it}^* =(x_{it,1},\ldots, x_{it,p^*})'$, and $\beta_0^*(z) = (\beta_{01}(z),\ldots, \beta_{0p^*}(z))'$. Throughout this study, we always define the variables or functions corresponding to the sets $\mathcal{A}^*$ and $\mathcal{A}^\dagger$ with super-indices $^*$ and $^\dagger$ respectively. Thus, identifying growth determinants is equivalent to distinguishing $\mathcal{A}^*$ and $\mathcal{A}^\dagger$, which will be achieved by a LASSO estimator presented in the following section. Finally, regarding the dimension of regressors, we consider two cases where (1) $p$ is fixed, and (2) $p$ diverges as the sample size increases. We refer to them as the low dimensional (LD) case and the high dimensional (HD) case, respectively. In terms of econometric methodology, both cases with the sparsity setting have not been studied in the literature to the best of our knowledge.

\medskip

As discussed in the Introduction, failure to perform variable selection may result in spurious inference, failure to allow parameters to differ across countries is inconsistent with the increasing body of research that find cross-country parameter heterogeneity, and failure to account for cross-sectional dependence can lead to biased estimates and spurious inference. These possible consequences thus necessitate an integrated approach to simultaneously addressing the three issues.

In the following section, we introduce a LASSO estimator that is designed specifically for performing variable selection on  the extended growth regression model in (\ref{model1}). The combination of varying coefficients, factor structure, and the LASSO
estimator provides us an integrated framework that is capable of simultaneously addressing the three problems mentioned in Section \ref{Section1}  --- variable selection, parameter heterogeneity, and cross-sectional dependence.

\section{Estimation}\label{Method}

In this section, we describe a procedure for estimating the model in (\ref{model1}) and derive the associated asymptotic properties. Specifically, we propose a LASSO estimator to select the appropriate variables, adopt a sieve method to recover the functional components, and employ the principle component analysis (PCA) technique to estimate the unobservable factor structure.   

Before proceeding further, it is convenient to introduce some notations. We let $Y_i = (y_{i1},\ldots, y_{iT})'$, $X_i  = (x_{i1},\ldots, x_{iT})'$, $Z_i  = (z_{i1},\ldots, z_{iT})'$, $ \mathcal{E}_i  = (\varepsilon_{i1},\ldots, \varepsilon_{iT})'$, $F_0  =(f_{01},\ldots, f_{0T})'$, and $\Gamma_0 = (\gamma_{01}, \ldots, \gamma_{0N})'$. $\| \cdot\|$ denotes the Euclidean norm of a vector or the Frobenius norm of a matrix; for a square matrix $W$, let $\eta_{\text{min}}(W)$ and $\eta_{\normalfont\text{max}}(W)$ stand for the minimum and maximum eigenvalues of $W$ respectively; $M_W=I_T - P_W$ denotes the orthogonal projection matrix generated by matrix $W$, where $P_W=W(W'W)^{-1}W'$, and $W$ is a matrix with full column rank.

\bigskip

We adopt the sieve method (e.g., \citealp{DongLinton}) to estimate the functional component of (\ref{model1}). Specifically, assume that  $\beta_{0\ell}\in L^2(V_z)$ for $\ell =1,\ldots, p$, where $L^2(V_z)= \{g \, | \, \int_{V_z} g^2(z) dz <\infty\}$ is a Hilbert space. Suppose that there exists an orthonormal function sequence $\{h_j(z)\, | \, j\ge 0\}$ in $L^2(V_z)$ such that  $\sup_{z\in V_z}\sup_{j\ge 0} |h_j(z)| <\infty$. Then, for $\forall g(z)\in L^2 (V_z)$, we have an orthogonal series expansion $g(z) := g_m(z) + \delta_{m}(z)$, where $g_m(z)= \sum_{j=0}^{m-1} c_j h_j(z)$, $\delta_{m}(z)= \sum_{j=m}^{\infty} c_j h_j(z)$, $c_{j}=\int_{V_z} g(z) h_j(z)dz$, and $m$ is the so-called truncation parameter. By the Parseval equality, the norm can be expressed as $\| g\|_{L^2} = \{\int_{V_z} g^2(z)dz  \}^{1/2} = \{\sum_{j=0}^\infty c_j^2 \}^{1/2}$. For a vector of functions $G(z) = (g_1(z),\ldots, g_d (z))'$,  its norm is defined by $\| G\|_{L^2} = \{\sum_{\ell =1}^{d}\|g_\ell\|_{L^2}^2 \}^{1/2}$. 

Without loss of generality, truncating the expansions of all the elements of $\beta_0(z)$ by the same $m$ gives

\begin{eqnarray}\label{Eq25}
\beta_0(z)  =  \beta_{0,m}(z) + \Delta_{m}(z)  ,
\end{eqnarray}
where $\beta_{0,m}(z) =C_{\beta_0}H_{m}(z)$, $H_m(z) = (h_0(z),h_1(z)\ldots, h_{m-1}(z) )'$, $C_{\beta_0} = ({C_{\beta_0}^*}', 0_{(p-p^*)\times m}')'$, and $\Delta_m (z) = (\Delta_m^* (z)', 0_{ (p-p^*)\times 1}')'$. Thus, the first $p^*$ elements of $\beta_0(z)$ can be expressed by $\beta_0^*(z) =\beta_{0,m}^*(z)  +\Delta_m^* (z) $, where $\beta_{0,m}^*(z) =C_{\beta_0}^* H_m(z)$.

We are now ready to rewrite (\ref{model1}) as
\begin{eqnarray*} 
M_{F_0}Y_i = M_{F_0}\phi_i[\beta_{0,m}] + M_{F_0}\phi_i[\Delta_{m}]  +M_{F_0}\mathcal{E}_i \nonumber
\end{eqnarray*}
by projecting out the factor structure, where $\phi_i[\beta] = (x_{i1}'\beta(z_{i1}),\ldots, x_{iT}'\beta(z_{iT}))' $ for any $p\times 1$ vector of functions $\beta(z)$. The objective function is then defined by

\begin{eqnarray} \label{obj1}
Q_\lambda(C_\beta, F) = \sum_{i=1}^N \left( Y_i - \phi_i[\beta_m]  \right)'M_F\left( Y_i -\phi_i[\beta_m]  \right) + \sum_{j=1}^p \lambda_j \|C_{\beta,j} \|,
\end{eqnarray}
where $\beta_{m}(w) =C_{\beta}H_{m}(w)$, $C_{\beta,j}$ stands for the $j^{th}$ row of $C_{\beta }$, and $\{\lambda_1,\ldots, \lambda_p\}$ are the regularizers of the coefficient functions and are to be determined by data. The estimators of $C_{\beta_0}$ and $F_0$ for both LD and HD cases are always obtained by

\begin{eqnarray}\label{est1}
(\widehat{C}_\beta,\widehat{F}) =\argmin_{C_\beta, F\in \mathsf{D}_F} Q_\lambda(C_\beta, F),
\end{eqnarray}
where $\mathsf{D}_F =\{F\, | \, \frac{F'F}{T} =I_r\}$. In what follows, we always partition $\widehat{C}_\beta$, according to the partition $\mathcal{A}^*$ and $\mathcal{A}^\dagger$, as $\widehat{C}_\beta= ({\widehat{C}_\beta^*} \,^{\prime}, {\widehat{C}_\beta^\dagger} \,^{\prime} )'$ wherever necessary. 

At this point it is convenient to state some fundamental assumptions that are needed for the derivation of the asymptotic results for both LD and HD cases. 

\medskip

\begin{assumption}\label{Assumption1}
\item
\begin{enumerate}
\item Let $\mathcal{F}_{-\infty}^0$ and $\mathcal{F}_\tau^\infty$ denote the $\sigma$-algebras generated by $\{(x_t,z_t, \varepsilon_t, f_{0t} ) \ | \ t \le  0\}$ and $\{(x_t,z_t, \varepsilon_t, f_{0t}) \ | \ t \ge  \tau\}$ respectively, where $x_t=(x_{1t},\ldots,x_{Nt})'$, $z_t=(z_{1t},\ldots,z_{Nt})'$, $\varepsilon_t=(\varepsilon_{1t},\ldots,\varepsilon_{Nt})'$. Let $\alpha(\tau) = \sup_{A\in \mathcal{F}_{-\infty}^0, B\in \mathcal{F}_\tau^\infty} \left|\Pr(A)\Pr(B) -\Pr(AB) \right| $ be the mixing coefficient.

\begin{enumerate}
\item $\{X_i, Z_i, \mathcal{E}_i, \gamma_{0i}\}$ is identically distributed over $i$. $\{x_t,z_t, \varepsilon_t, f_{0t}  \}$ is strictly stationary and $\alpha$-mixing such that for some $\nu_1>0$, $E[\|\varepsilon_{11}\| + \|x_{11}\|]^{4+\nu_1}<\infty$, and the mixing coefficient satisfies $  \sum_{t=1}^\infty  [\alpha(t)]^{\nu_1/(2+\nu_1)}<\infty$.

\item $E[\varepsilon_{11}]=0$, $E[\varepsilon_{11}^2]=\sigma_\varepsilon^2$, and $\{\varepsilon_{it}\}$ is independent of the other variables. Let $E[\varepsilon_{i1}\varepsilon_{j1}] =\sigma_{ij}$ for $ i\ne j$, $ \sum_{i\ne j}|\sigma_{ij}|=O(N)$, and $\sum_{i,j=1}^N\sum_{t,s=1}^T|E[\varepsilon_{it}\varepsilon_{js}]|=O(NT)$.
\end{enumerate}

\item Let $\left\|\frac{1}{T}F_0'F_0 -\Sigma_f\right\| = O_P\left( \frac{1}{\sqrt{T}}\right)$ and $\left\|\frac{1}{N}\Gamma_0'\Gamma_0-\Sigma_{\gamma}\right\|=O_P\left( \frac{1}{\sqrt{N}}\right)$, where $\Sigma_f$ and  $\Sigma_\gamma$ are deterministic and positive definite. Moreover, $E\|f_{01} \|^4<\infty$ and $E\|\gamma_{01} \|^4<\infty$.
\end{enumerate}
\end{assumption}

\begin{assumption}\label{Assumption2}
\item
\begin{enumerate}
\item Suppose that $\sup_{z\in V_z}\|\Delta_m(z) \| = O(m^{-\mu/2})$, and $\eta_{\normalfont\text{max}}( \frac{1}{NT}\sum_{i=1}^N  \mathcal{Z}_i'  \mathcal{Z}_i' )<\infty$ with probability approaching one, where $ \mathcal{Z}_i = (\mathcal{Z}_{i1},\ldots, \mathcal{Z}_{iT})'$ and $\mathcal{Z}_{it}=H_m(z_{it})\otimes x_{it}$.

\item Let  $\Omega_1(F)=\frac{1}{NT}\sum_{i=1}^{N}\mathcal{Z}_i'M_{F}\mathcal{Z}_i$, $ \Omega_2(F)=\frac{1}{NT}\sum_{i=1}^{N}\gamma_{0i}\otimes (M_F \mathcal{Z}_i)$ and $\Omega_3 = \frac{\Gamma_0'\Gamma_0}{NT}\otimes I_{T}$. Suppose $\inf _{F\in \mathsf{D}_F} \eta_{\normalfont\text{min}} ( \Omega(F)) >0$, where $\Omega (F)= \Omega_1(F)- \Omega_2'(F)\Omega_3 ^{-1} \Omega_2(F)$.
\end{enumerate}
\end{assumption}

\medskip

Assumption \ref{Assumption1} is standard in the literature. The mixing conditions are similar to Assumption C of \cite{Bai} and Assumption 3.4 of \cite{FanLiaoWang}. In Assumption \ref{Assumption2}.1, the condition $\sup_{z\in V_z}\|\Delta_m(z) \| = O(m^{-\mu/2})$ is the same as Assumption 3 of \cite{Newey1997}, and essentially requires certain smoothness of the elements of $\beta_0(\cdot)$. The condition $\eta_{\normalfont\text{max}}( \frac{1}{NT}\sum_{i=1}^N  \mathcal{Z}_i'  \mathcal{Z}_i' )<\infty$ of Assumption \ref{Assumption2}.1 is similar to Assumption 3.1 of \cite{FanLiaoWang}.  Now, consider a special case where $\{z_{it}\}$, $\{x_{it}\}$ and $\{f_{0t}\}$ are mutually independent, $E[x_{it}]=0$ and $E[x_{it}x_{it}']=I_p$. Under this setting, it is easy to see that $\eta_{\normalfont\text{max}}( \frac{1}{NT}\sum_{i=1}^N  \mathcal{Z}_i'  \mathcal{Z}_i' )<\infty$ holds true for both of the LD and HD cases by some standard analyses. Assumption \ref{Assumption2}.2 ensures that the estimators given in (\ref{est1}) are well defined, and is equivalent to Assumption A of \cite{Bai}. 

Based on the above setting, we move on to investigate the asymptotic results associated with (\ref{est1}) under the LD setting.

\subsection{Low Dimensional Case}\label{SectionLD}

In order to identify $\mathcal{A}^*$ and $\mathcal{A}^\dagger$ and establish the asymptotic distribution, we further make the following assumptions.

\medskip

\begin{assumption}\label{ConsisLD}
\item 
\begin{enumerate}
\item $\frac{m^2}{T}\to 0$ and $\frac{\lambda_{\normalfont\text{max}}^*}{N^{\frac{3}{4}}T}\to 0 $, where $\lambda_{\normalfont\text{max}}^* = \max\{\lambda_1,\ldots,\lambda_{p^*} \}$. 
\item $\frac{N}{T}\to \kappa_0$  and $\frac{\lambda_{\normalfont\text{min}}^\dagger}{m^{\frac{1}{2}}N^{\frac{7}{8}}T}\to \kappa_1 $, where $\lambda_{\normalfont\text{min}}^\dagger = \min\{\lambda_{p^*+1},\ldots,\lambda_{p} \}$, $0\le \kappa_0<\infty$, and $\kappa_1>0$.
\end{enumerate}
\end{assumption}

\begin{assumption}\label{Assumption3}

\item

\begin{enumerate}
\item Suppose that for $t\ge s$, $E[f_{0t}'f_{0s}\, | \,  \mathcal{X}_{Nt} ] = a_{ts}$, and $\sum_{t=1}^T\sum_{s=1}^t |a_{ts}|=O(T)$, where $\mathcal{X}_{Nt} := \{(x_{1t}, z_{1t}),\ldots, (x_{Nt}, z_{Nt})\}$. Moreover, $\frac{NT}{m^{\mu +1}}\to 0$, $\frac{mN}{T}\to 0$, $\frac{T}{N^2}\to 0$ and $\frac{m\lambda_{\normalfont\text{max}}^*}{\sqrt{NT}}\to 0 $.

\item Let $\Sigma_{\mathcal{Z}}^*=E[\mathcal{Z}_{11}^* {\mathcal{Z}_{11}^*}']$ and $\displaystyle\Omega_\star = \lim_{(N,T)\to (\infty,\infty)} E[\Psi_1 \Psi_1']$ for $\forall z\in V_z$, where 

\begin{eqnarray*}
\Psi_1 &=& \sqrt{\frac{NT}{m}} \left[H_m'(z) \otimes I_{p^*}\right]  \Psi_2^{-1}  \Sigma_{\mathcal{Z}}^{*\, -1}\cdot\frac{1}{NT } \sum_{i=1}^N  \left\{ \mathcal{Z}_i'+  \frac{1}{N}\sum_{j=1}^N{\mathcal{Z}_j^*}' \gamma_{0j}'  \Sigma_\gamma^{-1}\gamma_{0i}\right\} \mathcal{E}_{i},\nonumber\\
\Psi_2 &=&I_{m p^* } - \Sigma_{\mathcal{Z}}^{*\, -1} E[ \mathcal{Z}_{11}^*  \gamma_{01}' ] \Sigma_\gamma^{-1}E[\gamma_{01} {\mathcal{Z}_{11}^*}'],\quad \text{and}\quad\mathcal{Z}_{it}^*=H_m(z_{it})\otimes x_{it}^*.\nonumber
\end{eqnarray*}
Suppose that for $\forall z\in V_z$, as $(N,T)\to (\infty,\infty)$, $\Psi_1\to_D N(0, \Omega_\star)$.
\end{enumerate}
\end{assumption}

\medskip

The conditions of Assumption \ref{ConsisLD}, though seemingly complicated, can be easily satisfied. For example, let $N = \lfloor T^{b_0}\rfloor $, $m =\lfloor T^{b_1} \rfloor $, $\lambda_{\text{max}}^* = T^{b_2}$, and $\lambda_{\text{min}}^\dagger = T^{b_3}$, where $\lfloor a\rfloor$ means the largest integer part of a real number $a$. Then Assumption \ref{ConsisLD} essentially requires that $0<b_0\le 1$, $0<b_1<\frac{1}{2}$, $b_2<\frac{3}{4}b_0+1$ and $b_3\ge  \frac{b_1}{2} + \frac{7b_0}{8}+1$.

The current requirements of Assumption  \ref{Assumption3}.1 are in the same spirit as \citet[Eq. 3 and Eq. 20]{CHL2012} and \citet[pp. 21-22]{JYGH}. Without this assumption, some other types of conditions would be needed to achieve asymptotic normality. For example, one can require $N/T\to \rho $ with $0<\rho <\infty$ and establish the normality with biases as in Theorem 3 of \cite{Bai}. Assumption  \ref{Assumption3}.2 is equivalent to Assumption E of \cite{Bai}. It is worth mentioning that deriving the rates of convergence in Lemma A.3 and Lemma A.5 of the supplementary Appendix A does not require Assumption \ref{Assumption3} at all. For better presentation and in order not to deviate from our main goal, we present these lemmas in the supplementary Appendix A instead of the main text.

\begin{theorem}\label{TheoremMain42}
Let Assumptions \ref{Assumption1}-\ref{ConsisLD} hold.  
\begin{enumerate}
\item $\Pr (\| \widehat{C}_{\beta}^\dagger\|=0 )\to 1$.

\item Suppose Assumption \ref{Assumption3} also holds. Then $\sqrt{\frac{NT}{m}}  (\widehat{\beta}_m^*(z)-\beta_0^*(z)  ) \to_D N(0, \Omega_\star)$ for $\forall z\in V_z$, as $(N, T)\to (\infty,\infty)$, where $\widehat{\beta}_m^*(z) = \widehat{C}_\beta^* H_m(z)$.
\end{enumerate}
\end{theorem}

The first result of Theorem \ref{TheoremMain42} indicates that we are able to distinguish $\mathcal{A}^*$ and $\mathcal{A}^\dagger$; while the second result of Theorem \ref{TheoremMain42} establishes the asymptotic distribution of the coefficient functions associated with the variables which truly drive economic growth.

 To complete our discussion on the LD case, we propose the following BIC type criteria in order to select $\lambda$ practically.

\begin{eqnarray}\label{BIC}
\text{BIC}_{\lambda} =\ln \text{RSS}_{\lambda} + \text{df}_{\lambda} \frac{\ln N} {\sqrt[4]{N}},
\end{eqnarray}
where $\text{RSS}_{\lambda} = \frac{1}{NT} \sum_{i=1}^N \big( Y_i -\phi_i[\widehat{\beta}_{m}^\lambda]  \big)'M_{\widehat{F}^\lambda}\big( Y_i - \phi_i[\widehat{\beta}_{m}^\lambda]  \big)$, $\widehat{\beta}_{m}^\lambda(z) = \widehat{C}_\beta^\lambda H_m(z)$, $(\widehat{C}_\beta^\lambda, \widehat{F}^\lambda)$ are obtained by implementing (\ref{est1}) using $\lambda$ as the weight vector, and $\text{df}_{\lambda}$ is the number of nonzero coefficient functions identified by $\widehat{C}_\beta^\lambda$. The penalty term  $ \frac{\ln N} {\sqrt[4]{N}}$ of (\ref{BIC}) is constructed in view of the slow rate documented in Lemma A.2 of the supplementary Appendix A. We select $\lambda$ by

\begin{eqnarray}\label{estlam}
\widehat{\lambda} =\argmin_{\lambda} \text{BIC}_{\lambda}.
\end{eqnarray}
Further let $S_{\widehat{\lambda}} = \{j\, | \, \|\widehat{C}_{\beta,j}^{\widehat{\lambda}} \|>0, 1\le j\le p \}$ indicate the set of relevant variables identified by $\widehat{C}_{\beta}^{\widehat{\lambda}}$. Then the next result follows.

\begin{theorem}\label{theorem3}
Let Assumptions \ref{Assumption1}-\ref{ConsisLD} hold. $\Pr(S_{\widehat{\lambda}} = \mathcal{A}^*)\to 1$ as $(N,T)\to (\infty,\infty)$. 
\end{theorem}
Again, Assumption \ref{Assumption3} is unnecessary for establishing Theorem \ref{theorem3}.

\subsection{High Dimensional Case}\label{SectionHD}

In this subsection, we allow the dimension of $x_{it}$ to diverge as the sample size increases. The following assumptions are crucial for deriving the asymptotic results for the HD case.

\medskip

\begin{assumption}\label{Assumption4}
\item
\begin{enumerate}
\item  $\|\mathcal{E}\| _{\normalfont\text{sp}}=O_{P}(\max\{\sqrt{N},\sqrt{T} \})$, where $\| \cdot\|_{\normalfont\text{sp}}$ denotes the spectral norm of a matrix and $\mathcal{E}= (\mathcal{E}_1,\ldots, \mathcal{E}_N)'$;

\item $\frac{p^* \lambda_{\normalfont \text{max}}^* \sqrt[4]{\xi_{NT}}}{NT}\to 0$, $( \frac{\xi_{NT}+mp}{NT} + p^* m^{-\mu} ) \sqrt{\xi_{NT}}\to \kappa_2$,  $\frac{\sqrt[8]{\xi_{NT}} \lambda_{\normalfont\text{min}}^\dagger  }{ NT }\to \kappa_3$, where $\xi_{NT} = \min\{N, T \}$, $0\le \kappa_2<\infty$ and $\kappa_3>0$.
\end{enumerate}
\end{assumption}

\medskip

Assumption \ref{Assumption4}.1 is identical to Assumption iii of \cite{LiQianSu} and Assumption A.1.v of \cite{LuSu}. Assumption \ref{Assumption4}.2 further imposes bounds on some parameters, and can be verified in exactly the same way as shown under Assumption \ref{ConsisLD}. 

With regard to the selection of $\lambda$, we still use the BIC criterion with a minor modification:

\begin{eqnarray} \label{BICLP}
\text{BIC}_{\lambda} =\ln \text{RSS}_{\lambda} + \text{df}_{\lambda}\Upsilon_{NT},
\end{eqnarray}
where $\Upsilon_{NT}$ is a penalty term satisfying $\Upsilon_{NT}\to 0$ as $(N,T)\to (\infty,\infty)$; and all other notations are defined in exactly the same way as in the LD case. Select $\lambda$ by $\widehat{\lambda} =\argmin_{\lambda} \text{BIC}_{\lambda}$.

Then the next theorem holds.

\begin{theorem} \label{theorem5}
Let Assumptions \ref{Assumption1}, \ref{Assumption2} and \ref{Assumption4} hold.  As $(N,T)\to (\infty,\infty)$, 

\begin{enumerate}
\item $\Pr (\| \widehat{C}_{\beta}^\dagger\|=0 )\to 1$.

\item Additionally, let $\Upsilon_{NT}\xi_{NT}^{1/8}\to \kappa_4>0$. Then $\Pr(S_{\widehat{\lambda}} = \mathcal{A}^*)\to 1$.
\end{enumerate}
\end{theorem}

Once the zero coefficient functions are identified, the rest of the analysis (such as the investigation of the rate of convergence) will be similar to that done in Section \ref{SectionLD} except that one needs to account for the divergence of both $m$ and $p^*$. To avoid repetition, we will not present the analysis here again. 

\bigskip

In summary, in either of the two cases (LD and HD), when $\Pr(S_{\widehat{\lambda}} = \mathcal{A}^*)\to 1$, all zero coefficient functions can be identified. In the growth regression context, this is equivalent to saying that when $\Pr(S_{\widehat{\lambda}} = \mathcal{A}^*)\to 1$, all variables not driving economic growth can be identified and thus removed from the growth regression. In the meantime, the varying coefficients can be recovered using the sieve method, while the factor structure can be estimated by the PCA technique. Thus, all the three aforementioned issues that are prominent in the empirical growth literature (i.e., variable selection, parameter heterogeneity, and cross-sectional dependence) can be addressed simultaneously within a single, integrated framework. Before moving on to the empirical analysis, we next describe the data employed in this study.

\section{Data}\label{Section4}

Of the many variables that have been found to be significantly correlated with growth in the literature, we choose a total of 60 (including the dependent variable, the growth rate of per capita GDP). The choice of these variables is based on previous studies (e.g., \citealp{Martin, Benito}) and data availability. Our final dataset covers 89 countries over the period 1960 - 2014. It contains countries in different stages of development and with a wide geographic dispersion. The explanatory variables cover a wide range of factors, including stage of development, social issues, health, geography, politics, education and more. The variable names, their means, and standard deviations are presented in Table Table \ref{Table_Var}. Table \ref{Tabel_Country} provides a list of the included countries. 

A common practice in the literature is to take a five-year simple moving average of both dependent and independent variables\footnote{Another popular method of looking at annual data in empirical growth literature is to use averaged five-year period data. But, as is stressed by \cite{Soto2003} and \cite{Attanasio2000}, the use of n-year averages is not suitable because of the lost of information that it implies, and attempting to use data on averaged five-year periods severely limited the number of observations to draw from in the data.}. This  technique has the advantages of reducing the potential effects of short-term fluctuations and maintaining a high number of time series observations. Despite these advantages, this technique may still suffer from reverse causality or simultaneity, because causality between regressors and growth could go the other way as well or some regressors and growth may be simultaneously determined (e.g., \citealp{Bils}). To mitigate this problem, we deviate from the common practice by measuring dependent and independent variables differently. Specifically, while the dependent variable is measured as a five-year moving average of economic growth, all explanatory variables are measured at the beginning of each five-year period, with the exception of the variables related to war, geography, and terms of trade\footnote{Specifically, these variables include: fraction spent in war (each five-year period); number of war participation (each five-year period); number of revolutions (each five-year period); coups d'etat and coup attempts within (each five-year period); time of independence; East Asian dummy; African dummy; European dummy; Latin American dummy; British colony dummy; Spanish colony dummy; landlocked country dummy; percentage of land area in Koeppen-Geiger tropics; percentage of land area within 100 km of ice-free coast; terms of trade; and terms of trade growth.} (\citealp{Salimans}). These latter explanatory variables are expected to be truly independent of contemporaneous economic growth, and thus also are measured as five-year moving averages (as with the dependent variable). This treatment further alleviates endogeneity, which is already mitigated by the use of multi-factor error structure as discussed in Section \ref{Section2}.

\section{Empirical Results}\label{Section5}

\subsection{Choices of the Number of Factors  and the Development Index}

In Section \ref{Method}, we assume that the number of factors $r$ is known. In practice, $r$ is unknown and has to be estimated. The main tool for estimating the number of factors of large dimensional datasets is the use of information criteria. In view of the fact that there are 59 observable explanatory variables in our case, we follow \cite{Ando} to choose the number of the factors by minimizing the next criteria function: 

\begin{eqnarray}\label{PIC}
\text{PIC}(r) =\widehat{\sigma}_\varepsilon^2 \cdot\left(1 + r\cdot\frac{N+T}{NT}\log(NT)\right),
\end{eqnarray}
where $\widehat{\sigma}_\varepsilon^2= \frac{1}{NT}\sum_{i=1}^N\sum_{t=1}^T\left(y_{it}-x_{it}'\widehat{\beta}_m-\widehat{f}_t'\widehat{\gamma}_i\right)^2$, and for $\forall r$, $\widehat{\beta}_m$, $\widehat{f}_t$ and $\widehat{\gamma}_i$ are the corresponding estimates using the approach of Section \ref{Section2}. 

We now turn to the choice of the development index, $z_{it}$. In Section 2 we have specified a varying coefficient growth regression model capable of capturing parameter heterogeneity by means of a development index. Of the possible development indexes, output and human capital are believed to be the most important ones in previous studies (e.g., \citealp{DurlaufJohnson, Liu1999, Minier2007, Salimans}). Following those studies, we consider four alternative development indexes in log form: (1) initial GDP per capita, (2) initial primary schooling enrolment rate, (3) initial secondary schooling enrolment, and (4) initial higher education enrolment rate.

When choosing among the four alternative development indices, we use the in-sample root mean squared error (RMSE) which is consistent with the criterion function used in estimation. Specifically, for each development index, we first choose the number of factors and select the regressors. Then we run post selection regression as documented in Section A.1 of the supplementary Appendix A without including the weight parameters to calculate the corresponding RMSE. Table \ref{RMSE_z} presents the chosen number of factors and in-sample RMSE for each of the four development indices. In addition, we also consider the homogeneous parameter growth regression model where $\beta_0$ (i.e., the coefficients of growth determinants) is homogeneous across countries, and the results are reported in the first column of Table \ref{RMSE_z}. This table shows that the model with initial GDP per capita as the development index has the lowest in-sample RMSE and thus fits the data best.

In summary, the model with six factors and initial GDP per capita as the development index (i.e., $r=6$ and $z=\text{initial GDP per capita}$) receives the most support from the data. Hence, in what follows we concentrate on the results obtained from this model.

\subsection{Estimates of the Common Factors and Their Associated Loadings}

Figure \ref{Fig_Factors} and Figure \ref{Fig_Loading} plot the estimates of the factors identified above and their corresponding loadings respectively. The former shows that all the common factors varies considerably over time with the exception of the first factor which exhibits a relatively small amount of variation during the sample period, while the latter shows that all the factor loadings vary substantially across countries. 

We are also interested in the importance of each common factor in explaining the total variance of the error terms $e_{it}$'s of \eqref{b3}. Table \ref{CV_Factors} shows the proportion of the total variance attributed to each common factor. As the table shows, the first common factor accounts for 87.86\% percent of the total variance, and the other five factors account for 7.01\%, 1.78\%, 1.45\%, 0.61\%, 0.30\% respectively. Overall these six common factors account for 99.01\% of the total variance, indicating a fairly parsimonious description of the data.

\subsection{Estimates of the Coefficient Functions of Selected Variables}

\subsubsection{General Findings}

Our results are broadly consistent with those of \cite{Fernandez} and \cite{Martin} in that we have identified a number of robust growth determinants (variables) that are also found to be significant in the previous studies. In this sense, our results broadly support the more ``optimistic'' conclusion of \cite{Sala-i-Martin1997}, that is, some variables are important regressors for explaining cross-country growth patterns. Specifically, we have identified 31 robust growth determinants, providing evidentiary support for the canonical neoclassical growth variables; i.e., initial income, investment, and population growth, as well as macroeconomic policies, geography, institutions, religion and ethnic fractionalization. Table \ref{Tabel_Est} reports the estimates of the coefficients of each of the 31 robust growth determinants for $\ln ($initial GDP per capita$)=3.98$ (minimum), $5$, $6$, $7$, $8$, and $8.81$ (maximum), together with their associated 95\% bootstrapped confidence intervals  (CI)\footnote{Note that these confidence intervals need to be interpreted carefully. As well understood, one cannot establish the confidence intervals for the estimates under HD case unless certain transformation is further employed  (e.g., \citealp{HuangHorowitzMa, DGL2017}). However, if one regards 31 (the number of selected variables) as a relatively small number, then one can treat our regression as a LD case and employ the same bootstrap procedure  as in \cite{SuJinZhang}. In order to ensure the validity of the bootstrap procedure, stronger assumptions on the error terms are needed. For example, one can employ the martingale difference type of assumptions (see Assumption A.4 of \citealp{SuJinZhang}), or simply assume that the error terms are i.i.d. over both $i$ and $t$. Generally speaking, when the error term exhibits both cross-sectional and serial correlation, the bootstrap results are not reliable or incorrect.}. To see these coefficients more clearly, we also plot them against initial GDP per capita in Figure A.5 of the supplementary file.

Despite the similarity, there are at least three differences between the results of this study and those of the previous studies. First, our set of robust growth determinants differs from those identified in the previous studies, in spite of many overlaps between them. Specifically, some variables appear to be robust in our study but not in the previous (such as secondary school enrolment rate and terms of trade growth) or vice versa (such as primary school enrolment rate and fraction GDP of mining). There are at least three possible reasons for this difference: (1) we use a different model specification that allows for both parameter heterogeneity and cross-sectional dependence; (2) we use a different variable selection procedure (i.e., a LASSO estimator); and (3) we use a different dataset that spans a longer time period and covers a slightly different set of countries.

Second, our estimates of the coefficients of the robust growth determinants vary considerably across countries according to their level of development, while those in most previous studies are identical across countries. Specifically, we find that some coefficients have the same sign but different values across different levels of initial GDP per capita (such as civil liberty, terms of trade growth, and percentage of land area in tropics), while other coefficients not only have different signs but also different magnitudes across different levels of initial GDP per capita (such as consumption share of government, life expectancy, military expenditure, and OPEC dummy). These findings suggest that it is inappropriate to apply a growth regression with homogeneous parameters to all countries.

Third, our estimates of the coefficients of the robust growth determinants reveal some cross-country patterns not found in previous studies. Taking the coefficient of initial GDP per capita for example, we find that its estimate is positive for countries with GDP per capita between \$1,780 and \$2,117 in 1960 U.S. dollars (between \$13,166 and \$15,665 in 2014 U.S. dollars) while being negative for all other countries. This finding is in accordance with the ``middle income trap hypothesis", which refers to countries that have experienced rapid growth and thus quickly reached middle-income status but then failed to overcome that income range to further catch up to the developed countries (\citealp{Gill2007}). To give another example, our estimate of the oil reserve coefficient increases monotonically with GDP per capita and eventually becomes positive for economies with initial GDP per capital above \$2,175 in 1960 U.S. dollars (\$16,094 in 2014 U.S. dollars). This latter finding is consistent with recent studies (e.g., \citealp{Leite1999}) which suggest that in developed economies where economic institutions are generally well-developed, natural resources tend to promote economic growth; whereas in developing economies where economic institutions are generally weak, natural resources tend to hamper economic growth. We will discuss these two examples in more details below where it is appropriate. 

\subsubsection{Specific Findings}

We now analyse some of the variables that are ``significantly" related to growth in more details. Figure \ref{Fig_Examples}.1 presents our estimate of the coefficient of initial GDP per capita. This figure reveals three findings. First, this estimate is negative for most GDP per capita levels, thus being largely consistent with findings in the existing conditional convergence literature as well as previous studies that have employed model averaging methods to growth. Second, the coefficient has an inverse U-shaped relationship with initial GDP per capita. This finding is consistent with those reported by previous studies. For example, \cite{Durlauf2001} find that the coefficient of initial GDP per capita does not exhibit any sort of monotonicity with respect to level of development; \cite{Salimans} finds that the coefficient of initial GDP per capita first increases with level of development up to a point and then declines afterwards; and \cite{DurlaufJohnson} find that the coefficient of initial GDP per capita is not monotonic with respect either GDP per capita or literacy rate. Third, the coefficient is positive for countries with GDP per capita between \$319 and \$1,812 in 1960 U.S. dollars (or between \$2,554 and \$14,510 in 2014 U.S. dollars), suggesting that some of these countries have been unable to catch up with more developed countries. This finding is in line with the ``middle-income trap" hypothesis, which refers to the phenomenon of hitherto rapidly growing economies stagnating at middle-income levels and failing to graduate into the ranks of high-income countries (e.g., \citealp{Eichengreen}). In our sample South Africa and Columbia are two example countries that have never been able leave the ``middle-income range" over the entire sample period since their GDP per capita fell into this range at the beginning of the sample period (i.e., 1960).

Figure \ref{Fig_Examples}.2 shows our estimate of the coefficient of price for investment goods. Three findings emerge from this figure. First, this estimate is negative for countries with initial GDP per capita up to \$543 in 1960 U.S. dollars (or \$4,348 in 2014 U.S. dollars), suggesting that for these countries a relative low price of investment goods in the first year of each five-year period is strongly and positively related to subsequent income growth. This finding is not surprising because a low investment price stimulates investment (including investment in machinery and equipment), which further spurs economic growth (\citealp{LongSummers, LongSummers1992}). Second, the estimated coefficient falls in absolute value as initial GDP per capita increases, meaning that the marginal effect of investment price on growth is stronger for poor countries than for rich countries. This latter finding is consistent with \cite{Temple} who find that the growth-spurring effects of investment is greater for developing countries, because total investment includes machinery embodying well-established technologies and developing countries may be able to take advantage of new and old equipment because they have little of any technology. Third, for countries with initial GDP per capita above \$543 in 1960 U.S. dollars (or \$4,348 in 2014 U.S. dollars) the estimated coefficient of investment goods price is positive but insignificant, because the associated confidence intervals contain zero. This suggests that investment goods price has no growth effects for these countries. A possible reason for this latter finding is that the data from the Penn World Table is not disaggregated enough to distinguish price of equipment investment, which has strong growth effects, and price of other forms of investment, which have little growth effects (\citealp{LongSummers}). 

Figure \ref{Fig_Examples}.3 shows our estimate of the coefficient of secondary schooling enrolment. As this figure shows, this estimate is positive for most levels of initial GDP per capita. This is not surprising because secondary education is a vital part of a virtuous circle of economic growth within the context of a globalized knowledge economy. Many studies have documented that a large pool of workers with secondary education is indispensable for knowledge spillover to take place and for attracting imports of technologically advanced goods and foreign direct investment (\citealp{Borensztein, CaselliColeman}). That said, we also note that the estimated coefficient for secondary schooling is negative for middle income countries with initial GDP\ per capita between \$272 and \$1,192 in 1960 U.S. dollars (or between \$2178 and \$9,545 in 2014 U.S. dollars). This result suggests that a possible reason for the ``middle income trap" discussed above is that these middle income countries, unlike other countries, fail to take advantage of the benefits brought by secondary schooling.

We also note from Figure \ref{Fig_Examples}.4 that the estimate of the coefficient of higher education is negative for almost all countries with the exception of middle income countries. This finding is consistent with \cite{Martin} and \cite{Salimans}, both of which find that the higher schooling coefficient is negative for most of their sample countries but positive for the rest. It also parallels the concavity argument that the earnings function is concave in education, meaning that returns are higher for lower levels of education (\citealp{Psacharopoulos, PsacharopoulosPatrinos}).

Figure \ref{Fig_Examples}.5 shows our estimate of the coefficient of oil reserve. Two findings stand out from this figure. First, this estimate is negative for countries with initial GDP per capita below \$2,373 in 1960 U.S. dollars (or \$19,002 in 2014 U.S. dollars). This finding is consistent with the ``natural resources curse hypothesis" (e.g., \citealp{Sachs}) and can be explained by the rent-seeking behaviour of countries with large endowments of natural resources. Second, the oil reserve coefficient increases monotonically with GDP per capita and eventually becomes positive for economies with initial GDP per capita above \$2,373 in 1960 U.S. dollars (or \$19,002 in 2014 U.S. dollars). This latter finding accords with recent studies (e.g., \citealp{Leite1999}) on the nexus between natural resources and economic growth. Specifically, these studies suggest that the contribution of natural resources to a country's economy does not take place in isolation, but rather in the overall context of the country's economic management and institutions. It is thus the quality and competency of these policies and institutions that will determine whether natural resources can promote economic growth, or whether revenues generated by the sector might impede development. Therefore, in developed economies where economic institutions are generally well-developed, natural resources tend to promote economic growth; whereas in developing economies where economic institutions are generally weak, natural resources tend to hamper economic growth.

Figure \ref{Fig_Examples}.6 presents the estimate of the coefficient of terms of trade growth. As this figure shows, our estimate of the terms of trade growth coefficient is positive for all countries, suggesting that growth tends to be faster in countries where the rate of change of terms of trade is higher. This finding is consistent with previous studies (\citealp{Mendoza1995, Mendoza1997, Kose, Bleaney}) that find that an improvement in the terms of trade leads to higher levels of investment, and hence long-run economic growth. In addition, this figure shows that the terms of trade growth coefficient increases with GDP per capita, indicating that the marginal effect of terms of trade growth is larger in richer countries than in poorer ones. This latter finding is also consistent with previous studies (e.g., \citealp{Blattman}). Specifically, those studies argue that higher volatility in the terms of trade reduces investment and hence growth because of aversion to risk, and that rich countries with more sophisticated institutions and markets are likely to have cheaper ways to insure against price volatility than poor countries, so terms of trade instability is likely to have a smaller negative impact on rich countries.

Here we note that as in \cite{Martin}, trade openness (defined as exports plus imports as a share of GDP) is insignificant, presumably reflecting the crudity of this measure, and perhaps the distinction between opening to international trade generating a one-time step increase in income as factors are reallocated according to comparative advantage versus an ongoing growth impact associated with greater openness.

Figures \ref{Fig_Examples}.7-\ref{Fig_Examples}.9 show the estimates of the coefficients of fraction of Christian, Muslim, and Jewish respectively. These coefficients are negative at all levels of development or nearly all levels of development. This finding is consistent with that of \cite{BarroMcCleary} who find that religion works via belief, not practice. They argue that higher church attendance uses up time and resources and eventually runs into diminishing returns. The ``religion sector", as they call it, can consume more than it yields.

\section{Conclusion}\label{Section6}

A rigorous cross-country growth regression analysis should simultaneously account for three major problems identified in the literature --- variable selection, parameter heterogeneity, and cross-sectional dependence. Though these three problems have received individual attention, little or no research has sought to integrate them into a single, comprehensive framework. The purpose of this study is to fill this void by proposing a new, integrated framework that is capable of dealing with parameter heterogeneity and cross-sectional dependence, while simultaneously performing variable selection. Specifically, parameter heterogeneity is allowed for by means of a varying coefficient growth regression model, while cross-sectional dependence is introduced into the model via a multi-factor structure. For simplicity, we refer to the resulting growth regression model as the ``varying coefficient growth regression model with factor structure and sparsity". We then propose a LASSO estimator that is capable of performing variable selection on this model. In addition, we have established the associated asymptotic results for this estimator and further investigate the performance of the estimator by conducting extensive simulations.

We apply the above framework to a new dataset that covers 89 countries over the period from 1960 to 2014. We have identified 31 robust growth determinants, providing evidentiary support for the canonical neoclassical growth variables; i.e., initial income, investment, and population growth, as well as macroeconomic policies, geography, institutions, religion and ethnic fractionalization. Moreover, we find that all the coefficients of the robust growth determinants vary considerably across countries according to their level of development, which reveals some interesting cross-country patterns not found in previous studies. For example, we find that the coefficient of the initial GDP\ per capita is positive for countries with GDP per capita between \$319 and \$1,812 in 1960 U.S. dollars (or between \$2,554 and \$14,510 in 2014 U.S. dollars)\textbf{, }suggesting that some of these countries have fallen into the so-called ``middle income trap". As another example, we find that the oil reserve coefficient increases monotonically with GDP per capita and eventually becomes positive for economies with initial GDP per capita above \$2,373 in 1960 U.S. dollars (or \$19,002 in 2014 U.S. dollars), thus being consistent with recent studies that stress the role of institutions in determining how natural resources affect economic growth. 

{\footnotesize
\bibliography{Refs}
}

\newpage

\begin{landscape}

\footnotesize

\renewcommand{\arraystretch}{0.68}
\begin{longtable}{lllrr}
\caption{Definitions of All Variables in the Regression}\label{Table_Var}\\
\hline \hline
Variables & Description & Formula & Mean & Std \\ \hline      
EG & Economic growth rate & ln(rgdpo$_t$/rgdpo$_{t-1}$) & 0.0363 & 0.0649 \\
log(GPC) & log GDP per capita &   & 6.0642 & 0.9861 \\
csh\_g & Government consumption share &  & 0.2074 & 0.1163 \\
Openness & Openness measure & csh\_x + csh\_m & -0.0322 & 0.1274 \\
IP & Investment price, i.e., price level of capital formation &  & 0.4213 & 0.3220 \\
PGR & Population growth rates & ln(pop$_t$/pop$_{t-1}$) & 0.0197 & 0.0127 \\
Sch\_P & Primary school enrollment &  & 0.7396 & 0.2170 \\
Sch\_S & Secondary school enrollment &  & 0.4672 & 0.3204 \\
Sch\_H & Higher education School Enrollment &  & 0.1336 & 0.1581 \\
LE & Life Expecancy &  & 0.5932 & 0.1102 \\
PESS & Public education spending share in GDP &  & 0.0399 & 0.0296 \\
PIS & Public investment share & GFCF - GFCF\_PS & 0.0713 & 0.0501 \\
Land & Land area (sq. km / 1,000,000) &  & 0.8719 & 2.1518 \\
Exports & Percentage of Primary Export & Exports\_OM + Exports\_ARM & 0.1978 & 0.2085 \\
Mining & Fraction GDP in mining &  & 0.0733 & 0.0874 \\
Fertility & Fertility rate, total (births per woman) &  & 4.7412 & 1.9724 \\
Military & Military expenditure share in GDP &  & 0.0295 & 0.0302 \\
PCS & Public consumption share & GGFCE - PESS - Military & 0.0809 & 0.0501 \\
Malaria & Malaria prevalence: Incidence of malaria &  & 155.1712 & 245.5145 \\
 & (per 1,000 population at risk) &  &  &  \\
Inflation & Inflation rate &  & 1.3854 & 7.0448 \\
Political & Political rights &  & 4.2100 & 1.9429 \\
Civil & Civil liberties &  & 4.1496 & 1.6628 \\
Cap & Degree of capitalism &  & 3.2697 & 1.7275 \\
Trade & Terms of trade &  & 1.3475 & 2.2807 \\
Tra\_Gro & Terms of trade growth &  & 0.0073 & 0.0974 \\
Locked & Landlocked country dummy (1, yes; 0, no) &  & 0.2697 & 0.4438 \\
Ind\_Year & Time of independence &  & 1.4382 & 1.0382 \\
 & \textless{}=1914 = 0; 1915-1945 = 1; 1946-1989 = 2; \textgreater{}= 1990 = 3 &  &  &  \\
kgatr & Percentage of land area in Koeppen-Geiger tropics &  & 0.4017 & 0.4205 \\
kgptr & Percentage of population in Koeppen-Geiger tropics &  & 0.3915 & 0.4217 \\
lcr100km & Percentage of Land area within 100 km of ice-free coast &  & 0.3788 & 0.3640 \\
pop100cr & Ratio of population within 100 km of ice-free &  & 0.4520 & 0.3728 \\
 & coast/navigable river to total population &  &  &  \\
cen\_lat & latitude of country centroid &  & 0.1522 & 0.2197 \\
Bri\_Col & British colony dummy (1, yes; 0, no) &  & 0.2584 & 0.4378 \\
Spa\_Col & Spanish colony dummy (1, yes; 0, no) &  & 0.1910 & 0.3931 \\
Oil\_OPEC & Oil-producing country dummy (1, yes; 0, no) &  & 0.0674 & 0.2508 \\
Gas & proved reserves (cubic meters / 10\textasciicircum{}12) &  & 1.3789 & 6.1997 \\
Oil & proved reserves (bbl / 10\textasciicircum{}9) &  & 4.5521 & 19.1378 \\
Chris & Percentage of Christian &  & 0.5369 & 0.3807 \\
Mus & Percentage of Muslim &  & 0.3046 & 0.3825 \\
Hin & Percentage of Hindu &  & 0.0263 & 0.1211 \\
Bud & Percentage of Buddhist &  & 0.0410 & 0.1541 \\
Fol & Percentage of Folk religion &  & 0.0284 & 0.0628 \\
Oth & Percentage of other religion &  & 0.0037 & 0.0064 \\
Jew & Percentage of Jewish &  & 0.0019 & 0.0027 \\
GS & Government spending share of GDP &  & 0.1501 & 0.0686 \\
Distortion & Real exchange rate distortions &  & 129.6824 & 35.8479 \\
OO & Outward orientation &  & -2.7398 & 0.7542 \\
SIL & Ethnolinguistic fractionalization &  & 0.4886 & 0.3127 \\
ESP & English-speaking population in percentage &  & 0.1762 & 0.2692 \\
EA & East Asian dummy &  & 0.0225 & 0.1482 \\
AF & African dummy &  & 0.4270 & 0.4947 \\
EU & European dummy &  & 0.1124 & 0.3158 \\
LA & Latin American dummy &  & 0.1573 & 0.3641 \\
WarFrac & Fraction spent in war (1960-2014) &  & 0.3265 & 0.4343 \\
NoWars & No. of war participation (1960-2014) &  & 0.8028 & 1.2609 \\
Coup & coups d'etat and coup attempts within (1960-2014) &  & 0.1870 & 0.4964 \\
Revolution & Number of revolutions (1960-2014) &  & 0.1941 & 0.5038 \\
Pop\_Dens & Population Density/1000 &  & 0.0812 & 0.1135 \\
WorkIR & Growth rate of work force & ln(WP$_t$/WP$_{t-1}$) & 0.0210 & 0.0132\\ \hline \hline
\multicolumn{5}{l}{\scriptsize rgdpo --- Size of economy (GDP in million)}\\
\multicolumn{5}{l}{\scriptsize pop --- Population (in million)}\\
\multicolumn{5}{l}{\scriptsize csh\_x --- Share of merchandise exports}\\
\multicolumn{5}{l}{\scriptsize csh\_m --- Share of merchandise imports}\\
\multicolumn{5}{l}{\scriptsize WP ---  Fraction population of work force (1-A65-U15)} \\
\multicolumn{5}{l}{\scriptsize A65 --- Fraction population over 65 years old} \\
\multicolumn{5}{l}{\scriptsize U15 --- Fraction population under 15 years old} \\
\multicolumn{5}{l}{\scriptsize GFCF --- Gross fixed capital formation} \\
\multicolumn{5}{l}{\scriptsize GFCF\_PS --- Gross fixed capital formation, private sector} \\
\multicolumn{5}{l}{\scriptsize Exports\_OM --- Percentage of Ores and metals exports}  \\
\multicolumn{5}{l}{\scriptsize Exports\_ARM --- Percentage of Agricultural raw materials exports}  \\
\multicolumn{5}{l}{\scriptsize GGFCE --- General government final consumption expenditure share in GDP}
\end{longtable}
\end{landscape}

\begin{landscape}
\renewcommand{\arraystretch}{0.71}\small
\begin{longtable}{llllll}
\caption{Sample Countries and Their Associated ISO 3166-1 alpha-3 Codes}\label{Tabel_Country}\\ 
\hline \hline
AGO & Angola & HND & Honduras & PAK & Pakistan \\
ALB & Albania & HRV & Croatia & PAN & Panama \\
ARM & Armenia & HTI & Haiti & PER & Peru \\
AZE & Azerbaijan & IND & India & PHL & Philippines \\
BDI & Burundi & IRN & Iran, Islamic Republic of & POL & Poland \\
BEN & Benin & JAM & Jamaica & PRY & Paraguay \\
BFA & Burkina Faso & JOR & Jordan & RUS & Russian Federation \\
BGD & Bangladesh & JPN & Japan & RWA & Rwanda \\
BGR & Bulgaria & KAZ & Kazakhstan & SDN & Sudan \\
BLR & Belarus & KEN & Kenya & SEN & Senegal \\
BRA & Brazil & KGZ & Kyrgyzstan & SLE & Sierra Leone \\
BWA & Botswana & KHM & Cambodia & SLV & El Salvador \\
CAF & Central African Republic & LAO & Lao People's Democratic Republic & SWZ & Swaziland \\
CIV & Côte d'Ivoire & LBN & Lebanon & SYR & Syrian Arab Republic \\
CMR & Cameroon & LKA & Sri Lanka & TCD & Chad \\
COG & Congo & LSO & Lesotho & TGO & Togo \\
COL & Colombia & MDA & Moldova, Republic of & THA & Thailand \\
DOM & Dominican Republic & MDG & Madagascar & TTO & Trinidad and Tobago \\
DZA & Algeria & MEX & Mexico & TUN & Tunisia \\
ECU & Ecuador & MKD & Macedonia & TUR & Turkey \\
EGY & Egypt & MLI & Mali & TZA & Tanzania, United Republic of \\
ETH & Ethiopia & MNG & Mongolia & UGA & Uganda \\
GAB & Gabon & MOZ & Mozambique & UKR & Ukraine \\
GBR & United Kingdom & MWI & Malawi & URY & Uruguay \\
GEO & Georgia & MYS & Malaysia & USA & United States \\
GHA & Ghana & NAM & Namibia & VEN & Venezuela, Bolivarian Republic of \\
GIN & Guinea & NER & Niger & YEM & Yemen \\
GMB & Gambia & NIC & Nicaragua & ZAF & South Africa \\
GNB & Guinea-Bissau & NPL & Nepal & ZWE & Zimbabwe \\
GTM & Guatemala & OMN & Oman &  & \\ \hline \hline
\end{longtable}
\end{landscape}

\begin{table}[h] \small
\caption{Comparison among Alternative Development Indexes ($z$)}\label{RMSE_z}
\begin{tabular}{lcrrrr}
\hline\hline
               & \multirow{2}{*}{Parametric} & \multicolumn{4}{c}{$z$ of Varying Coefficient} \\ \cline{3-6} 
               &                             & $\ln$(GPC)  & School\_P  & School\_S  & School\_H  \\ \cline{2-6} 
RMSE           & \multicolumn{1}{r}{0.022}   & 0.017    & 0.019     & 0.027     & 0.020     \\
No. of factors & \multicolumn{1}{r}{6}       & 6        & 6         & 3         & 5        \\
\hline\hline
\end{tabular}

\bigskip
\bigskip

\caption{Cumulative Variation of the Residuals Explained by the Factors}\label{CV_Factors}
\begin{tabular}{lrrrrrr}
\hline \hline
No. Factors & 1 & 2 & 3 & 4 & 5 & 6 \\
Cumulative Variation& 87.86\% & 94.87\% & 96.65\% & 98.10\% & 98.71\% & 99.01\% \\
\hline \hline
\end{tabular}
\end{table}

\begin{landscape}
\small
\renewcommand{\arraystretch}{0.71}
\begin{longtable}{lcccccccccccc}
\caption{Estimates of Coefficients at log(GPC)=3.98, 5, 6, 7, 8 and 8.81}\label{Tabel_Est}\\ 
\hline \hline
 & \multicolumn{2}{c}{ log(GPC)=3.98}  & \multicolumn{2}{c}{log(GPC)=5}  & \multicolumn{2}{c}{log(GPC)=6}  &\multicolumn{2}{c}{log(GPC)=7}  & \multicolumn{2}{c}{log(GPC)=8}  & \multicolumn{2}{c}{log(GPC)=8.81}   \\ \cline{2-13}
log(GPC) & \multicolumn{2}{c}{-0.0766}  &  \multicolumn{2}{c}{-0.0447}  &  \multicolumn{2}{c}{0.0117}  &  \multicolumn{2}{c}{0.0255}  &  \multicolumn{2}{c}{-0.0531}  & \multicolumn{2}{c}{-0.2055}  \\
 & \multicolumn{2}{c}{(-0.1145, -0.0429)} & \multicolumn{2}{c}{(-0.0618, -0.0305)} & \multicolumn{2}{c}{(-0.0038, 0.0249)} & \multicolumn{2}{c}{(0.0080, 0.0469)} & \multicolumn{2}{c}{(-0.0886, -0.0034)} & \multicolumn{2}{c}{(-0.2883, -0.0898)} \\
csh\_g & \multicolumn{2}{c}{0.3341} & \multicolumn{2}{c}{0.1483}  & \multicolumn{2}{c}{-0.0587}  & \multicolumn{2}{c}{-0.2143}  & \multicolumn{2}{c}{-0.2619}  & \multicolumn{2}{c}{-0.1964}  \\
 & \multicolumn{2}{c}{(0.1077, 0.5534)} & \multicolumn{2}{c}{(0.0706, 0.2248)} & \multicolumn{2}{c}{(-0.1110, -0.0107)} & \multicolumn{2}{c}{(-0.3141, -0.1454)} & \multicolumn{2}{c}{(-0.5782, 0.0082)} & \multicolumn{2}{c}{(-0.9416, 0.4733)} \\
IP & \multicolumn{2}{c}{-0.1284}  & \multicolumn{2}{c}{-0.0545}  & \multicolumn{2}{c}{-0.0093}  & \multicolumn{2}{c}{0.0160}  & \multicolumn{2}{c}{0.0282} & \multicolumn{2}{c}{0.0325}  \\
 & \multicolumn{2}{c}{(-0.1971, -0.0621)} & \multicolumn{2}{c}{(-0.0686, -0.0422)} & \multicolumn{2}{c}{(-0.0191, 0.0059)} & \multicolumn{2}{c}{(-0.0046, 0.0406)} & \multicolumn{2}{c}{(-0.0146, 0.0638)} & \multicolumn{2}{c}{(-0.0895, 0.1209)} \\
PGR & \multicolumn{2}{c}{ 3.2821}  & \multicolumn{2}{c}{0.5353}  & \multicolumn{2}{c}{0.8485}  & \multicolumn{2}{c}{1.5689}  & \multicolumn{2}{c}{0.6502}  & \multicolumn{2}{c}{-2.2853}  \\
 & \multicolumn{2}{c}{(1.7325, 4.7072)} & \multicolumn{2}{c}{(0.0426, 0.9486)} & \multicolumn{2}{c}{(0.4054, 1.2542)} & \multicolumn{2}{c}{(1.1371, 1.9781)} & \multicolumn{2}{c}{(-0.1228, 1.3620)} & \multicolumn{2}{c}{(-3.5441, -0.7206)} \\
School\_S & \multicolumn{2}{c}{0.1322}  & \multicolumn{2}{c}{0.0823}  & \multicolumn{2}{c}{-0.0420}  & \multicolumn{2}{c}{-0.0147}  & \multicolumn{2}{c}{0.3303}  & \multicolumn{2}{c}{0.9119}  \\
 & \multicolumn{2}{c}{(-0.1505, 0.4845)} & \multicolumn{2}{c}{(0.0108, 0.1495)} & \multicolumn{2}{c}{(-0.0780, -0.0081)} & \multicolumn{2}{c}{(-0.0548, 0.0177)} & \multicolumn{2}{c}{(0.1989, 0.4201)} & \multicolumn{2}{c}{(0.5495, 1.1719)} \\
School\_H & \multicolumn{2}{c}{-1.5869}  & \multicolumn{2}{c}{-0.1175}  & \multicolumn{2}{c}{0.2220}  & \multicolumn{2}{c}{0.0326}  & \multicolumn{2}{c}{-0.2066}  & \multicolumn{2}{c}{-0.1808}  \\
 & \multicolumn{2}{c}{(-2.4454, -0.7631)} & \multicolumn{2}{c}{(-0.3214, 0.1119)} & \multicolumn{2}{c}{(0.1562, 0.3020)} & \multicolumn{2}{c}{(-0.0157, 0.0921)} & \multicolumn{2}{c}{(-0.2810, -0.1039)} & \multicolumn{2}{c}{(-0.3986, 0.1401)} \\
LE & \multicolumn{2}{c}{0.3148}  & \multicolumn{2}{c}{0.2935}  & \multicolumn{2}{c}{-0.0597}  & \multicolumn{2}{c}{-0.2370}  & \multicolumn{2}{c}{0.1455}  & \multicolumn{2}{c}{1.0373}  \\
 & \multicolumn{2}{c}{(-0.0171, 0.6403)} & \multicolumn{2}{c}{(0.1811, 0.3938)} & \multicolumn{2}{c}{(-0.1739, 0.0389)} & \multicolumn{2}{c}{(-0.3918, -0.0990)} & \multicolumn{2}{c}{(-0.2875, 0.5411)} & \multicolumn{2}{c}{(0.0000, 1.9922)} \\
PESS & \multicolumn{2}{c}{-1.2399}  & \multicolumn{2}{c}{-0.5737}  & \multicolumn{2}{c}{-0.2476}  & \multicolumn{2}{c}{-0.0158}  & \multicolumn{2}{c}{0.3070}  & \multicolumn{2}{c}{0.7260}  \\
 & \multicolumn{2}{c}{(-2.4950, 0.0107)} & \multicolumn{2}{c}{(-0.7969, -0.3513)} & \multicolumn{2}{c}{(-0.4210, -0.0944)} & \multicolumn{2}{c}{(-0.2163, 0.1716)} & \multicolumn{2}{c}{(-0.6249, 1.3302)} & \multicolumn{2}{c}{(-1.5776, 3.4530)} \\
Military & \multicolumn{2}{c}{-1.1748}  & \multicolumn{2}{c}{-0.1436}  & \multicolumn{2}{c}{-0.0428}  & \multicolumn{2}{c}{-0.0020}  & \multicolumn{2}{c}{0.6417}  & \multicolumn{2}{c}{1.9197}  \\
 & \multicolumn{2}{c}{(-2.4253, -0.0639)} & \multicolumn{2}{c}{(-0.3588, 0.1047)} & \multicolumn{2}{c}{(-0.1667, 0.0719)} & \multicolumn{2}{c}{(-0.2609, 0.1609)} & \multicolumn{2}{c}{(0.0134, 1.1085)} & \multicolumn{2}{c}{(0.1982, 3.4369)} \\
Inflation & \multicolumn{2}{c}{-0.0055}  & \multicolumn{2}{c}{0.0006}  & \multicolumn{2}{c}{-0.0011}  & \multicolumn{2}{c}{-0.0031}  & \multicolumn{2}{c}{0.0007}  & \multicolumn{2}{c}{0.0107}  \\
 & \multicolumn{2}{c}{(-0.0094, -0.0012)} & \multicolumn{2}{c}{(0.0000, 0.0012)} & \multicolumn{2}{c}{(-0.0016, -0.0005)} & \multicolumn{2}{c}{(-0.0039, -0.0023)} & \multicolumn{2}{c}{(-0.0043, 0.0048)} & \multicolumn{2}{c}{(-0.0020, 0.0209)} \\
Civil & \multicolumn{2}{c}{0.0183}  & \multicolumn{2}{c}{0.0050}  & \multicolumn{2}{c}{-0.0004}  & \multicolumn{2}{c}{0.0033}  & \multicolumn{2}{c}{0.0166}  & \multicolumn{2}{c}{0.0340}  \\
 & \multicolumn{2}{c}{(0.0048, 0.0308)} & \multicolumn{2}{c}{(0.0011, 0.0085)} & \multicolumn{2}{c}{(-0.0027, 0.0024)} & \multicolumn{2}{c}{(0.0000, 0.0064)} & \multicolumn{2}{c}{(0.0027, 0.0282)} & \multicolumn{2}{c}{(-0.0036, 0.0661)} \\
Tra\_Gro & \multicolumn{2}{c}{0.0009}  & \multicolumn{2}{c}{0.0304}  & \multicolumn{2}{c}{0.0048}  & \multicolumn{2}{c}{0.0018}  & \multicolumn{2}{c}{0.0796}  & \multicolumn{2}{c}{0.2279}  \\
 & \multicolumn{2}{c}{(-0.1138, 0.1100)} & \multicolumn{2}{c}{(0.0117, 0.0458)} & \multicolumn{2}{c}{(-0.0162, 0.0252)} & \multicolumn{2}{c}{(-0.0248, 0.0351)} & \multicolumn{2}{c}{(-0.0062, 0.1727)} & \multicolumn{2}{c}{(-0.0124, 0.4799)} \\
kgatr & \multicolumn{2}{c}{-0.3099}  & \multicolumn{2}{c}{-0.0815}  & \multicolumn{2}{c}{-0.0763}  & \multicolumn{2}{c}{-0.1422}  & \multicolumn{2}{c}{-0.1597}  & \multicolumn{2}{c}{-0.0776}  \\
 & \multicolumn{2}{c}{(-0.4367, -0.1693)} & \multicolumn{2}{c}{(-0.1281, -0.0391)} & \multicolumn{2}{c}{(-0.1161, -0.0447)} & \multicolumn{2}{c}{(-0.1887, -0.0969)} & \multicolumn{2}{c}{(-0.2602, -0.0333)} & \multicolumn{2}{c}{(-0.3270, 0.2396)} \\
lcr100km & \multicolumn{2}{c}{0.6793}  & \multicolumn{2}{c}{0.1930}  & \multicolumn{2}{c}{0.0511}  & \multicolumn{2}{c}{-0.0621}  & \multicolumn{2}{c}{-0.3850}  & \multicolumn{2}{c}{-0.9133}  \\
 & \multicolumn{2}{c}{(0.5005, 0.8498)} & \multicolumn{2}{c}{(0.1396, 0.2461)} & \multicolumn{2}{c}{(0.0127, 0.0940)} & \multicolumn{2}{c}{(-0.1305, -0.0005)} & \multicolumn{2}{c}{(-0.5977, -0.2168)} & \multicolumn{2}{c}{(-1.4617, -0.5189)} \\
cen\_lat & \multicolumn{2}{c}{0.5894}  & \multicolumn{2}{c}{0.0293}  & \multicolumn{2}{c}{0.0360}  & \multicolumn{2}{c}{0.1693}  & \multicolumn{2}{c}{0.0873}  & \multicolumn{2}{c}{-0.3058}  \\
 & \multicolumn{2}{c}{(0.3011, 0.8198)} & \multicolumn{2}{c}{(-0.0723, 0.1294)} & \multicolumn{2}{c}{(-0.0315, 0.1114)} & \multicolumn{2}{c}{(0.0985, 0.2655)} & \multicolumn{2}{c}{(-0.0526, 0.2459)} & \multicolumn{2}{c}{(-0.6344, 0.0789)} \\
Spa\_Col & \multicolumn{2}{c}{-0.5174} & \multicolumn{2}{c}{-0.1564}  & \multicolumn{2}{c}{-0.0204}  & \multicolumn{2}{c}{0.0548}  & \multicolumn{2}{c}{0.1945}  & \multicolumn{2}{c}{0.4167}  \\
 & \multicolumn{2}{c}{(-0.6748, -0.3222)} & \multicolumn{2}{c}{(-0.2111, -0.1080)} & \multicolumn{2}{c}{(-0.0503, 0.0091)} & \multicolumn{2}{c}{(0.0171, 0.0950)} & \multicolumn{2}{c}{(0.1040, 0.3131)} & \multicolumn{2}{c}{(0.1867, 0.6800)} \\
Oil\_OPEC & \multicolumn{2}{c}{-0.1254}  & \multicolumn{2}{c}{-0.0460}  & \multicolumn{2}{c}{-0.0947}  & \multicolumn{2}{c}{-0.0815}  & \multicolumn{2}{c}{0.1358}  & \multicolumn{2}{c}{0.5247}  \\
 & \multicolumn{2}{c}{(-1.0835, 0.6246)} & \multicolumn{2}{c}{(-0.2325, 0.1181)} & \multicolumn{2}{c}{(-0.1496, -0.0423)} & \multicolumn{2}{c}{(-0.1383, -0.0344)} & \multicolumn{2}{c}{(-0.0124, 0.2413)} & \multicolumn{2}{c}{(0.1090, 0.8785)} \\
Oil & \multicolumn{2}{c}{-0.0100} & \multicolumn{2}{c}{-0.0021}  & \multicolumn{2}{c}{-0.0008}  & \multicolumn{2}{c}{-0.0011}  & \multicolumn{2}{c}{0.0008}  & \multicolumn{2}{c}{0.0057}  \\
 & \multicolumn{2}{c}{(-0.0217, 0.0018)} & \multicolumn{2}{c}{(-0.0051, 0.0009)} & \multicolumn{2}{c}{(-0.0017, 0.0000)} & \multicolumn{2}{c}{(-0.0022, -0.0001)} & \multicolumn{2}{c}{(-0.0010, 0.0023)} & \multicolumn{2}{c}{(0.0004, 0.0099)} \\
Chris & \multicolumn{2}{c}{0.2689}  & \multicolumn{2}{c}{0.1112}  & \multicolumn{2}{c}{-0.0056}  & \multicolumn{2}{c}{-0.1262}  & \multicolumn{2}{c}{-0.2818}  & \multicolumn{2}{c}{-0.4469}  \\
 & \multicolumn{2}{c}{(0.1164, 0.4267)} & \multicolumn{2}{c}{(0.0650, 0.1711)} & \multicolumn{2}{c}{(-0.0535, 0.0461)} & \multicolumn{2}{c}{(-0.2001, -0.0629)} & \multicolumn{2}{c}{(-0.4824, -0.1333)} & \multicolumn{2}{c}{(-0.8908, -0.0694)} \\
Mus & \multicolumn{2}{c}{-0.0736}  & \multicolumn{2}{c}{0.0130}  & \multicolumn{2}{c}{-0.0292}  & \multicolumn{2}{c}{-0.1685}  & \multicolumn{2}{c}{-0.3744}  & \multicolumn{2}{c}{-0.5692}  \\
 & \multicolumn{2}{c}{(-0.1864, 0.0735)} & \multicolumn{2}{c}{(-0.0383, 0.0599)} & \multicolumn{2}{c}{(-0.0774, 0.0195)} & \multicolumn{2}{c}{(-0.2344, -0.1004)} & \multicolumn{2}{c}{(-0.5710, -0.2335)} & \multicolumn{2}{c}{(-1.0085, -0.2474)} \\
Oth & \multicolumn{2}{c}{-3.7386}  & \multicolumn{2}{c}{0.6890}  & \multicolumn{2}{c}{-1.1133}  & \multicolumn{2}{c}{2.9016}  & \multicolumn{2}{c}{21.6126}  & \multicolumn{2}{c}{51.3537}  \\
 & \multicolumn{2}{c}{(-13.0771, 4.1928)} & \multicolumn{2}{c}{(-1.3860, 2.4081)} & \multicolumn{2}{c}{(-3.0584, 0.7782)} & \multicolumn{2}{c}{(0.2919, 6.3889)} & \multicolumn{2}{c}{(11.6269, 32.3911)} & \multicolumn{2}{c}{(27.3474, 75.2005)} \\
Jew & \multicolumn{2}{c}{-32.4452}  & \multicolumn{2}{c}{-10.5477}  & \multicolumn{2}{c}{-2.0163}  & \multicolumn{2}{c}{-0.9826}  & \multicolumn{2}{c}{-2.7502}  & \multicolumn{2}{c}{-3.6114}  \\
 & \multicolumn{2}{c}{(-55.7335, -10.0679)} & \multicolumn{2}{c}{(-18.6022, -3.8609)} & \multicolumn{2}{c}{(-7.6269, 3.6243)} & \multicolumn{2}{c}{(-6.9105, 4.0461)} & \multicolumn{2}{c}{(-12.1471, 5.8912)} & \multicolumn{2}{c}{(-27.3730, 18.2231)} \\
GS & \multicolumn{2}{c}{-0.2516}  & \multicolumn{2}{c}{-0.2079}  & \multicolumn{2}{c}{-0.0476}  & \multicolumn{2}{c}{-0.0738}  & \multicolumn{2}{c}{-0.5101}  & \multicolumn{2}{c}{-1.2583}  \\
 & \multicolumn{2}{c}{(-0.5709, 0.0529)} & \multicolumn{2}{c}{(-0.3047, -0.1254)} & \multicolumn{2}{c}{(-0.1142, 0.0215)} & \multicolumn{2}{c}{(-0.1676, 0.0172)} & \multicolumn{2}{c}{(-0.8230, -0.1881)} & \multicolumn{2}{c}{(-2.1023, -0.4396)} \\
Distortion & \multicolumn{2}{c}{0.0231}  & \multicolumn{2}{c}{0.0884} & \multicolumn{2}{c}{0.1085}  & \multicolumn{2}{c}{0.0482}  & \multicolumn{2}{c}{-0.1148}  & \multicolumn{2}{c}{-0.3280}  \\
 & \multicolumn{2}{c}{(-0.0449, 0.1048)} & \multicolumn{2}{c}{(0.0580, 0.1187)} & \multicolumn{2}{c}{(0.0701, 0.1435)} & \multicolumn{2}{c}{(-0.0064, 0.1025)} & \multicolumn{2}{c}{(-0.2793, 0.0061)} & \multicolumn{2}{c}{(-0.6718, -0.0148)} \\
OO & \multicolumn{2}{c}{1.0569}  & \multicolumn{2}{c}{4.1811}  & \multicolumn{2}{c}{5.1496}  & \multicolumn{2}{c}{2.3016}  & \multicolumn{2}{c}{-5.4384}  & \multicolumn{2}{c}{-15.5641}  \\
 & \multicolumn{2}{c}{(-2.1588, 4.9579)} & \multicolumn{2}{c}{(2.7557, 5.6164)} & \multicolumn{2}{c}{(3.3263, 6.8047)} & \multicolumn{2}{c}{(-0.2707, 4.8851)} & \multicolumn{2}{c}{(-13.1848, 0.3040)} & \multicolumn{2}{c}{(-31.9253, -0.7429)} \\
ESP & \multicolumn{2}{c}{0.3819} & \multicolumn{2}{c}{0.1039}  & \multicolumn{2}{c}{0.0581}  & \multicolumn{2}{c}{0.1010}  & \multicolumn{2}{c}{0.1193}  & \multicolumn{2}{c}{0.0577}  \\
 & \multicolumn{2}{c}{(0.2151, 0.5483)} & \multicolumn{2}{c}{(0.0575, 0.1465)} & \multicolumn{2}{c}{(0.0214, 0.0975)} & \multicolumn{2}{c}{(0.0494, 0.1631)} & \multicolumn{2}{c}{(-0.0059, 0.2319)} & \multicolumn{2}{c}{(-0.2260, 0.2967)} \\
EA & \multicolumn{2}{c}{0.2924}  & \multicolumn{2}{c}{0.0841}  & \multicolumn{2}{c}{0.0263}  & \multicolumn{2}{c}{-0.1256}  & \multicolumn{2}{c}{-0.5515}  & \multicolumn{2}{c}{-1.1753}  \\
 & \multicolumn{2}{c}{(-0.1574, 0.7975)} & \multicolumn{2}{c}{(-0.0589, 0.2081)} & \multicolumn{2}{c}{(-0.0668, 0.1185)} & \multicolumn{2}{c}{(-0.2975, 0.0293)} & \multicolumn{2}{c}{(-1.0307, -0.1357)} & \multicolumn{2}{c}{(-2.2484, -0.2191)} \\
EU & \multicolumn{2}{c}{-1.0237}  & \multicolumn{2}{c}{-0.2206}  & \multicolumn{2}{c}{-0.0275}  & \multicolumn{2}{c}{0.0738}  & \multicolumn{2}{c}{0.4780}  & \multicolumn{2}{c}{1.2174}  \\
 & \multicolumn{2}{c}{(-1.7313, -0.3070)} & \multicolumn{2}{c}{(-0.3896, -0.0438)} & \multicolumn{2}{c}{(-0.0797, 0.0197)} & \multicolumn{2}{c}{(0.0169, 0.1225)} & \multicolumn{2}{c}{(0.3182, 0.6845)} & \multicolumn{2}{c}{(0.7527, 1.7964)} \\
WarFrac & \multicolumn{2}{c}{-0.0044}  & \multicolumn{2}{c}{-0.0035}  & \multicolumn{2}{c}{-0.0120}  & \multicolumn{2}{c}{-0.0130}  & \multicolumn{2}{c}{0.0058}  & \multicolumn{2}{c}{0.0411} \\
 & \multicolumn{2}{c}{(-0.0372, 0.0351)} & \multicolumn{2}{c}{(-0.0121, 0.0072)} & \multicolumn{2}{c}{(-0.0179, -0.0040)} & \multicolumn{2}{c}{(-0.0208, -0.0036)} & \multicolumn{2}{c}{(-0.0157, 0.0235)} & \multicolumn{2}{c}{(-0.0227, 0.0882)} \\
Coup & \multicolumn{2}{c}{-0.0190}  & \multicolumn{2}{c}{-0.0004}  & \multicolumn{2}{c}{-0.0028}  & \multicolumn{2}{c}{-0.0070}  & \multicolumn{2}{c}{0.0016}  & \multicolumn{2}{c}{0.0251}  \\
 & \multicolumn{2}{c}{(-0.0287, -0.0089)} & \multicolumn{2}{c}{(-0.0031, 0.0020)} & \multicolumn{2}{c}{(-0.0051, -0.0005)} & \multicolumn{2}{c}{(-0.0102, -0.0038)} & \multicolumn{2}{c}{(-0.0086, 0.0113)} & \multicolumn{2}{c}{(-0.0005, 0.0502)} \\
Revolution & \multicolumn{2}{c}{-0.0052}  & \multicolumn{2}{c}{0.0044}  & \multicolumn{2}{c}{-0.0016}  & \multicolumn{2}{c}{-0.0066}  & \multicolumn{2}{c}{0.0021}  & \multicolumn{2}{c}{0.0252}  \\
 & \multicolumn{2}{c}{(-0.0198, 0.0090)} & \multicolumn{2}{c}{(0.0007, 0.0075)} & \multicolumn{2}{c}{(-0.0042, 0.0011)} & \multicolumn{2}{c}{(-0.0106, -0.0020)} & \multicolumn{2}{c}{(-0.0054, 0.0084)} & \multicolumn{2}{c}{(-0.0034, 0.0464)} \\
 \hline \hline 
\end{longtable}
\end{landscape}

\begin{landscape}

\begin{figure}[H]\caption{Estimates of Common Factors} \label{Fig_Factors}
\centering
\hspace*{-2cm}\includegraphics[scale=0.67]{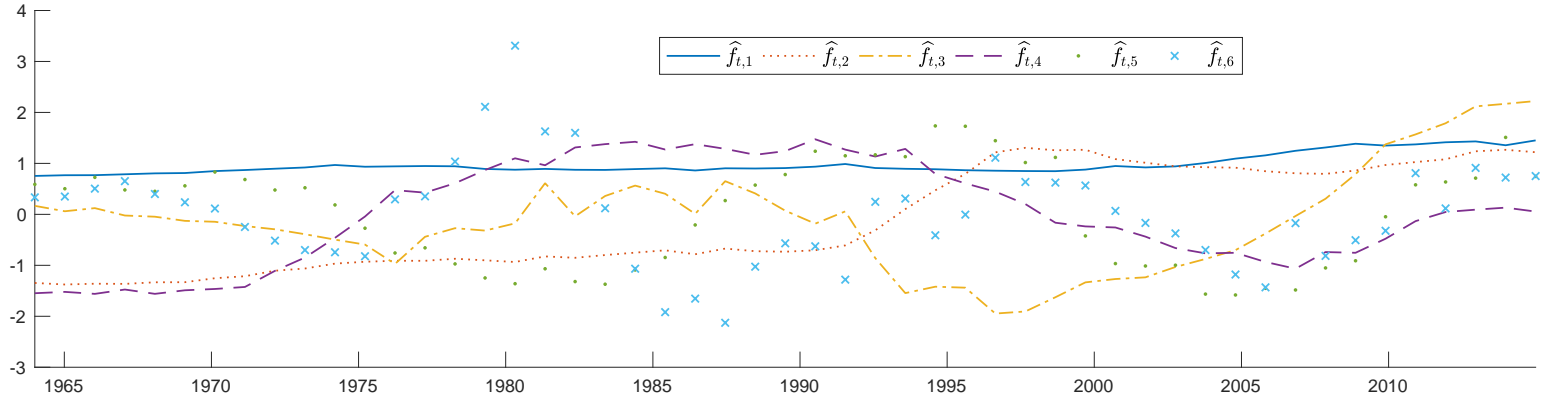}
\\$\widehat{f}_{t,j}$ stands for the estimate of the $j^{th}$ factor, where $j=1\ldots,6$.
\end{figure}

\bigskip\bigskip

\begin{figure}[H] \caption{Estimates of Factor Loadings}\label{Fig_Loading}
\centering
\hspace*{-2cm} \includegraphics[scale=0.67]{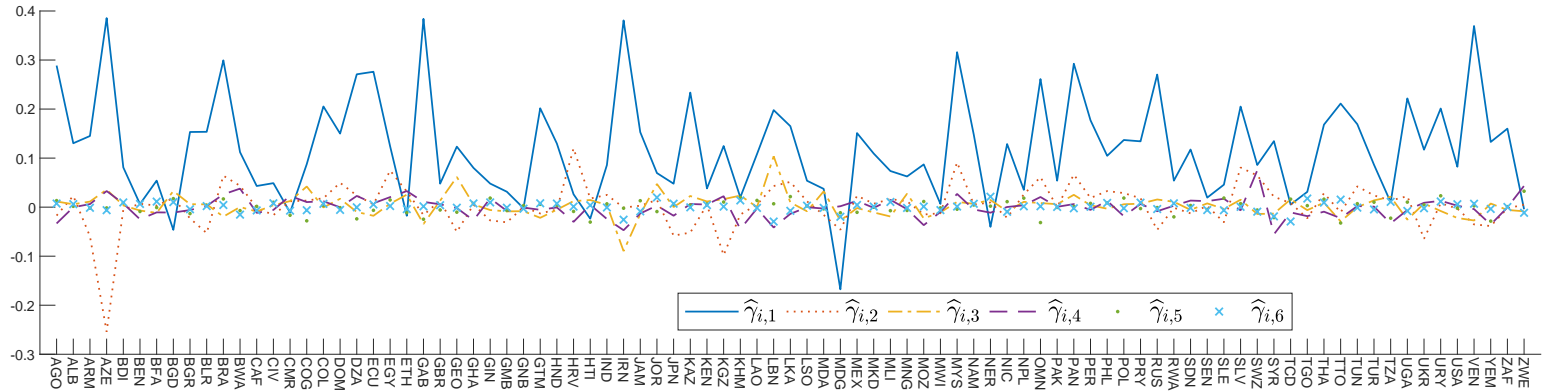}
\\$\widehat{\gamma}_{i,j}$ stands for the estimate of the $j^{th}$ factor loading, where $j=1\ldots,6$.
\end{figure} 
\end{landscape}

\begin{figure}[h]\caption{Estimates of Selected Coefficient Functions} \label{Fig_Examples}
\centering
\hspace*{-1cm}  \includegraphics[scale=0.45]{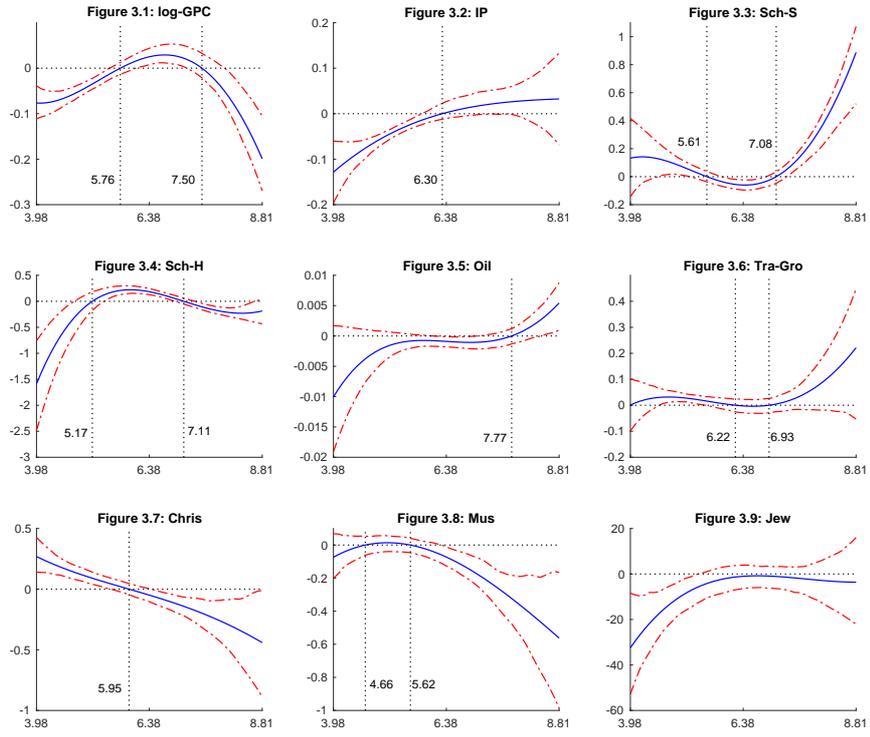}
\end{figure}

\newpage
\setcounter{page}{1}
\renewcommand{\theequation}{A.\arabic{equation}}
\renewcommand{\thesection}{A.\arabic{section}}
\renewcommand{\thefigure}{A.\arabic{figure}}
\renewcommand{\thetable}{A.\arabic{table}}
\renewcommand{\thelemma}{A.\arabic{lemma}}

\setcounter{equation}{0}
\setcounter{lemma}{0}
\setcounter{section}{0}
\setcounter{table}{0}
\setcounter{figure}{0}
\numberwithin{equation}{section}

\begin{center}
{\Large \bf Supplementary Appendix A to \\``An Integrated Panel Data Approach to \\Modelling Economic Growth"}

\medskip

{\sc Guohua Feng$^{\ast}$, Jiti Gao$^\sharp$ and Bin Peng$^{\dag}$ }
\medskip

$^{\ast}$University of North Texas, $^\sharp$Monash University and $^{\dag}$University of Bath
\end{center}

{\small 

Appendix A is divided into five sections. Section \ref{SectionA2} provides the numerical algorithm. Section \ref{Simulation} examines the asymptotic results of Section 3 through several simulations. Section \ref{SectionA4} presents the preliminary lemmas and the proofs of the main theorems. Section \ref{TimeTrend} explains why our method can partially solve the issue of time trend that has found limited attention in the empirical growth literature. In Section \ref{SectionA1}, we provide auxiliary tables and figures of the empirical study.

Recall that in the main text, we have let $\xi_{NT} = \min\{N, T \}$,  $\| \cdot\|_{\textrm{sp}}$ be the spectral norm of a matrix, and $\lfloor a\rfloor$ stand for the largest integer part of a real number $a$. Here we further define some notations, which will be used throughout this file. Let $\phi_i^*[\beta^*]=({x_{i1}^*} '\beta^*(z_{i1}),\ldots, {x_{iT}^*}'\beta^*(z_{iT}))'$, and $\phi_i^\dagger[\beta^\dagger]=({x_{i1}^\dagger} '\beta^\dagger(z_{i1}),\ldots, {x_{iT}^\dagger}'\beta^\dagger(z_{iT}))' $, where $\beta^*(\cdot)$ and $\beta^\dagger(\cdot)$ are $p^*\times 1$ and $(p-p^*)\times 1$ respectively. Moreover, $\textrm{diag}\{A_1,\ldots,A_k\}$ means constructing block diagonal matrix from matrices (or scalars) $A_1,\ldots,A_k$.

\section{Numerical Implementation}\label{SectionA2}

The following procedure essentially combines two algorithms discussed in \cite{Bai} and \cite{WangXia}  together. For each given $\lambda = (\lambda_1,\ldots,\lambda_p)'$, the estimates can be obtained using the following iteration procedure. Let $\widehat{C}_{\beta}^{(n)}$ and $\widehat{F}^{(n)}$ be the estimates obtained from the $n^{th}\ge 1$ iteration. Then, for the $(n+1)^{th}$ iteration, the estimates are obtained as

\begin{eqnarray*}
&&\text{Sub-step 1:} \quad\vect( \widehat{C}_\beta^{(n+1)})= \left(\sum_{i=1}^N \mathcal{Z}_i' M_{\widehat{F}^{(n)}} \mathcal{Z}_i+ \frac{D_{m,p}^{(n)}}{2}\right)^{-1} \sum_{i=1}^N  \mathcal{Z}_i' M_{\widehat{F}^{(n)}} Y_i,\nonumber\\
&&\text{Sub-step 2:} \quad\frac{1}{NT}\sum_{i=1}^N\left( Y_i - \phi_i[\widehat{\beta}_m^{(n+1)}] \right)\left( Y_i - \phi_i[\widehat{\beta}_m^{(n+1)}] \right)'\widehat{F} ^{(n+1)}= \widehat{F}^{(n+1)}V_{NT},\nonumber
\end{eqnarray*}
where $D_{m,p} = I_m\otimes \text{diag}\left\{\frac{\lambda_1}{ \|  \widehat{C}_{\beta,1}^{(n)}\| } ,\ldots, \frac{\lambda_{p}}{ \| \widehat{C}_{\beta,p}^{(n)}\| }\right\}$; and $V_{NT}$ is a diagonal matrix with the diagonal being the $r$ largest eigenvalues of

\begin{eqnarray*}
\frac{1}{NT}\sum_{i=1}^N\left( Y_i - \phi_i[\widehat{\beta}_m^{(n+1)}] \right)\left( Y_i - \phi_i[\widehat{\beta}_m^{(n+1)}] \right)'\nonumber
\end{eqnarray*}
arranged in descending order. We stop the iteration when the estimates reach certain criteria,  say $\|\widehat{C}_\beta^{(n+1)}-\widehat{C}_\beta^{(n)} \|\le \epsilon$. To start the above iteration, we randomly generate $\widehat{F}^{(0)}$, where each element of $\widehat{F}^{(0)}$ follows from $N(0,1)$. 

To choose the optimal $\lambda$, we follow \cite{WangXia} to simplify it as follows:

\begin{eqnarray}\label{lambda}
\lambda =\nu \left(\|\bar{C}_{\beta,1} \|^{-1},\ldots, \|\bar{C}_{\beta,p} \|^{-1} \right)',
\end{eqnarray}
where $\nu $ is a scalar, and $\bar{C}_{\beta,j}$ stands for the $j^{th}$ row of the unregularized estimator $\bar{C}_{\beta}$ (i.e., implementing (3.3) of the main text with $\lambda=0_{p\times 1}$). With the specification of \eqref{lambda}, the idea for choosing lambda becomes straightforward.  The unregularized estimator $\bar{C}_{\beta}$ is a consistent estimator. It provides information on how likely each row of $C_{\beta_0}$ is a zero row. In other words, smaller $\| \bar{C}_{\beta,j} \|$ implies that the $j^{th}$ row of $C_{\beta_0}$ is more likely to be zero and hence suggests a larger regularizer on $\| C_{\beta,j} \|$. Given (\ref{lambda}), the selection on the vector $\lambda$ reduces to the selection on the scalar $\nu$. 
Finally, we consider the possible value of $\nu$ over a sufficiently large interval of the real line. The optimal $\nu$ is chosen by minimizing the BIC type criteria proposed in the main text. For the HD case, $\Upsilon_{NT}$ is chosen as $\frac{\ln \xi_{NT}}{\sqrt[8]{\xi_{NT}}}$ in view of the development of Lemma \ref{theorem4} and Theorem 3.3.

\section{A Numerical Study}\label{Simulation}

In this section, we examine the performance of the methodology of Section 3 through several simulations. Consider the model (2.4) of the main text. For the factor structure, let $f_{0t}\sim \text{i.i.d. } N(0_{r\times 1}, I_r)$ and $\gamma_{0i}\sim \text{i.i.d. } N(0.5 \cdot 1_{r\times 1}, I_r)$.  In order to generate the regressors and univariate index variable, we firstly generate $v_{it} = 0.5 \cdot v_{i,t-1} + \xi_{it}$, where $\xi_{it}\sim \text{i.i.d. } N(0_{p\times 1}, I_p)$. Then let $x_{it} = v_{it}+|\gamma_{0i}'f_{0t}|$, and $z_{it} = |v_{it,1}| + \text{i.i.d. } N(0, 1)$, where $v_{it,1}$ stands for the first element of $v_{it}$. By doing so, we generate certain correlation between the regressors and the factor structure, and also introduce some correlation between $z_{it}$ and $x_{it}$. The error terms are generated as $\varepsilon_t = 0.5\cdot \varepsilon_{t-1} + \zeta_t$ in which $\zeta_t\sim \text{i.i.d. } N(0_{N\times 1}, \Sigma_\zeta)$ and $\Sigma_\zeta =\{ 0.5^{|i-j|}\}_{N \times N}$, so that the weak cross-sectional dependence among individuals, and serial correlation over time dimension are generated. For both LD and HD cases, the rest settings are as follows:

\begin{itemize}
\item \textbf{LD Case:} $p^*=2$, $p=5$, $r=3$, and let  $\beta_{01}(z) =\exp(-z^2/2) +0.4 $ and $\beta_{02}(z) =z\cdot\exp(-z^2/2) +0.7 $;

\item \textbf{HD Case:} $p^*=2\cdot\lfloor 1.2(NT)^{1/6}\rfloor$, $p=30$, $r=3$. For $j=1,\ldots,p^*$, $\beta_{0j}(z) =\exp(-z^2/2) +0.4 $ when $j$ is odd, and $\beta_{0j}(z) =z\cdot\exp(-z^2/2) +0.7 $ when $j$ is even. 
\end{itemize}

For each dataset generated, we implement the procedure of Section 3 to perform variable selection first. After identifying $\mathcal{A}^*$ and $\mathcal{A}^\dagger$, we implement post-selection estimation (i.e., remove the irrelevant regressors and then implement (3.3) of the main text with $\lambda=0_{p\times 1}$) for the coefficient functions. We adopt the Hermite functions of \cite{DongLinton} as the basis functions, repeat the above procedure 1000 times, and let\footnote{Note that the optimal choice of $m$ may not be the optimal one, but it satisfies all the requirements of our assumptions. Although the optimal choice of truncation parameter and the optimal bandwidth selection have been solved for some cross-sectional models and time series models (e.g., \citealp{Gao2007, HallLiRacine}) under the low dimensional cases, it is well understood that the question is still open even for the nonparametric panel data model with fixed effects (cf., \citealp{ChenGaoLiA, SuJin}). The question is even more daunting when the factor structure and variable selection procedure get involved.} $N\in \{40,\, 80,\, 120\}$, $T\in \{40,\, 80,\, 120\}$, and $m=\lfloor 1.2(NT)^{1/6}\rfloor$. 

To evaluate our simulation results, we firstly report two percentages: (1) the percentage of missed true regressors (i.e., false negative rate, FNR); and (2) the percentage of falsely selected noise regressors (i.e., false positive rate, FPR). Secondly, we evaluate the estimates on the components of $\beta_0(\cdot)$. Take $\beta_{01}(\cdot)$ as an example. For the $j^{th}$ replication, we obtain $\widehat{\beta}_{1j}(z)$ for $\forall z$ (given it is not identified as 0; otherwise, we record 0 as the estimate). For $\forall z$, we calculate $\widehat{\beta}_{1}(z) =\frac{1}{1000}\sum_{j=1}^{1000}\widehat{\beta}_{1j}(z)$, and also record the 95\% confidence bands  based on $\{\widehat{\beta}_{1j}(z) \, | \, j=1,\ldots, 1000\}$. We plot these values over a certain range of $z$. The values of $\beta_{01}(z)$ are plotted in solid black line, and the values of $\widehat{\beta}_{1}(z)$ are plotted in red dotted line, and the associated 95\% confidence bands are plotted in blue dashed curves.

Table \ref{table1} summarizes the FNR and FPR for the LD and HD cases respectively. It is clear that our method proposed in Section 3 works well, as both FNR and FPR are either 0 or very close to 0. It is worth mentioning that although FPR is slightly higher than zero for the HD case, over selecting the regressors will still yield consistent estimation.

\begin{table}[H]
\small
\centering
\caption{FNR \& FPR}
\label{table1}
\begin{tabular}{llrrrrrrr}
\hline \hline
   &     & \multicolumn{3}{c}{FNR}  & \multicolumn{1}{c}{} & \multicolumn{3}{c}{FPR}   \\ \cline{3-5} \cline{7-9} 
  & $N \setminus T$ & 40     & 80     & 120    &                      & 40      & 80     & 120    \\
LD & 40  & 0.00\% & 0.00\% & 0.00\% &                      & 0.00\%  & 0.00\% & 0.00\% \\
   & 40  & 0.00\% & 0.00\% & 0.00\% &                      & 0.00\%  & 0.00\% & 0.00\% \\
   & 120 & 0.00\% & 0.00\% & 0.00\% &                      & 0.00\%  & 0.00\% & 0.00\% \\
   &     &        &        &        &                      &         &        &        \\
 HD  & 40  & 0.00\% & 0.00\% & 0.00\% &                      & 9.90\% & 1.10\% & 1.60\% \\
   & 40  & 0.00\% & 0.00\% & 0.00\% &                      & 1.70\%  & 1.90\% & 1.20\% \\
   & 120 & 0.00\% & 0.00\% & 0.00\% &                      & 3.60\%  & 1.00\% & 0.80\%  \\
\hline \hline
\end{tabular}
\end{table}

For both the LD and HD cases, we plot $\beta_{01}(z)$ and $\beta_{02}(z)$  on $ [-1,2]$ in Figures \ref{fig1}-\ref{fig4}, as the majority of $z_{it}$'s lie in this range. Due to similarity, we do not report the estimates of the rest coefficient functions for HD case. It is easy to see that as the sample size increases, the 95\% confidence bands become much narrower and the mean estimate approaches the true curve. Also, the estimates from the HD case have wider 95\% confidence bands, and seem to be less accurate compared to the LD case as expected.

\begin{figure}[H]
\centering
\hspace*{-1cm}\includegraphics[scale=0.4]{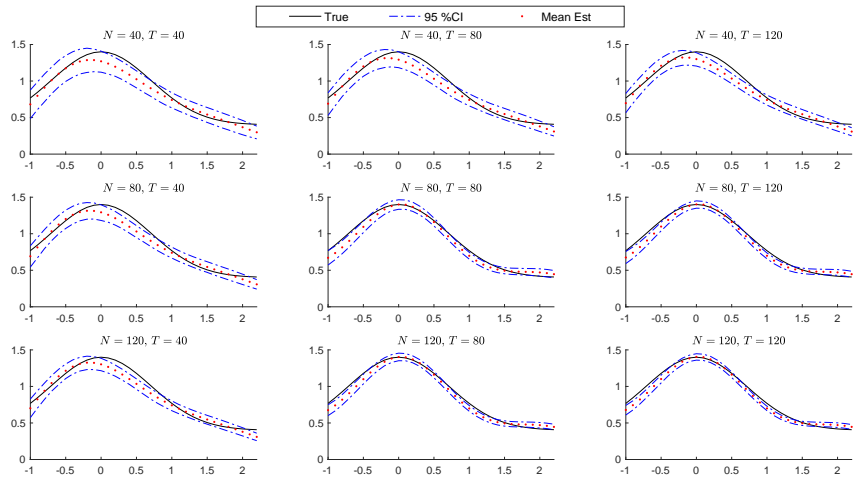}
\caption{LD: $\beta_{01}(z) =\exp(-z^2/2) +0.4 $}
\label{fig1}
\end{figure}

\begin{figure}[H]
\centering
\hspace*{-1cm}\includegraphics[scale=0.4]{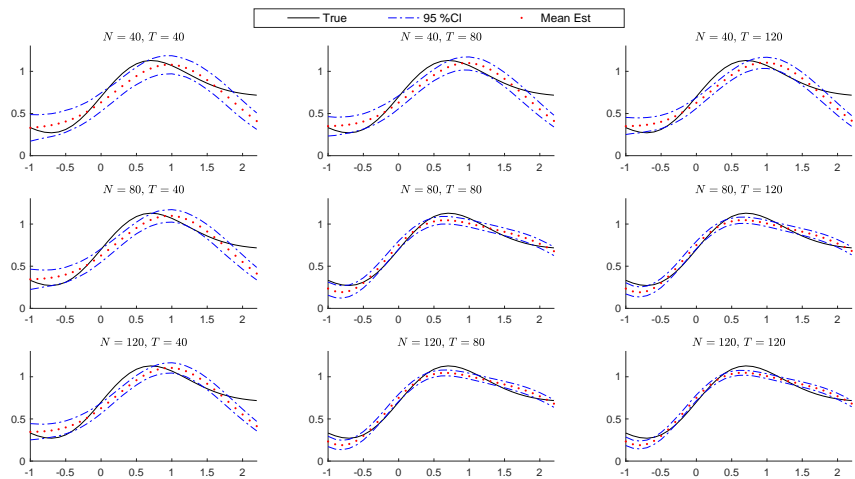}
\caption{LD: $\beta_{02}(z) =z\exp(-z^2/2) +0.7 $}
\label{fig2}
\end{figure}

\begin{figure}[H]
\centering
\hspace*{-1cm}\includegraphics[scale=0.4]{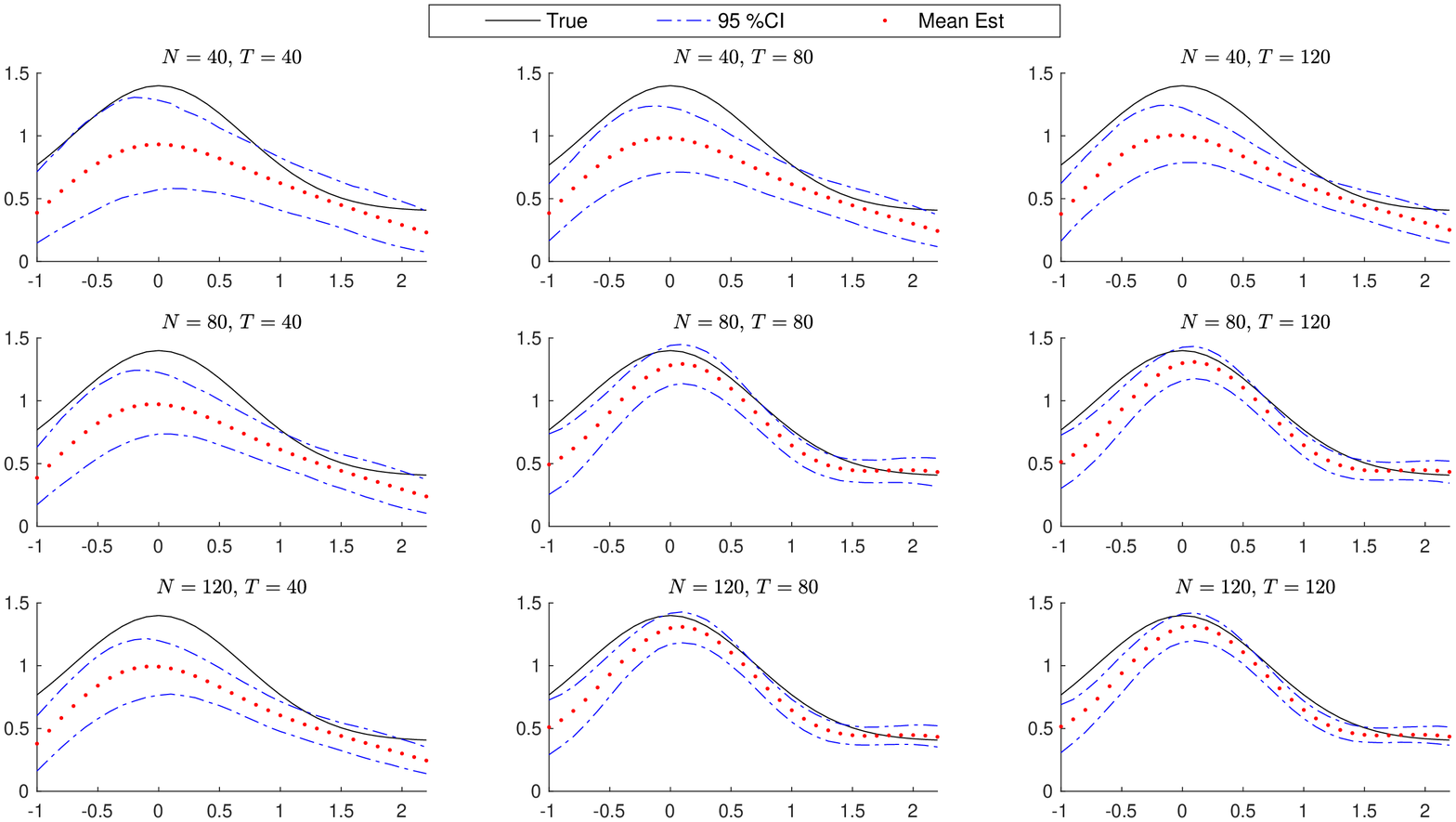}
\caption{HD: $\beta_{01}(z) =\exp(-z^2/2) +0.4 $}
\label{fig3}
\end{figure}

\begin{figure}[H]
\centering
\hspace*{-1cm}\includegraphics[scale=0.4]{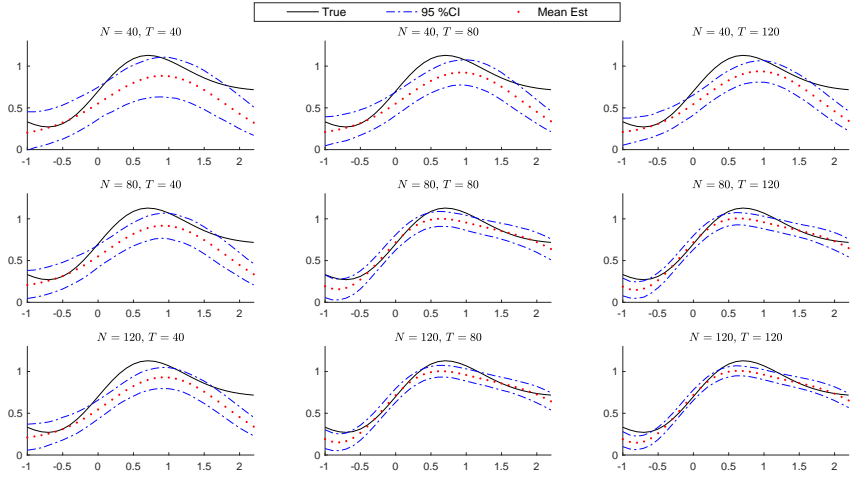}
\caption{HD: $\beta_{02}(z) =z\exp(-z^2/2) +0.7 $}
\label{fig4}
\end{figure}

\section{Proofs}\label{SectionA4}

Before proving the main theorems, we present the following preliminary lemmas.

\begin{lemma}\label{lemma3b}
Consider two non-singular symmetric matrices $A, B$ with the same dimensions $k\times k$, where $k$ tends to $\infty$. Suppose that their minimum eigenvalues satisfy that $\eta_{\normalfont \text{min}}(A)>0$ and $\eta_{\normalfont \text{min}}(B)>0$ uniformly in $k$. Then $\left\|A^{-1}-B^{-1}\right\| \le \eta_{\normalfont\text{min}}^{-1}\left(A\right)\cdot \eta_{\normalfont\text{min}}^{-1}\left\|A-B\right\|$.
\end{lemma}

\begin{lemma}\label{LemmaA5}
Let Assumptions 1 and 2 hold. As $(N,T)\to (\infty,\infty)$,

\begin{enumerate}
\item $\|\frac{1}{NT} \mathcal{E}'\mathcal{E}\| =O_P\left(\frac{1}{\sqrt{N}}\right)+O_P\left(\frac{1}{\sqrt{T}}\right)$ and $\|\frac{1}{NT}\mathcal{E}\mathcal{E}' \|=O_P\left(\frac{1}{\sqrt{N}}\right)+O_P\left(\frac{1}{\sqrt{T}}\right)$, where $\mathcal{E}= (\mathcal{E}_1,\ldots, \mathcal{E}_N)'$,

\item $\displaystyle\sup_{F\in \mathsf{D}_F}\frac{1}{NT}\sum_{i=1}^{N} \mathcal{E}_i^{\prime }P_{F} \mathcal{E}_i =O_P\left(\frac{1}{\sqrt{N}}\right)+O_P\left(\frac{1}{\sqrt{T}}\right)$,

\item $\displaystyle\sup_{F\in \mathsf{D}_F}\left\vert \frac{1}{NT}\sum_{i=1}^{N}\gamma_{0i}'F_0 ' M_F \mathcal{E}_i\right\vert =O_P\left(\frac{1}{\sqrt[4]{N}}\right)+O_P\left(\frac{1}{\sqrt[4]{T}}\right)$,

\item $\displaystyle\sup_{\|  C_\beta \|\le M, \, F\in \mathsf{D}_F}\left| \frac{1}{NT}\sum_{i=1}^N \left(\phi_i[\beta_{0,m}]-\phi_i[\beta_m] \right)' M_F \mathcal{E}_i\right| =O_P\left(\frac{1}{\sqrt[4]{N}}\right)+O_P\left(\frac{1}{\sqrt[4]{T}}\right)$,

\item $\displaystyle\sup_{F\in \mathsf{D}_F}\left|\frac{1}{NT}\sum_{i=1}^N\phi_i[\Delta_{ m}]^{\prime }M_F \phi_i[\Delta_{ m}]\right| = O_{P}\left(m^{-\mu}\right)$,

\item $\displaystyle\sup_{F\in \mathsf{D}_F}\left|\frac{1}{NT}\sum_{i=1}^N\phi_i[\Delta_{ m}]^{\prime }M_F F_0\gamma_{0i}\right| =O_P(m^{-\frac{\mu}{2}})$,

\item $\displaystyle\sup_{\|  C_\beta \|\le M, \, F\in \mathsf{D}_F}\left\vert \frac{1}{NT}\sum_{i=1}^{N}\phi _{i}\left[  \Delta_{m}\right] ^{\prime }M_{F}\left\{\phi _{i} \left[\beta_m\right] -\phi _{i} \left[\beta_{0,m}\right] \right\} \right\vert =O_P(m^{-\frac{\mu}{2}})$,
\end{enumerate}
where $M$ is a sufficiently large constant.
\end{lemma}

\bigskip

\noindent Let $\Pi_{NT}^{-1}=V_{NT}(F_0'\widehat{F} /T)^{-1}(\Gamma_0'\Gamma_0/N)^{-1}$, where $V_{NT}$ is a diagonal matrix with the diagonal being the $r$ largest eigenvalues of

\begin{eqnarray*}
\frac{1}{NT}\sum_{i=1}^N\left( Y_i - \phi_i[\widehat{\beta}_m] \right)\left( Y_i - \phi_i[\widehat{\beta}_m] \right)'\nonumber
\end{eqnarray*}
arranged in descending order.  

\begin{lemma}\label{LemmaA6}
Let Assumptions 1, 2 and 3.1 hold.  As $(N,T)\to (\infty,\infty)$, 

\begin{enumerate}
\item $\|\widehat{\beta}_m -\beta_0 \|_{L^2} =o_P(1)$;

\item $\|P_{\widehat{F}} -P_{F_0}\| =o_P(1)$;

\item $V_{NT}\to_P V$, where $V$ is an $r\times r$ diagonal matrix consisting of the eigenvalues of $\Sigma_f\Sigma_{\gamma}$;

\item $\frac{1}{\sqrt{T}}\|\widehat{F} \Pi_{NT}^{-1} - F_0 \| =O_P(\|\widehat{\beta}_m -\beta_0 \|_{L^2})+O_P\left( \frac{1}{\sqrt{N}}\right) +O_P\left( \frac{1}{\sqrt{T}}\right) $;

\item $\left\|\frac{1}{T}\widehat{F}  '(\widehat{F}  -F_0\Pi_{NT})\right\| = O_P(\|\widehat{\beta}_m   -\beta_0 \|_{L^2}) +O_P\left( \frac{1}{N}\right)+O_P\left( \frac{1}{T}\right)$;

\item $\|P_{\widehat{F}} -P_{F_0}\|^2 =O_P(\|\widehat{\beta}_m  -\beta_0 \|_{L^2}) +O_P\left( \frac{1}{N}\right)+O_P\left( \frac{1}{T}\right)$.
\end{enumerate}
\end{lemma}

\begin{lemma}\label{theorem1}
Let Assumptions 1, 2 and 3 hold.  As $(N,T)\to (\infty,\infty)$, $\Pr (\| \widehat{C}_{\beta}^\dagger\|=0 )\to 1$.
\end{lemma}

\begin{lemma}\label{rateCstar}
Let Assumptions 1-3 hold. As $(N,T)\to (\infty,\infty)$, 

\begin{eqnarray*}
\|\widehat{C}_\beta^* -C_{\beta_0}^* \| = O_P\left( \sqrt{\frac{m}{NT}}\right) + O_P(m^{-\frac{\mu}{2} }) +O_P\left(\frac{m \lambda_{\normalfont\text{max}}^* }{NT} \right).\nonumber
\end{eqnarray*}
\end{lemma}

\begin{lemma}\label{LemmaLP}
Let Assumptions 1, 2 and 5 hold. As $(N,T)\to (\infty,\infty)$,

\begin{enumerate}
\item $\displaystyle\sup_{\|  C_\beta \|\le a_0\sqrt{p}, \, F\in \mathsf{D}_F}\left| \frac{1}{NT}\sum_{i=1}^N \left(\phi_i[\beta_{0,m}]-\phi_i[\beta_m] \right)' M_F \mathcal{E}_i\right| =O_P\left( \sqrt{\frac{p(\xi_{NT}+mp)}{NT}}\right)$;

\item $\displaystyle\sup_{F\in \mathsf{D}_F}\left|\frac{1}{NT}\sum_{i=1}^N\phi_i[\Delta_{ m}]^{\prime }M_F \phi_i[\Delta_{ m}]\right| = O_{P}\left(p^* m^{-\mu}\right)$;

\item $\displaystyle\sup_{F\in \mathsf{D}_F}\left|\frac{1}{NT}\sum_{i=1}^N\phi_i[\Delta_{ m}]^{\prime }M_F F_0\gamma_{0i}\right| =O_P(\sqrt{p^*} m^{-\frac{\mu}{2}})$;

\item $\displaystyle\sup_{F\in \mathsf{D}_F}\left|\frac{1}{NT}\sum_{i=1}^N\phi_i[\Delta_{ m}]^{\prime }M_F \mathcal{E}_i\right| =O_P(\sqrt{p^*} m^{-\frac{\mu}{2}})$;

\item $\displaystyle\sup_{\|  C_\beta \|\le a_0\sqrt{p}, \, F\in \mathsf{D}_F}\left\vert \frac{1}{NT}\sum_{i=1}^{N}\phi _{i}\left[  \Delta_{m}\right] ^{\prime }M_{F}\left\{\phi _{i} \left[\beta_m\right] -\phi _{i} \left[\beta_{0,m}\right] \right\} \right\vert =O_P(\sqrt{p\, p^*}m^{-\frac{\mu}{2}})$,
\end{enumerate}
where $a_0$ is a sufficiently large constant.
\end{lemma}

\begin{lemma}\label{theorem4}
Let Assumptions 1, 2 and 5 hold. As $(N,T)\to (\infty,\infty)$, 

\begin{enumerate}
\item $\|\widehat{\beta}_m -\beta_0 \|_{L^2} =o_P(1)$;

\item $\|P_{\widehat{F}} -P_{F_0}\| =o_P(1)$.
\end{enumerate}
\end{lemma}

\bigskip

\noindent \textbf{Proof of Theorem 3.1:}

(1). The first result follows from Lemma \ref{theorem1}.

\medskip

(2). Based on the development of Lemma \ref{rateCstar}, the definition of $\widehat{\beta}_m^* $ and (3.1), we can write for $\forall z\in V_z$,

\begin{eqnarray*}
&&\sqrt{\frac{NT}{m}} \left(\widehat{\beta}_m^* (z) -\beta_0^*(z)\right) \nonumber \\
&=& \sqrt{\frac{NT}{m}} \left[H_m'(z) \otimes I_{p^*}\right] \left[\vect(\widehat{C}_{\beta}^* - C_{\beta_0}^* ) \right] +  \sqrt{\frac{NT}{m}}\Delta_{m}^*(z)\nonumber\\
&=&\sqrt{\frac{NT}{m}}\left[H_m'(z) \otimes I_{p^*}\right] \left[\vect(\widehat{C}_{\beta}^* - \widehat{C}_{\beta}^\sharp )+\vect(\widehat{C}_{\beta}^\sharp - C_{\beta_0}^* ) \right] +o_P(1)\nonumber\\
&=&\sqrt{\frac{NT}{m}}  \left[H_m'(z) \otimes I_{p^*}\right]  A_{1NT}^{-1} \Sigma_{\mathcal{Z}}^{*\, -1} \cdot\frac{1}{NT }\sum_{i=1}^N  \left\{{\mathcal{Z}_i^*}' M_{\widehat{F} }+A_{3,i}\right\}\mathcal{E}_{i}\nonumber\\
&&+ \sqrt{\frac{NT}{m}} \left[H_m'(z) \otimes I_{p^*}\right]  A_{1NT}^{-1} \Sigma_{\mathcal{Z}}^{*\, -1} \cdot J_{6NT,1}+O_P\left(\frac{m\| \lambda^*\|}{\sqrt{NT}} \right)  +o_P(1)\nonumber\\
&=&\sqrt{\frac{NT}{m}} \left[H_m'(z) \otimes I_{p^*}\right]  A_{1NT}^{-1} \Sigma_{\mathcal{Z}}^{*\, -1}\cdot \frac{1}{NT }\sum_{i=1}^N  \left\{ {\mathcal{Z}_i^*} ' M_{\widehat{F} }+A_{3,i}\right\}\mathcal{E}_{i}+o_P(1)\nonumber\\
&:=& \Lambda_1 +o_P(1),\nonumber
\end{eqnarray*}
where the second equality follows from $\| \Delta_{ m}(z)\|=O(m^{-\mu/2})$ and the condition $\frac{N T }{m^{\mu+1}}\to 0$; the third equality follows from the above development on $\vect(\widehat{C}_{\beta}^*  )-\vect(\widehat{C}_{\beta}^\sharp )$ and $\vect(\widehat{C}_{\beta}^\sharp  )-\vect(C_{\beta_0}^* ) $, and the fact that $\Sigma_{\mathcal{Z},f}^{*\, -1}$ reduces to $\Sigma_{\mathcal{Z}}^{*\, -1}$ using Assumption 4; and the fourth equality follows from (B.22) of Appendix B, $\frac{mN}{T}\to 0$, and $\frac{m\lambda_{\text{max}}^* }{\sqrt{NT}}\to 0$.

We next consider $\Lambda_1$ by starting with  $\frac{1}{NT }\sum_{i=1}^N  {\mathcal{Z}_i^*}'M_{\widehat{F} }\mathcal{E}_{i}$.

\begin{eqnarray*}
\frac{1}{NT }\sum_{i=1}^N {\mathcal{Z}_i^*}' M_{\widehat{F} }\mathcal{E}_{i} &=&\frac{1}{NT }\sum_{i=1}^N  {\mathcal{Z}_i^*}' M_{F_0}\mathcal{E}_{i} + \frac{1}{NT }\sum_{i=1}^N {\mathcal{Z}_i^*}'(M_{\widehat{F} } -M_{F_0})\mathcal{E}_{i}\nonumber\\
&=& \frac{1}{NT }\sum_{i=1}^N  {\mathcal{Z}_i^*}' M_{F_0}\mathcal{E}_{i} -\frac{1}{NT }\sum_{i=1}^N  {\mathcal{Z}_i^*}'( P_{\widehat{F} } - P_{F_0})\mathcal{E}_{i} \nonumber\\
&:=& D_1-D_2.\nonumber
\end{eqnarray*}
Firstly, we shall show $\left\| \sqrt{\frac{NT}{m}}  \left[H_m'(z) \otimes I_{p^*}\right]  A_{1NT}^{-1} \Sigma_{\mathcal{Z}}^{*\, -1}D_2\right\| = o_P(1)$. Let $\mathcal{Z}_{i,j}^*$ be the $j^{th}$ column of ${\mathcal{Z}_i^*}$, and let $\mathcal{Z}_{it,j}^*$ be the $t^{th}$ element of $\mathcal{Z}_{i,j}^*$. Write

\begin{eqnarray*}
D_2 &=& \frac{1}{NT }\sum_{i=1}^N  {\mathcal{Z}_i^*}'\left(\frac{\widehat{F} \widehat{F}'}{T}   -P_{F_0}\right)\mathcal{E}_{i} \nonumber\\
&=&\frac{1}{NT }\sum_{i=1}^N \frac{{\mathcal{Z}_i^*}'(\widehat{F}-F_0\Pi_{NT})}{T}\Pi_{NT}'F_0'\mathcal{E}_i
+\frac{1}{NT }\sum_{i=1}^N \frac{{\mathcal{Z}_i^*}'(\widehat{F}-F_0\Pi_{NT})}{T} (\widehat{F}-F_0\Pi_{NT})'\mathcal{E}_i\nonumber\\
&&+\frac{1}{NT }\sum_{i=1}^N \frac{{\mathcal{Z}_i^*}'F_0\Pi_{NT}}{T} (\widehat{F}-F_0\Pi_{NT})'\mathcal{E}_i +\frac{1}{NT }\sum_{i=1}^N \frac{{\mathcal{Z}_i^*}'F_0 }{T} [\Pi_{NT}\Pi_{NT}' -(F_0'F_0/T)^{-1}]F_0'\mathcal{E}_i\nonumber\\
&:=&D_{21}+D_{22}+D_{23}+D_{24},\nonumber
\end{eqnarray*}
where the definitions of $D_{21}$ to $D_{24}$ are obvious.

In the following, we let $D_{2\ell,j}$ be the $j^{th}$ row of $D_{2\ell}$ for $\ell = 1,2,3,4$. Thus, for $D_{21}$, consider

\begin{eqnarray*}
\| D_{21,j}\| &=&\left\|\frac{1}{NT }\sum_{i=1}^N \frac{{\mathcal{Z}_{i,j}^*}'(\widehat{F}-F_0\Pi_{NT})}{T}\Pi_{NT}'F_0'\mathcal{E}_i \right\|\nonumber\\
&\le &\left\| \frac{1}{NT }\sum_{i=1}^N  (\mathcal{E}_i'F_0)\otimes \frac{{\mathcal{Z}_{i,j}^*}'}{\sqrt{T}}  \right\| \cdot \left\|\frac{1}{\sqrt{T}}\vect\left[(\widehat{F}-F_0\Pi_{NT})\Pi_{NT}'\right]\right\| \nonumber\\
&=&O_P\left( \frac{1}{\sqrt{NT}}\right) \frac{1}{\sqrt{T}}\|\widehat{F}-F_0\Pi_{NT} \|.\nonumber
\end{eqnarray*}
Summing up over $j$ for $D_{21,j}$, we obtain that $\| D_{21}\| =O_P\left( \sqrt{\frac{m}{NT}}\right) \frac{1}{\sqrt{T}}\|\widehat{F}-F_0\Pi_{NT} \|.$

For $D_{22}$, write

\begin{eqnarray*}
\|D_{22,j} \| &=& \left\| \frac{1}{NT }\sum_{i=1}^N \frac{ {\mathcal{Z}_{i,j}^*}'(\widehat{F}-F_0\Pi_{NT})}{T} (\widehat{F}-F_0\Pi_{NT})'\mathcal{E}_i \right\|\nonumber\\
&\le &\left\| \frac{1}{NT}\sum_{i=1}^N  \mathcal{E}_i'\otimes {\mathcal{Z}_{i,j}^*}'  \right\| \cdot \left\|\frac{1}{T}\vect\left[(\widehat{F}-F_0\Pi_{NT})(\widehat{F}-F_0\Pi_{NT})'
\right]\right\| \nonumber\\
&=&O_P\left( \frac{1}{\sqrt{N}}\right)\frac{1}{T}\|\widehat{F}-F_0\Pi_{NT} \|^2.\nonumber
\end{eqnarray*}
Summing $D_{22,j}$ up over $j$, we obtain that $\| D_{22}\| =O_P\left(\sqrt{\frac{m}{N}}\right)\frac{1}{T}\|\widehat{F}-F_0\Pi_{NT} \|^2.$

For $D_{23}$, write

\begin{eqnarray*}
\|D_{23,j} \| &=& \left\|\frac{1}{NT }\sum_{i=1}^N \frac{{\mathcal{Z}_{i,j}^*}'F_0\Pi_{NT}}{T} (\widehat{F}-F_0\Pi_{NT})'\mathcal{e}_i \right\|\nonumber\\
&\le &\left\| \frac{1}{NT}\sum_{i=1}^N  \mathcal{E}_i'\otimes \frac{{\mathcal{Z}_{i,j}^*}'F_0}{\sqrt{T}} \right\| \cdot \left\|\frac{1}{\sqrt{T}}\vect\left[\Pi_{NT}(\widehat{F}-F_0\Pi_{NT})' \right]\right\| .\nonumber
\end{eqnarray*}

Note that

\begin{eqnarray*}
&&E\left\| \frac{1}{NT}\sum_{i=1}^N  \mathcal{E}_i'\otimes \frac{{\mathcal{Z}_{i,j}^*}'F_0}{\sqrt{T}} \right\|^2 \nonumber\\
&=&\frac{1}{N^2T^3} E\left\| \sum_{i=1}^N  \mathcal{E}_i'\otimes \sum_{t=1}^T{\mathcal{Z}_{it,j}^*}f_{0t}' \right\|^2=\frac{1}{N^2T^3} \sum_{s=1}^T E\left\| \sum_{i=1}^N  \varepsilon_{is}\sum_{t=1}^T {\mathcal{Z}_{it,j}^*} f_{0t}' \right\|^2 \nonumber\\
&=&\frac{1}{N^2T^3} \sum_{s=1}^T \sum_{i_1=1}^N\sum_{i_2=1}^NE\left[ \left(\sum_{t=1}^T  {\mathcal{Z}_{i_1t,j}^*} f_{0t}' \right)\left( \sum_{t=1}^T  {\mathcal{Z}_{i_2t,j}^*} f_{0t} \right) \right]E[\varepsilon_{i_1s}\varepsilon_{i_2s} ] \nonumber\\
&=&\frac{1}{N^2T^2} \sum_{i_1=1}^N\sum_{i_2=1}^NE\left[ \left(\sum_{t=1}^T  {\mathcal{Z}_{i_1t,j}^*}  f_{0t}' \right)\left( \sum_{t=1}^T  {\mathcal{Z}_{i_2t,j}^*} f_{0t} \right) \right] \sigma_{i_1i_2} \nonumber\\
&=&\frac{1}{N^2T^2} \sum_{t=1}^T \sum_{i_1=1}^N\sum_{i_2=1}^N E\left[ {\mathcal{Z}_{i_1t,j}^*}  {\mathcal{Z}_{i_2t,j}^*}  E\left[\|f_{0t}\|^2 \, | \, \mathcal{X}_{Nt} \right]\right]  \sigma_{i_1i_2}\nonumber\\
&&+\frac{2}{N^2T^2} \sum_{t_1> t_2} \sum_{i_1=1}^N\sum_{i_2=1}^N E\left[ {\mathcal{Z}_{i_1t_1,j}^*}  {\mathcal{Z}_{i_2t_2,j}^*}  E\left[f_{0t_1}'  f_{0t_2}\, | \, \mathcal{X}_{N t_1} \right]\right]  \sigma_{i_1i_2}\nonumber\\
&=&\frac{1}{N^2T^2} \sum_{t=1}^T \sum_{i_1=1}^N\sum_{i_2=1}^N E\left[ {\mathcal{Z}_{i_1t,j}^*}  {\mathcal{Z}_{i_2t,j}^*}  \right]  a_{tt}  \sigma_{i_1i_2} \nonumber\\
&&+\frac{2}{N^2T^2} \sum_{t_1> t_2} \sum_{i_1=1}^N\sum_{i_2=1}^N E\left[{\mathcal{Z}_{i_1t_1,j}^*}  {\mathcal{Z}_{i_2t_2,j}^*} \right]   a_{t_1t_2}\sigma_{i_1i_2}\nonumber\\
&\le &O(1)\frac{2}{N^2T^2} \sum_{t_1\ge t_2} \sum_{i_1=1}^N\sum_{i_2=1}^N |a_{t_1t_2}|  \cdot | \sigma_{i_1i_2}| =O(1)\frac{1}{N T},\nonumber
\end{eqnarray*}
where the fourth equality follows from Assumption 1.2; the sixth equality follows from Assumption 4; and the seventh equality follows from both Assumptions 1.1 and 4.1.

Thus, $\|D_{23,j} \| = O_P\left( \frac{1}{\sqrt{NT}}\right) \frac{1}{\sqrt{T}} \left\|\widehat{F}-F_0\Pi_{NT}\right\|$. Summing $D_{23,j}$ up  over $j$, we obtain that $\| D_{23}\| =O_P\left( \sqrt{\frac{m}{NT}}\right) \frac{1}{\sqrt{T}}\|\widehat{F}-F_0\Pi_{NT} \|.$

Similarly, write for $D_{24}$, 

\begin{eqnarray*}
\| D_{24,j}\| &=&\left\|\frac{1}{NT }\sum_{i=1}^N \frac{{\mathcal{Z}_{i,j}^*}'F_0 }{T} [\Pi_{NT}\Pi_{NT}' -(F_0'F_0/T)^{-1}]F_0'\mathcal{E}_i\right\|\nonumber\\
&\le &\left\| \frac{1}{NT}\sum_{i=1}^N  (\mathcal{E}_i'F_0)\otimes \frac{{\mathcal{Z}_{i,j}^*}'F_0}{T}  \right\| \cdot \left\|\Pi_{NT}\Pi_{NT}' -(F_0'F_0/T)^{-1}\right\| \nonumber\\
&=&O_P\left( \frac{1}{\sqrt{NT}}\right) \left\|\Pi_{NT}\Pi_{NT}' -(F_0'F_0/T)^{-1}\right\|.\nonumber
\end{eqnarray*}
Summing $D_{24,j}$ up over $j$, we obtain that $\| D_{24}\| = O_P\left( \sqrt{\frac{m}{NT}}\right) \left\|\Pi_{NT}\Pi_{NT}' -(F_0'F_0/T)^{-1}\right\|$, where $\left\|\Pi_{NT}\Pi_{NT}' -(F_0'F_0/T)^{-1}\right\|=o_P(1)$ by the development for the fourth result of Lemma \ref{LemmaA6}.

Based on the analyses of $D_{21}$ to $D_{24}$, we obtain

\begin{eqnarray*}
\sqrt{\frac{NT}{m}}\| D_2\| &=&O_P(1)\frac{1}{\sqrt{T}}\|\widehat{F}-F_0\Pi_{NT} \| +O_P(1)\left\|\Pi_{NT}\Pi_{NT}' -(F_0'F_0/T)^{-1}\right\|\nonumber\\
&&+O_P(1)\sqrt{T}\cdot\frac{1}{T}\|\widehat{F}-F_0\Pi_{NT} \|^2\nonumber
\end{eqnarray*}
which further gives $\left\|\sqrt{\frac{NT}{m}}  \left[H_m'(z) \otimes I_{d_x}\right]  A_{1NT}^{-1} \Sigma_{\mathcal{Z}}^{*\, -1} D_2\right\| = o_P(1)$ by Lemma \ref{LemmaA6} and the condition $\frac{T}{N^2}\to 0$.

Similarly, we obtain

\begin{eqnarray*}
&&\Big\| \sqrt{\frac{NT}{m}}  \left[H_m'(z) \otimes I_{p^*}\right]  A_{1NT}^{-1}\Sigma_{\mathcal{Z}}^{*\, -1} \frac{1}{NT}\sum_{i=1}^N  A_{3,i}\mathcal{E}_{i} \nonumber\\
&& -\sqrt{\frac{NT}{m}}  \left[H_m'(z) \otimes I_{p^*}\right]  A_{1NT}^{-1} \Sigma_{\mathcal{Z}}^{*\, -1} \frac{1}{NT}\sum_{i=1}^N  \widetilde{A}_{3,i}\mathcal{E}_{i} \Big\| =o_P(1),\nonumber
\end{eqnarray*}
where $\widetilde{A}_{3,i}=\frac{1}{N}  \sum_{j=1}^N {\mathcal{Z}_j^*} 'M_{F_0} \gamma_{0j}'  (\Gamma_0' \Gamma_0/N)^{-1} \gamma_{0i}.$

Finally, by Assumption 4 and after some simple algebra, we obtain

\begin{eqnarray*}
\Lambda_1 &=&\sqrt{\frac{NT}{m}} \left[H_m'(z) \otimes I_{p^*}\right]  \widetilde{A}_{1NT}^{-1} \Sigma_{\mathcal{Z}}^{*\, -1} \cdot\frac{1}{NT }\sum_{i=1}^N  \left\{{\mathcal{Z}_i^*}'M_{F_0}+ \widetilde{A}_{3,i}\right\}\mathcal{E}_{i} +o_P(1)\nonumber\\
&\to_D& N(0, \Omega_\star),\nonumber
\end{eqnarray*}
where $\widetilde{A}_{1NT}=I_{m p^*} - \Sigma_{\mathcal{Z}}^{*\, -1}\widetilde{A}_{2NT} $ and $\widetilde{A}_{2NT}=\frac{1}{N^2T} \sum_{i=1}^N\sum_{j=1}^N {\mathcal{Z}_i^*}'M_{F_0} {\mathcal{Z}_j^*} \gamma_{0j}'\Sigma_\gamma^{-1}\gamma_{0i} .$ The proof is then complete. \hspace*{\fill}{$\blacksquare$}

\bigskip

\noindent \textbf{Proof of Theorem 3.2:}

Recall that we have denoted the set $\mathcal{A}^*$. Before proceeding further, we introduce some variables to facilitate the development. For an arbitrary model $S$, we say it is under-fitted if it misses at least one variable with a nonzero coefficient\footnote{Under-fitted case allows for including redundant regressor.}; it is over-fitted if $S$ not only includes all relevant variables but also includes at least one redundant regressor (i.e., $\mathcal{A}^*\subset S$ but $\mathcal{A}^* \ne S$). Then, according to whether the model $S_{\lambda}$ is under fitted, correctly fitted, or over fitted, we create three mutually exclusive sets $A^- $, $A^0 = \left\{\lambda \in \mathbb{R}^p: S_{\lambda} =\mathcal{A}^*  \right\}$  and $A^+ = \left\{\lambda \in \mathbb{R}^p: S_{\lambda} \supset \mathcal{A}^*, S_{\lambda}\ne \mathcal{A}^* \right\}$. Suppose that there is a sequence $\{ \lambda_{NT} \}$ that ensures the conditions required by Lemma \ref{theorem1}. Let $(\widehat{C}_{\beta}^{\lambda_{NT}}, \widehat{F}^{\lambda_{NT}})$ denote the estimator obtained by implementing (3.3) using $\lambda_{NT}$. 

\bigskip

\textbf{Case 1: Under-fitted model.} Without loss of generality, we assume that only one variable is missing, and suppose that the first $p^* - 1$ rows of $\widehat{C}_\beta^{\lambda}$ are obtained from the under-fitted model and the $ {p^*} ^{th}$ row of $\widehat{C}_\beta^{\lambda}$ is a 0 row. Moreover, let $\text{RSS}_0 = \frac{1}{NT} \sum_{i=1}^N \big( Y_i -\phi_i[\beta_{0,m}]  \big)'M_{F_0}\big( Y_i - \phi_i[\beta_{0,m}]  \big)$.

We then write

\begin{eqnarray*}
\text{RSS}_{\lambda} -\text{RSS}_0&=& \frac{1}{NT} \sum_{i=1}^N \big( Y_i -\phi_i[\widehat{\beta}_{m}^\lambda ]  \big)'M_{\widehat{F}^\lambda}\big( Y_i - \phi_i[\widehat{\beta}_{m}^\lambda ]  \big) \nonumber\\ 
&&- \frac{1}{NT} \sum_{i=1}^N \big( Y_i -\phi_i[\beta_{0,m}]  \big)'M_{F_0}\big( Y_i - \phi_i[\beta_{0,m}]  \big) \nonumber\\
&\ge & \rho_1 \| C_{\beta_0, p^*}\|^2>\frac{\rho_1}{2}\|\beta_{0p^*} \|_{L^2}^2>0 ,\nonumber
\end{eqnarray*}
where the first inequality follows from the development given for (B.3) of Appendix B.

Again, using the development given for (B.3) of Appendix B, we have

\begin{eqnarray*}
&&\text{RSS}_{\lambda_{NT}}-\text{RSS}_0 \nonumber  \\
&=&\vect(C_{\beta_0}- \widehat{C}_{\beta}^{\lambda_{NT}} )'\frac{1}{NT}\sum_{i=1}^{N}\mathcal{Z}_i'M_{\widehat{F}^{\lambda_{NT}}}\mathcal{Z}_i\vect(C_{\beta_0}- \widehat{C}_{\beta}^{\lambda_{NT}} )\nonumber\\
&&+\frac{1}{NT}\text{tr}\left( M_{\widehat{F}^{\lambda_{NT}}} F_0\Gamma _0'\Gamma _0F_0'M_{\widehat{F}^{\lambda_{NT}}}\right)\nonumber \\
&&+2\vect(C_{\beta_0}- \widehat{C}_{\beta}^{\lambda_{NT}} )'\frac{1}{NT}\sum_{i=1}^{N}\mathcal{Z}_i'M_{\widehat{F}^{\lambda_{NT}}}F_0\gamma_{0i} +o_P(1) \nonumber  \\
&=&o_P(1),\nonumber
\end{eqnarray*}
where the second equality follows from the development of Lemma \ref{LemmaA6}.

Thus, we can conclude that $\Pr\left( \inf_{\lambda\in A^-}\text{BIC}_{\lambda}  >  \text{BIC}_{\lambda_{NT}}\right) \to 1$.

\bigskip

\textbf{Case 2: Over-fitted model.} Consider $\forall \lambda\in A^+$ and recall that $\widehat{C}_\beta^{\lambda}$ determines a model $S_{\lambda}$. Under such a model $S_{\lambda}$, we can define another unpenalized estimator  as

\begin{eqnarray}
(\check{C}_\beta,\check{F}) = \argmin_{C_\beta, F}   \frac{1}{NT} \sum_{i=1}^N \big( Y_i -\phi_i[\beta_m]  \big)'M_{F}\big( Y_i - \phi_i[\beta_m]  \big)
\end{eqnarray}
subject to  $F\in \mathsf{D}_F$, where $\| C_{\beta,j}\| = 0$ with $\forall j \notin S_{\lambda}$. In other words, $(\check{C}_\beta,\check{F})$ is the unpenalized estimator under the model determined by $\widehat{C}_\beta^{\lambda}$. By definition, we obtain immediately that $\text{RSS}_{\lambda} \ge \text{RSS}_{S_{\lambda}} $, where $\text{RSS}_{S_{\lambda}}  =   \frac{1}{NT} \sum_{i=1}^N \big( Y_i -\phi_i[\check{\beta}_m]  \big)'M_{\check{F}}\big( Y_i - \phi_i[\check{\beta}_m]  \big).$

Write

\begin{eqnarray*}
\ln \text{RSS}_{S_{\lambda}} -\ln  \text{RSS}_{\lambda_{NT}} &=&\ln \left(1+ \frac{\text{RSS}_{S_{\lambda}} - \text{RSS}_{\lambda_{NT}} }{ \text{RSS}_{\lambda_{NT}}}\right)\nonumber\\
&\ge &- \frac{\text{RSS}_{S_{\lambda}} - \text{RSS}_{\lambda_{NT}} }{ \text{RSS}_{\lambda_{NT}}} .\nonumber
\end{eqnarray*}
In view of the proof of Lemma \ref{LemmaA6}, it is easy to see that $\text{RSS}_{\lambda_{NT}}$ converges to a positive constant. With regard to $\text{RSS}_{S_{\lambda}} - \text{RSS}_{\lambda_{NT}} $, we have

\begin{eqnarray*}
\text{RSS}_{S_{\lambda}} - \text{RSS}_{\lambda_{NT}}   &=&\frac{1}{NT} \sum_{i=1}^N \big( Y_i -\phi_i[\check{\beta}_m]  \big)'M_{\check{F}}\big( Y_i - \phi_i[\check{\beta}_m]  \big)\nonumber\\
&&-\frac{1}{NT} \sum_{i=1}^N \big( Y_i -\phi_i[\widehat{\beta}_m^{\lambda_{NT}}]  \big)'M_{\widehat{F}^{\lambda_{NT}}}\big( Y_i - \phi_i[\beta_m^{\lambda_{NT}}]  \big) .\nonumber
\end{eqnarray*}

By Lemmas \ref{LemmaA5} and \ref{rateCstar}, it is not hard to see $\left|\text{RSS}_{S_{\lambda}} - \text{RSS}_{\lambda_{NT}}   \right| \le O_P(1)  \frac{1}{\sqrt[4]{N}}$, so we can further write

\begin{eqnarray*}
\ln \text{RSS}_{S_{\lambda}} -\ln  \text{RSS}_{\lambda_{NT}} &\ge &- \frac{\text{RSS}_{S_{\lambda}} - \text{RSS}_{\lambda_{NT}} }{ \text{RSS}_{\lambda_{NT}}}\ge  -\left|O_P(1)  \frac{1}{\sqrt[4]{N}} \right|. \nonumber
\end{eqnarray*}
We then write

\begin{eqnarray*}
&&\inf_{\lambda\in A^+}\text{BIC}_{\lambda} - \text{BIC}_{\lambda_{NT}}=\inf_{\lambda\in A^+} \ln \text{RSS}_{S_{\lambda}} -\ln  \text{RSS}_{\lambda_{NT}} +(\text{df}_{\lambda} -\text{df}_{\lambda_{NT}})\frac{\ln N}{\sqrt[4]{N}}. \nonumber
\end{eqnarray*}
By Lemma \ref{theorem1}, we know that $\Pr (\text{df}_{\lambda_{NT}} = p^*)\to 1$. Since $\lambda\in A^+$, we must have that $\Pr (\text{df}_{\lambda} \ge p^*+1)\to 1$. Then it is clear that $\Pr\left( \inf_{\lambda\in A^+}\text{BIC}_{\lambda}  >  \text{BIC}_{\lambda_{NT}}\right) \to 1$.

Combining Cases 1 and 2, we obtain that $\Pr\left( \inf_{\lambda\in A^- \cup A^+} \text{BIC}_{\lambda}  >  \text{BIC}_{\lambda_{NT}}\right)\to 1$. This further indicates that $\Pr(S_{\widehat{\lambda}} = \mathcal{A}^*)\to 1$. The proof is now complete. \hspace*{\fill}{$\blacksquare$}

\bigskip

\noindent \textbf{Proof of Theorem 3.3:}

(1). By (B.26) of Appendix B, and the development of (1) and (5) of Lemma \ref{LemmaLP}, we now can further conclude that

\begin{eqnarray*}
&&\left| \frac{1}{NT}\sum_{i=1}^N \left(\phi_i[\beta_{0,m}]-\phi_i[\widehat{\beta}_m] \right)' M_{\widehat{F}} \mathcal{E}_i\right| =o_P\left(  \sqrt{\frac{\xi_{NT}+mp}{NT}}\right),\nonumber \\
&& \left\vert \frac{1}{NT}\sum_{i=1}^{N}\phi _{i}\left[  \Delta_{m}\right] ^{\prime }M_{\widehat{F}}\left\{\phi _{i} \left[\widehat{\beta}_m\right] -\phi _{i} \left[\beta_{0,m}\right] \right\} \right\vert =o_P(\sqrt{p^*}m^{-\frac{\mu}{2}}),\nonumber
\end{eqnarray*}
which allows us to improve the rate of (B.26) of Appendix B as follows.

\begin{eqnarray}\label{HDconsis2}
\|C_{\beta_0}-\widehat{C}_\beta \|^2  =o_P\left( \sqrt{\frac{\xi_{NT}+mp}{NT}} + \sqrt{p^*} m^{-\frac{\mu}{2}}\right)+O_P\left(\frac{1}{\sqrt[4]{\xi_{NT}}} + \frac{p^*\lambda_{\text{max}}^* }{NT}\right).
\end{eqnarray}
Then we can conclude $ \|C_{\beta_0}-\widehat{C}_\beta \| = O_P ( h_{NT}^{1/2})$ by Assumption 5.2, where $h_{NT}=\frac{1}{\sqrt[4]{\xi_{NT}}}$. 

Thus, for a large constant $A$, $\widehat{C}_\beta$ lies in a ball $\left\{C_\beta\, | \, \|C_\beta - C_{\beta_0}\| \le A h_{NT}^{1/2} \right\}$ with probability approaching 1. Corresponding to $\widehat{C}_\beta^*$ and $\widehat{C}_\beta^\dagger$, construct $C_\beta$ and $U$ as $C_\beta= ( C_\beta^{*\prime}, C_\beta^{\dagger\prime})'$ and $U= ( U^{*\prime}, U^{\dagger\prime})'$, where $C_\beta^* = C_{\beta_0}^*+ h_{NT}^{1/2} U^*$ and $C_\beta^\dagger= C_{\beta_0}^\dagger+ h_{NT}^{1/2} U^\dagger=h_{NT}^{1/2} U^\dagger$ with $ \|U\|^2\le A^2$. Further define
\begin{eqnarray}
V_{\lambda}(U^*, U^\dagger, F)  &=& \frac{1}{NT}Q_{\lambda}\big( C_\beta, F\big) = \frac{1}{NT}Q_{\lambda}\big( (C_{\beta_0}^{*\prime}+ h_{NT}^{1/2} U^{*\prime}, h_{NT}^{1/2} U^{\dagger\prime})', F\big)  ,
\end{eqnarray} 
so $\widehat{C}_\beta^{*}$ and $\widehat{C}_\beta^{\dagger}$ can be obtained by minimizing $V_{\lambda}(U^*, U^\dagger,F)$ over $\|U \|^2\le A^2$ except on an event with probability converging to 0. 

It suffices to show that for any $U= (U^*, U^\dagger)$ with $\|U \|^2\le A^2$ and $\| U^\dagger\|>0$, $V_{\lambda}(U^*, U^\dagger,F)> V_{NT}(U^*, 0_{(p-p^*)\times m},F)$ with probability converging to 1 regardless the value of $F$. Recall that some notations used below have been defined in the beginning of the supplementary file. Further denote that $\beta_m^\dagger(\cdot) = C_\beta^\dagger H_m(\cdot)$. Then write

\begin{eqnarray*}
&&V_{\lambda}(U^*, U^\dagger,F)- V_{NT}(U^*, 0_{(p-p^*)\times m},F) \nonumber\\
&= & \frac{1}{NT}\sum_{i=1}^N\Delta \phi_i^*[\beta_m^*]' M_F \Delta \phi_i^*[\beta_m^*]+ \frac{1}{NT}\sum_{i=1}^N\Delta \phi_i^\dagger[\beta_m^\dagger]' M_F \Delta \phi_i^\dagger[\beta_m^\dagger] \nonumber\\
&&+ \frac{1}{NT}\sum_{i=1}^N \gamma_{0i}' F_0' M_F F_0 \gamma_{0i} +\frac{1}{NT}\sum_{i=1}^N \mathcal{E}_i' M_F \mathcal{E}_i \nonumber\\
&&+ \frac{2}{NT}\sum_{i=1}^N\Delta \phi_i^*[\beta_m^*]' M_F\mathcal{E}_i+\frac{2}{NT}\sum_{i=1}^N\Delta \phi_i^\dagger[\beta_m^\dagger]' M_F\mathcal{E}_i + \frac{2}{NT}\sum_{i=1}^N \gamma_{0i}' F_0' M_F \mathcal{E}_i \nonumber\\
&&+\frac{2}{NT}\sum_{i=1}^N \Delta \phi_i^*[\beta_m^*]'  M_F F_0 \gamma_{0i}+\frac{2}{NT}\sum_{i=1}^N \Delta \phi_i^\dagger[\beta_m^\dagger]'  M_F F_0 \gamma_{0i}\nonumber \\
&&+\frac{2}{NT}\sum_{i=1}^N \Delta \phi_i^*[\beta_m^*]'  M_F \phi_i^\dagger[\beta_m^\dagger]+\sum_{j=1}^{p^*} \frac{\lambda_j}{NT} \|C_{\beta ,j} \| +\sum_{j=p^*+1}^{p} \frac{\lambda_j}{NT} \|C_{\beta ,j}\| \nonumber\\
&&-\frac{1}{NT}\sum_{i=1}^N\Delta \phi_i^*[\beta_m^*]' M_F \Delta \phi_i^*[\beta_m^*] - \frac{1}{NT}\sum_{i=1}^N \gamma_{0i}' F_0' M_F F_0 \gamma_{0i} -\frac{1}{NT}\sum_{i=1}^N \mathcal{E}_i' M_F \mathcal{E}_i\nonumber \\
&&- \frac{2}{NT}\sum_{i=1}^N\Delta \phi_i^*[\beta_m^*]' M_F\mathcal{E}_i - \frac{2}{NT}\sum_{i=1}^N \gamma_{0i}' F_0' M_F \mathcal{E}_i\nonumber \\
&& -\frac{2}{NT}\sum_{i=1}^N \Delta \phi_i^*[\beta_m^*]'  M_F F_0 \gamma_{0i}-\sum_{j=1}^{p^*} \frac{\lambda_j}{NT} \|C_{\beta ,j} \| \nonumber \\
&= & \frac{1}{NT}\sum_{i=1}^N\Delta \phi_i^\dagger[\beta_m^\dagger]' M_F \Delta \phi_i^\dagger[\beta_m^\dagger] +\frac{2}{NT}\sum_{i=1}^N \Delta \phi_i^*[\beta_m^*]'  M_F \phi_i^\dagger[\beta_m^\dagger] \nonumber\\
&&+\frac{2}{NT}\sum_{i=1}^N\Delta \phi_i^\dagger[\beta_m^\dagger]' M_F\mathcal{E}_i +\frac{2}{NT}\sum_{i=1}^N \Delta \phi_i^\dagger[\beta_m^\dagger]'  M_F F_0 \gamma_{0i}+\sum_{j=p^*+1}^{p} \frac{\lambda_j}{NT} \|C_{\beta ,j}\| \nonumber\\
&=&B_{1NT}+2B_{2NT}+2B_{3NT}+2B_{4NT}+B_{5NT},\nonumber
\end{eqnarray*}
where the definitions of $B_{jNT}$ for $j=1,\ldots,5$ are obvious.

For $B_{1NT}+2B_{2NT}$, write

\begin{eqnarray*}
B_{1NT}+2B_{2NT} &\ge & \frac{1}{NT}\sum_{i=1}^N\Delta \phi_i^\dagger[\beta_m^\dagger]' M_F \Delta \phi_i^\dagger[\beta_m^\dagger] \nonumber\\
&&- \frac{1}{NT}\sum_{i=1}^N \Delta \phi_i^*[\beta_m^*]'  M_F  \Delta \phi_i^*[\beta_m^*] - \frac{1}{NT}\sum_{i=1}^N \phi_i^\dagger[\beta_m^\dagger]'  M_F \phi_i^\dagger[\beta_m^\dagger]\nonumber\\
&\ge &- \frac{1}{NT}\sum_{i=1}^N \Delta \phi_i^*[\beta_m^*]'  M_F  \Delta \phi_i^*[\beta_m^*] \nonumber\\
&\ge &- h_{NT} \frac{1}{NT}\sum_{i=1}^N{U^*}'{\mathcal{Z}_i^*}' M_F \mathcal{Z}_i^* U^*\nonumber\\
&\ge &-h_{NT}\rho_1 \| U^*\|^2\ge -h_{NT}\rho_1 A^2\nonumber
\end{eqnarray*}
where the third inequality follows by construction.

For $B_{3NT}+B_{4NT}$, it is easy to know that

\begin{eqnarray*}
|B_{3NT}+B_{4NT}| &\le & \left\{\frac{1}{NT}\sum_{i=1}^N\| \Delta \phi_i^\dagger[\beta_m^\dagger] \|^2 \right\}^{1/2}\cdot  \left\{\frac{1}{NT}\sum_{i=1}^N\mathcal{E}_i' M_F\mathcal{E}_i  \right\}^{1/2}\nonumber\\
&&+ \left\{\frac{1}{NT}\sum_{i=1}^N\| \Delta \phi_i^\dagger[\beta_m^\dagger] \|^2 \right\}^{1/2}\cdot  \left\{\frac{1}{NT}\sum_{i=1}^N\gamma_{0i}'F_0' M_F F_0 \gamma_{0i} \right\}^{1/2}\nonumber\\
&=&O_P(1)h_{NT}^{1/2}.\nonumber
\end{eqnarray*}

For $B_{5NT}$, we have

\begin{eqnarray*}
B_{5NT} =\sum_{j=p^*+1}^{p} \frac{\lambda_j}{NT} \|C_{\beta ,j}\| \ge \frac{\lambda_{\text{min}}^\dagger}{NT} \left( \sum_{j=p^*+1}^{p}  \|C_{\beta ,j}\|^2\right)^{1/2} = \frac{\lambda_{\text{min}}^\dagger}{NT} \| U^\dagger\|.\nonumber
\end{eqnarray*}

In view of the above development and the condition $\frac{\lambda_{\text{min}}^\dagger}{NT h_{NT}^{1/2}}\to \kappa_3$ with $\kappa_3$ being sufficiently large, the first result follows.

\medskip

(2). As in the case of Theorem 3.2, we define three mutually exclusive sets $A^-$, $A^0 = \left\{\lambda \in \mathbb{R}^p: S_{\lambda} =\mathcal{A}^*  \right\}$  and $A^+ = \left\{\lambda \in \mathbb{R}^p: S_{\lambda} \supset \mathcal{A}^*, S_{\lambda}\ne \mathcal{A}^* \right\}$ according to whether the model $S_{\lambda}$ is under fitted, correctly fitted, or over fitted respectively. Suppose that there is a sequence $\{ \lambda_{NT} \}$ that satisfies he conditions required by the first result of this theorem. Let $(\widehat{C}^{\lambda_{NT}}  ,\widehat{F}^{\lambda_{NT}})$ denote the estimator obtained by implementing (3.3) using $\lambda_{NT}$. 

\textbf{Case 1: Under-fitted model.}  Without loss of generality, we assume that only one variable is missing, so suppose that the first $p^* - 1$ rows of $\widehat{C}_\beta^{\lambda}$ are obtained from the under-fitted model and the $ {p^*} ^{th}$ row of $\widehat{C}_\beta^{\lambda}$ is a 0 row.  Moreover, let $\text{RSS}_0 = \frac{1}{NT} \sum_{i=1}^N \big( Y_i -\phi_i[\beta_{0,m}]  \big)'M_{F_0}\big( Y_i - \phi_i[\beta_{0,m}]  \big)$.

We then write

\begin{eqnarray*}
\text{RSS}_{\lambda} -\text{RSS}_0&=& \frac{1}{NT} \sum_{i=1}^N \big( Y_i -\phi_i[\widehat{\beta}_{m}^\lambda ]  \big)'M_{\widehat{F}^\lambda}\big( Y_i - \phi_i[\widehat{\beta}_{m}^\lambda ]  \big) \nonumber\\ 
&&- \frac{1}{NT} \sum_{i=1}^N \big( Y_i -\phi_i[\beta_{0,m}]  \big)'M_{F_0}\big( Y_i - \phi_i[\beta_{0,m}]  \big) \nonumber\\
&\ge & \rho_1 \| C_{\beta_0, p^*}\|^2>\frac{\rho_1}{2}\|\beta_{0p^*} \|_{L^2}^2>0 ,\nonumber
\end{eqnarray*}
where the first inequality follows from the development of Lemma \ref{theorem4}. Similarly, we have

\begin{eqnarray*}
&&\text{RSS}_{\lambda_{NT}}-\text{RSS}_0 \nonumber \\
&=&\vect(C_{\beta_0}- \widehat{C}_{\beta}^{\lambda_{NT}} )'\frac{1}{NT}\sum_{i=1}^{N}\mathcal{Z}_i'M_{\widehat{F}^{\lambda_{NT}}}\mathcal{Z}_i\vect(C_{\beta_0}- \widehat{C}_{\beta}^{\lambda_{NT}} )\nonumber \\
&&+\frac{1}{NT}\text{tr}\left( M_{\widehat{F}^{\lambda_{NT}}} F_0\Gamma _0'\Gamma _0F_0'M_{\widehat{F}^{\lambda_{NT}}}\right)\nonumber \\
&&+2\vect(C_{\beta_0}- \widehat{C}_{\beta}^{\lambda_{NT}} )'\frac{1}{NT}\sum_{i=1}^{N}\mathcal{Z}_i'M_{\widehat{F}^{\lambda_{NT}}}F_0\gamma_{0i} +o_P(1) \nonumber  \\
&=&o_P(1).\nonumber
\end{eqnarray*}

Thus, we can conclude that $\Pr\left( \inf_{\lambda\in A^-}\text{BIC}_{\lambda}  >  \text{BIC}_{\lambda_{NT}}\right) \to 1$.

\bigskip

\textbf{Case 2: Over-fitted model.} Consider $\forall \lambda\in A^+$ and recall that $\widehat{C}_\beta^{\lambda}$ determines a model $S_{\lambda}$. Under such a model $S_{\lambda}$, we can define another unpenalized estimator  as

\begin{eqnarray}
(\check{C}_\beta,\check{F}) = \argmin_{C_\beta, F}   \frac{1}{NT} \sum_{i=1}^N \big( Y_i -\phi_i[\beta_m]  \big)'M_{F}\big( Y_i - \phi_i[\beta_m]  \big)
\end{eqnarray}
subject to $\| C_\beta\|\le a_0\sqrt{p}$ and $F\in \mathsf{D}_F$, where, for $j=1,\ldots,p$, $\| C_{\beta,j}\| = 0$ with $\forall j \notin S_{\lambda}$. In other words, $(\check{C}_\beta,\check{F})$ is the unpenalized estimator under the model determined by $\widehat{C}_\beta^{\lambda}$. By definition, we obtain immediately that $\text{RSS}_{\lambda} \ge \text{RSS}_{S_{\lambda}} $, where $\text{RSS}_{S_{\lambda}}  =   \frac{1}{NT} \sum_{i=1}^N \big( Y_i -\phi_i[\check{\beta}_m]  \big)'M_{\check{F}}\big( Y_i - \phi_i[\check{\beta}_m]  \big).$

Write

\begin{eqnarray*}
\ln \text{RSS}_{S_{\lambda}} -\ln  \text{RSS}_{\lambda_{NT}} &=&\ln \left(1+ \frac{\text{RSS}_{S_{\lambda}} - \text{RSS}_{\lambda_{NT}} }{ \text{RSS}_{\lambda_{NT}}}\right)\nonumber\\
&\ge &- \frac{\text{RSS}_{S_{\lambda}} - \text{RSS}_{\lambda_{NT}} }{ \text{RSS}_{\lambda_{NT}}} .\nonumber
\end{eqnarray*}
In view of the proof of Lemma \ref{theorem4}, it is easy to see that $\text{RSS}_{\lambda_{NT}}$ converges to a positive constant. As for $\text{RSS}_{S_{\lambda}} - \text{RSS}_{\lambda_{NT}} $, we obtain 

\begin{eqnarray*}
\text{RSS}_{S_{\lambda}} - \text{RSS}_{\lambda_{NT}}   &=&\frac{1}{NT} \sum_{i=1}^N \big( Y_i -\phi_i[\check{\beta}_m]  \big)'M_{\check{F}}\big( Y_i - \phi_i[\check{\beta}_m]  \big)\nonumber\\
&&-\frac{1}{NT} \sum_{i=1}^N \big( Y_i -\phi_i[\widehat{\beta}_m^{\lambda_{NT}}]  \big)'M_{\widehat{F}^{\lambda_{NT}}}\big( Y_i - \phi_i[\beta_m^{\lambda_{NT}}]  \big) .\nonumber
\end{eqnarray*}

Using the development of Lemma \ref{theorem4}, it is not hard to see $\left|\text{RSS}_{S_{\lambda}} - \text{RSS}_{\lambda_{NT}}   \right| \le O_P(1)  h_{NT}^{1/2}.$ Thus, we can further write

\begin{eqnarray*}
\ln \text{RSS}_{S_{\lambda}} -\ln  \text{RSS}_{\lambda_{NT}} &\ge &- \frac{\text{RSS}_{S_{\lambda}} - \text{RSS}_{\lambda_{NT}} }{ \text{RSS}_{\lambda_{NT}}}\ge  -\left|O_P(1)  h_{NT}^{1/2} \right|.\nonumber
\end{eqnarray*}
We then write

\begin{eqnarray*}
&&\inf_{\lambda\in A^+}\text{BIC}_{\lambda} - \text{BIC}_{\lambda_{NT}}=\inf_{\lambda\in A^+} \ln \text{RSS}_{S_{\lambda}} -\ln  \text{RSS}_{\lambda_{NT}} +(\text{df}_{\lambda} -\text{df}_{\lambda_{NT}})\Upsilon_{NT}.\nonumber
\end{eqnarray*}
By the first result of this theorem, we know that $\Pr (\text{df}_{\lambda_{NT}} = p^*)\to 1$. Since $\lambda\in A^+$, we must have $\Pr (\text{df}_{\lambda} \ge p^*+1)\to 1$. Given $\Upsilon_{NT} h_{NT}^{-1/2}\to \infty$, it is clear $\Pr\left( \inf_{\lambda\in A^+}\text{BIC}_{\lambda}  >  \text{BIC}_{\lambda_{NT}}\right) \to 1$.

Combining Cases 1 and 2, we obtain that $\Pr\left( \inf_{\lambda\in A^- \cup A^+} \text{BIC}_{\lambda}  >  \text{BIC}_{\lambda_{NT}}\right)\to 1$. This further indicates that $\Pr(S_{\widehat{\lambda}} = \mathcal{A}^*)\to 1$. The proof is now complete. \hspace*{\fill}{$\blacksquare$}

\section{Discussion on Time Trends}\label{TimeTrend}

As mentioned in Section 1, our primary focus is on proposing an integrated framework to tackle the three issues of variable selection, parameter heterogeneity, and cross-sectional dependence in the context of cross-country growth regressions. We do not attempt to address other interesting topics such as convergence of countries or any further model specification issues. That said, we would like to briefly discuss the issue of time trend that has found limited attention in the empirical growth literature (\citealp{Eberhardt}) but can be partially solved under our setting.

As pointed out by \cite{Eberhardt}, any macro production function is likely to contain at least some countries with certain time trends in the input and output variables, and these time-series properties need to be taken into account in the empirical analysis. The question in fact has somewhat been addressed by recent developments in econometrics. We provide two examples below.

Though macro variables are likely to have certain time trends, they do not have to be as strong as polynomial terms $t$ or $t^2$ (\citealp{DongLinton}). More often than not, it can simply be captured by a structure like

\begin{eqnarray} \label{newX1}
x_{it} =A_i' d_{0t}+ B_i' f_{0t} + u_{it}, 
\end{eqnarray}
where $f_{0t}$ is the same as that of (2.4) of the main text, $d_{0t}$ includes other unobservable common shocks, both $A_i$ and $B_i$ are the unknown factor loadings, and $u_{it}$ stands for an error term. This structure has been well discussed in \cite{Pesaran2006} and \cite{Kapetanios}.

In the second case, as in \cite{Pedroni} we assume that different countries have different types of time trends. This case has been partially discussed in \cite{ChenGaoLiA} and \cite{GaoXiaZhu}. In particular, \cite{GaoXiaZhu} allow the regressors $x_{it}$ to have the following form 

\begin{eqnarray}\label{newX2}
 x_{it} = g_i(t/T) + v_i +u_{it},
\end{eqnarray}
where $g_i(\cdot)$ is some trending function and varies across $i$, $v_i$ is an individual effect, and $u_{it}$ stands for an error term. Using (\ref{newX2}), $g_i(t/T)$ can mimic different types of time trends for each individual country.

For either case of (\ref{newX1}) and (\ref{newX2}), our methodology including the estimator and its asymptotic results remains valid with some minor modifications regarding the proof. 

\section{Additional Tables and Figures for the Empirical Results}\label{SectionA1}

\begin{figure}[H]\caption{Estimates of Coefficient Functions of All Selected Variables }\label{Fig_Est}
\centering
\hspace*{-2cm} \includegraphics[scale=0.65]{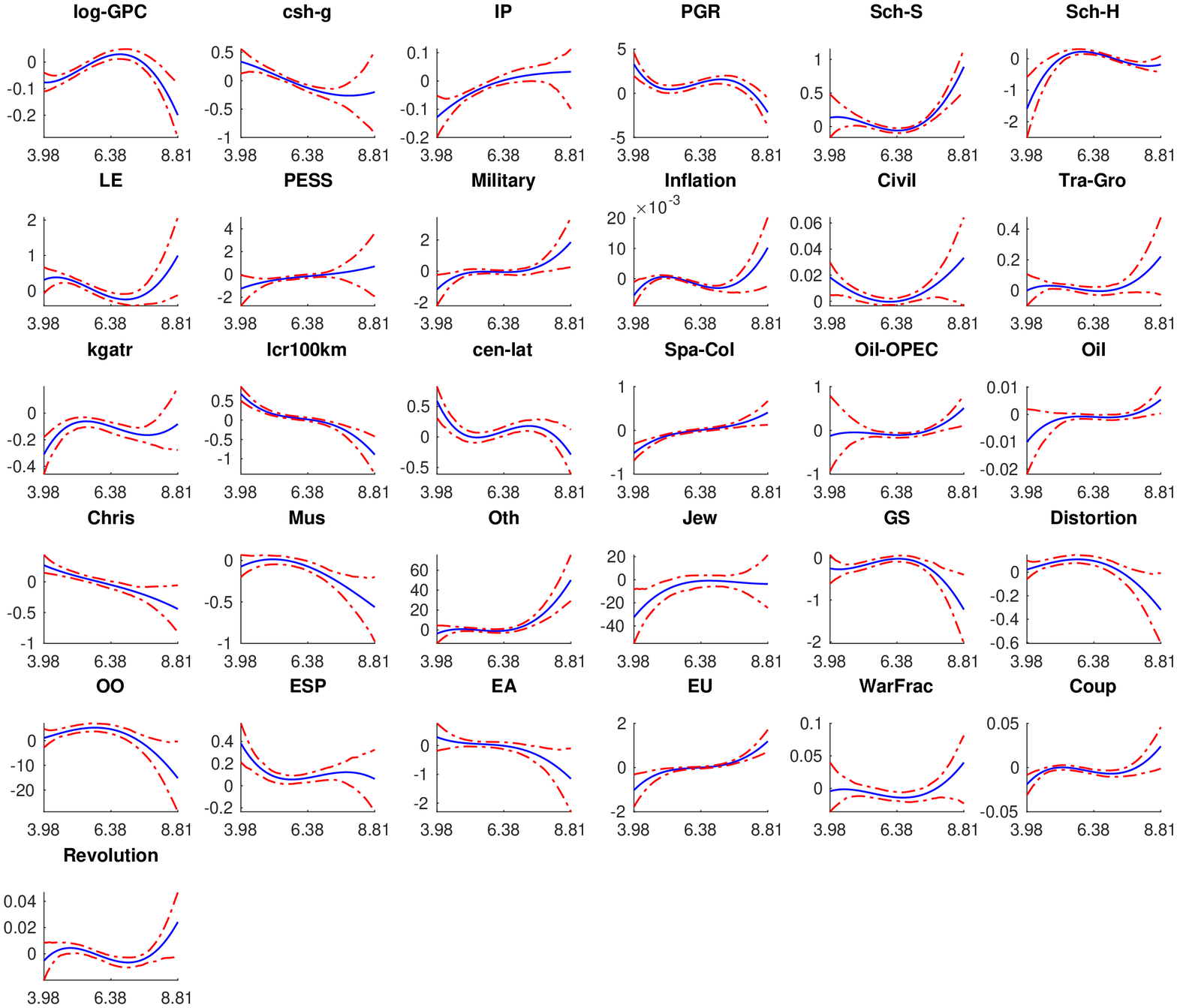}
\caption*{\footnotesize As discussed in the context of Figure 1 in the main text, these confidence intervals need to be interpreted with caution. As well understood, one cannot establish the confidence intervals for the estimates under HD case unless certain transformation is further employed  (e.g., \citealp{HuangHorowitzMa, DGL2017}). However, if one regards 31 as a relatively small number after selection, we can then employ a procedure similar to the relevant literature by considering our regression under LD framework. In order to ensure the validity of the bootstrap procedure, stronger assumptions on the error terms are needed. For example, one can employ the martingale difference type of assumptions (see Assumption A.4 of \citealp{SuJinZhang}), or simply assume that the error terms are i.i.d. over both $i$ and $t$. Generally speaking, when the error term exhibits both cross-sectional and serial correlation, the bootstrap results are not reliable or incorrect.}
\end{figure}

\newpage
\setcounter{page}{1}
\renewcommand{\theequation}{B.\arabic{equation}}
\renewcommand{\thesection}{B.\arabic{section}}

\setcounter{equation}{0}
\setcounter{lemma}{0}
\setcounter{section}{0}
\setcounter{table}{0}
\setcounter{figure}{0}

\begin{center}
{\Large \bf Supplementary Appendix B to \\``An Integrated Panel Data Approach to \\Modelling Economic Growth"}

\bigskip

(NOT for publication)

\bigskip

\medskip

{\sc Guohua Feng$^{\ast}$, Jiti Gao$^\sharp$ and Bin Peng$^{\dag}$ }
\medskip

$^{\ast}$University of North Texas, $^\sharp$Monash University and $^{\dag}$University of Bath
\end{center}

\small

\bigskip
\bigskip

Appendix B provides proofs of the preliminary lemmas stated in the supplementary Appendix A.

\noindent \textbf{Proof of Lemma A.1:}

Write
\begin{eqnarray*}
\left\|A^{-1}-B^{-1}\right\| &=&
\left\|B^{-1}\left(B-A\right)A^{-1}\right\| =\left\|\vect\left(B^{-1}\left(B-A\right)A^{-1}\right)\right\| \nonumber \\
&=&\left\|\left(A^{-1}\otimes B^{-1}\right)\vect\left(B-A\right)\right\| \le \eta_{\normalfont\text{min}}^{-1}\left(A\otimes B\right)\left\|\vect\left(B-A\right)\right\| \nonumber\\
&=& \eta_{\normalfont\text{min}}^{-1}\left(A\right)\cdot \eta_{\normalfont\text{min}}^{-1}\left( B\right)\left\|A-B\right\|.\nonumber
\end{eqnarray*}
The proof is then complete.
\hspace*{\fill}{$\blacksquare$}

\bigskip

\noindent \textbf{Proof of Lemma A.2:}

(1). Firstly, write

\begin{eqnarray}\label{EqA9}
&&\frac{1}{N^2T^2}E\| \mathcal{E}'\mathcal{E}\| ^2 =E\left\|\frac{1}{NT}\sum_{i=1}^N \mathcal{E}_i \mathcal{E}_i'\right\|^2=\frac{1}{N^2 T^2} \sum_{t=1}^T \sum_{s=1}^T \left(\sum_{i=1}^NE[\varepsilon_{it}^2\varepsilon_{is}^2] +\sum_{i\ne j}
E[\varepsilon_{it}\varepsilon_{is}\varepsilon_{jt}\varepsilon_{js}]\right)\nonumber \\
&=& \frac{1}{N^2 T^2}\sum_{t=1}^T \sum_{s=1}^T \left(\sum_{i=1}^N E[\varepsilon_{it}^2 \varepsilon_{is}^2] +\sum_{i\ne j}
E[(\varepsilon_{it}\varepsilon_{jt}-\sigma_{ij})(\varepsilon_{is}\varepsilon_{js}-\sigma_{ij})] +\sum_{i\ne j}\sigma_{ij}^2\right)\nonumber\\
&=& \frac{1}{N^2 T^2}\sum_{t=1}^T \left(\sum_{i=1}^N E[\varepsilon_{it}^4] +\sum_{i\ne j}
E[(\varepsilon_{it}\varepsilon_{jt}-\sigma_{ij})^2]  \right)\nonumber\\
&&+ \frac{1}{N^2 T^2}\sum_{t\ne s}\left(\sum_{i=1}^N E[\varepsilon_{it}^2 \varepsilon_{is}^2] +\sum_{i\ne j}
E[(\varepsilon_{it}\varepsilon_{jt}-\sigma_{ij})(\varepsilon_{is}\varepsilon_{js}-\sigma_{ij})] \right)+\frac{1}{N^2} \sum_{i\ne j}\sigma_{ij}^2\nonumber\\
&=&O(1)\frac{1}{N}+O(1)\frac{1}{T},
\end{eqnarray}
where the fifth equality follows from using the mixing condition on $e_{it}e_{jt}$ across $t$. Thus, $\frac{1}{NT}\|  \mathcal{E}' \mathcal{E}\| =O_P\left(\frac{1}{\sqrt{N}}\right)+O_P\left(\frac{1}{\sqrt{T}}\right)$.

Secondly, note that

\begin{eqnarray*}
&&E\left\| \frac{1}{NT}\mathcal{E}\mathcal{E}'\right\|^2 = \left\{E \left[\frac{1}{NT} \mathcal{E}_i '\mathcal{E}_j\right]^2 \right\}_{N\times N}= \sum_{i=1}^N \sum_{j=1}^N\frac{1}{N^2T^2} \sum_{t=1}^T \sum_{s=1}^T E \left[ \varepsilon_{it} \varepsilon_{jt} \varepsilon_{is} \varepsilon_{js}\right] \nonumber \\
&=& \sum_{t=1}^T \sum_{s=1}^T \frac{1}{N^2 T^2}\left(\sum_{i=1}^N
E[\varepsilon_{it}^2 \varepsilon_{is}^2] +\sum_{i\ne j}
E[\varepsilon_{it}\varepsilon_{is}\varepsilon_{jt}\varepsilon_{js}]\right)= O\left(\frac{1}{N}\right)+O\left(\frac{1}{T}\right),\nonumber
\end{eqnarray*}
where the last step follows from (\ref{EqA9}). Thus, $\frac{1}{NT}\|  \mathcal{E} \mathcal{E}'\| =O_P\left(\frac{1}{\sqrt{N}}\right)+O_P\left(\frac{1}{\sqrt{T}}\right)$.

\medskip

(2). Write

\begin{eqnarray*}
\sup_{F\in \mathsf{D}_F}\frac{1}{NT}\sum_{i=1}^{N} \mathcal{E}_i^{\prime }P_{F} \mathcal{E}_i &=&\sup_{F\in \mathsf{D}_F}\frac{1}{NT}\text{tr}\left( P_{F} \mathcal{E}' \mathcal{E} \right) \le \sup_{F\in \mathsf{D}_F}\frac{r}{NT}\|P_F \|_{\text{sp}}\|\mathcal{E}'\mathcal{E} \|_{\text{sp}} \nonumber \\
&\le & \sup_{F\in \mathsf{D}_F}\frac{r}{NT}\|P_F \|_{\text{sp}}\|\mathcal{E}'\mathcal{E}\| =O_P\left(\frac{1}{\sqrt{N}}\right)+O_P\left(\frac{1}{\sqrt{T}}\right),\nonumber
\end{eqnarray*}
where the first inequality follows from the fact that $\left\vert \text{tr}\left( A\right) \right\vert \leq \text{rank}\left( A\right)\left\Vert A\right\Vert _{\text{sp}}$; and the second equality follows from (1) of this lemma.

\medskip

(3). Write

\begin{eqnarray*}
&&\sup_{F\in \mathsf{D}_F} \left\vert \frac{1}{NT}\sum_{i=1}^{N}\gamma_{0i}'F_0 ' M_F \mathcal{E}_i\right\vert = \sup_{F\in \mathsf{D}_F}\left\vert \frac{1}{NT}\text{tr}\left(F_0' M_{F}\mathcal{E}'\Gamma_0\right) \right\vert \leq \sup_{F\in \mathsf{D}_F}\frac{r}{NT}\left\Vert F_0' M_F\mathcal{E}'\Gamma_0\right\Vert _{\text{sp}}  \nonumber\\
&\leq &\sup_{F\in \mathsf{D}_F} \frac{r}{NT}\left\Vert F_0\right\Vert_{\text{sp}}\left\Vert M_F\right\Vert _{\text{sp}}\left\Vert  \Gamma_0'\mathcal{E}\right\Vert _{\text{sp}} =\sup_{F\in \mathsf{D}_F} \frac{r}{NT}\left\Vert F_0\right\Vert _{\text{sp}}\left\Vert  \Gamma_0' \mathcal{E}\mathcal{E}' \Gamma_0\right\Vert _{\text{sp}}^{1/2}   \nonumber \\
&=&\sup_{F\in \mathsf{D}_F} \frac{r}{\sqrt{NT}}\left\Vert F_0\right\Vert _{\text{sp}} \left\Vert \Gamma_0\right\Vert _{\text{sp}}\left(\frac{1}{NT}\left\Vert \mathcal{E}\mathcal{E}' \right\Vert\right)^{1/2}   =O_P\left(\frac{1}{NT}\left\Vert \mathcal{E}\mathcal{E}' \right\Vert\right)^{1/2} \nonumber\\
&=&O_P\left(\frac{1}{\sqrt[4]{N}}\right)+O_P\left(\frac{1}{\sqrt[4]{T}}\right),\nonumber
\end{eqnarray*}
where  the first inequality follows from the fact that $\left\vert \text{tr}\left( A\right) \right\vert \leq \text{rank}\left( A\right)\left\Vert A\right\Vert _{\text{sp}}$; the second equality follows from Fact 5.10.18 of \cite{Bernstein}; and the last equality follows from (1) of this lemma.

\medskip

(4). Write

\begin{eqnarray*}
&&\frac{1}{NT}\sum_{i=1}^N \left(\phi_i[\beta_{0,m}]-\phi_i[\beta_m] \right)' M_F \mathcal{E}_i \nonumber\\
&=&\frac{1}{NT}\sum_{i=1}^N \left(\phi_i[\beta_{0,m}]-\phi_i[\beta_m] \right)'\mathcal{E}_i+\frac{1}{NT}\sum_{i=1}^N \left(\phi_i[\beta_{0,m}]-\phi_i[\beta_m] \right)' P_F \mathcal{E}_i \nonumber\\
&:=&\Lambda_1 +\Lambda_2. \nonumber
\end{eqnarray*}

For $\Lambda_1$, write

\begin{eqnarray*}
&&\sup_{\|  C_\beta \|\le M }\left|\frac{1}{NT}\sum_{i=1}^N \left(\phi_i[\beta_{0,m}]-\phi_i[\beta_m] \right)'\mathcal{E}_i\right|\le\sup_{\|  C_\beta \|\le M } \| \vect(C_{\beta_0} - C_\beta)\|\cdot \left\|\frac{1}{NT}\sum_{i=1}^N\mathcal{Z}_i' \mathcal{E}_i\right\|\nonumber \\
&=&\sup_{\|  C_\beta \|\le M }\|C_{\beta_0} - C_\beta \| \cdot O_P\left(\sqrt{\frac{m}{NT}}\right)=O_P\left(\sqrt{\frac{m}{NT}}\right),\nonumber
\end{eqnarray*}
where the first equality follows from some standard analysis on the term $\frac{1}{NT}\sum_{i=1}^N\mathcal{Z}_i' \mathcal{E}_i$ using Assumption 1.1; and the last equality follows from $\|C_\beta\|\le M$. 

In order to consider $\Lambda_2$, let $\Delta b= (\phi_1[\beta_{0,m}]-\phi_1[\beta_m],\ldots ,\phi_N[ \beta_{0,m}]-\phi_N[ \beta_m])$. Note that

\begin{eqnarray}\label{EqA10}
\sup_{\|  C_\beta \|\le M}\frac{1}{NT} \left\|\Delta b\right\|^2  &=&\sup_{\|  C_\beta \|\le M}\frac{1}{NT}\sum_{i=1}^N  (C_\beta -C_{\beta_0})'\mathcal{Z}_i'  \mathcal{Z}_i (C_\beta -C_{\beta_0})\nonumber \\
&\le &O_P(1)\sup_{\|  C_\beta \|\le M} \| C_\beta -C_{\beta_0} \|^2 =O_P(1),
\end{eqnarray}
where the inequality follows from Assumptions 2.1, and $\left\|\frac{1}{NT}\sum_{i=1}^N  \mathcal{Z}_i'  \mathcal{Z}_i' -\Sigma_{\mathcal{Z}}\right\|=o_P\left(1\right)$ by some standard analysis using Assumption 1.1; and the last equality follows from $\|  C_\beta \|\le M$.

Then we are able to write

\begin{eqnarray*}
&&\sup_{\|  C_\beta \|\le M, \, F\in \mathsf{D}_F}\left| \frac{1}{NT}\sum_{i=1}^N \left(\phi_i[\beta_{0,m}]-\phi_i[\beta_m] \right)' P_F \mathcal{E}_i\right| =\sup_{\|  C_\beta \|\le M, \, F\in \mathsf{D}_F}   \left| \frac{1}{NT}  \textrm{tr}\left( P_F  \mathcal{E}' \Delta b'\right)\right| \nonumber\\
&\le&  \frac{r}{NT} \sup_{\|  C_\beta \|\le M, \, F\in \mathsf{D}_F} \|P_F  \mathcal{E}' \Delta b' \|_{\textrm{sp}} \le   \sup_{\|  C_\beta \|\le M, \, F\in \mathsf{D}_F}\frac{r}{NT}  \|P_F\|_{\textrm{sp}} \| \mathcal{E}\|_{\textrm{sp}}\|  \Delta b \|_{\textrm{sp}}\nonumber\\
&=&\sup_{\|  C_\beta \|\le M, \, F\in \mathsf{D}_F} r  \|P_F\|_{\textrm{sp}}\left(\frac{1}{NT} \| \mathcal{E} \mathcal{E}'\|\right)^{1/2} \left( \frac{1}{\sqrt{NT}}\|  \Delta b \|\right)=O_P\left(\frac{1}{\sqrt[4]{N}}\right)+O_P\left(\frac{1}{\sqrt[4]{T}}\right),\nonumber
\end{eqnarray*}
where the second equality follows from Fact 5.10.18 of \cite{Bernstein}; and the last step follows from (1) of this lemma and (\ref{EqA10}).

Based on the above development on $\Lambda_1$ and $\Lambda_2$, the result follows.

\medskip

(5). Write

\begin{eqnarray*}
\sup_{F\in  \mathsf{D}_F} \left\vert \frac{1}{NT}\sum_{i=1}^{N}\phi _{i}\left[  \Delta_{ m}\right]' M_{F}\phi _{i}\left[ \Delta_{ m}\right]\right\vert &\leq &\left\vert \frac{1}{NT}\sum_{i=1}^{N}\phi _{i}\left[ \Delta_{ m}\right]'\phi _{i}\left[ \Delta_{ m}\right] \right\vert =O_P\left(m^{-\mu}\right), \nonumber
\end{eqnarray*}
where the last equality follows from the development of \cite{DongLinton} using Assumption 2.1.

\medskip

(6). Write
\begin{eqnarray*}
&&\sup_{F\in  \mathsf{D}_F} \left\vert \frac{1}{NT}\sum_{i=1}^{N}\phi _{i}\left[  \Delta_{ m}\right] ^{\prime }M_{F}F_0\gamma _{0i}\right\vert =\sup_{F\in  \mathsf{D}_F} \left\vert \frac{1}{NT}\text{tr}\left( M_{F}F_0\Gamma _0^{\prime }\Delta^{\prime }\right) \right\vert \nonumber\\
&\leq &\frac{r}{NT}\sup_{F\in  \mathsf{D}_F} \left\Vert M_{F}\right\Vert _{\text{sp}}\left\Vert
F_0\right\Vert _{\text{sp}}\left\Vert \Gamma _0\right\Vert _{\text{sp}}\left\Vert \Delta \right\Vert _{\text{sp}} =O_P(m^{-\frac{\mu}{2}}), \nonumber
\end{eqnarray*}
where $\Delta = (\phi_1(\Delta_m),\ldots, \phi_N(\Delta_m))$; and the second equality follows from $\frac{1}{NT}\left\Vert \Delta  \right\Vert ^2=O_P\left(m^{-\mu}\right)$ as in (5) of this lemma.

\medskip

(7). Write
\begin{eqnarray*}
&&\sup_{\|  C_\beta \|\le M, \, F\in \mathsf{D}_F}\left\vert \frac{1}{NT}\sum_{i=1}^{N}\phi _{i}\left[  \Delta_{m}\right] ^{\prime }M_{F}\left\{\phi _{i} \left[\beta_m\right] -\phi _{i} \left[\beta_{0,m}\right] \right\} \right\vert \nonumber\\
&\leq & \left\{ \frac{1}{NT}\sum_{i=1}^{N}\|\phi _{i}\left[  \Delta_{m}\right] \|^2\right\} ^{1/2}  \cdot\sup_{\|  C_\beta \|\le M}\left\{ \frac{1}{NT}\sum_{i=1}^{N}\|\phi _{i} \left[\beta_m\right] -\phi _{i} \left[\beta_{0,m}\right] \|^2 \right\}^{1/2}= O_P(m^{-\frac{\mu}{2}}),\nonumber
\end{eqnarray*}
where the last equality follows from (5) of this lemma. The proof is now complete. \hspace*{\fill}{$\blacksquare$}

\bigskip

\noindent \textbf{Proof of Lemma A.3:}

(1). For notational simplicity, let $\Delta\phi_i[\beta_m] = \phi_i[\beta_{0,m}] - \phi_i[\beta_{m}]$. Let $\xi_F =\text{vec}\left( M_F F_0\right) ,$ $A_{1F}=\frac{1}{NT}\sum_{i=1}^{N}\mathcal{Z}_i'M_{F}\mathcal{Z}_i$, $A_{2}=\frac{1}{NT}\left( \Gamma_0'\Gamma_0\right) \otimes I_{T},$ and $A_{3F}=\frac{1}{NT}\sum_{i=1}^{N}\gamma_{0i} \otimes (M_F\mathcal{Z}_i)$, where $\mathcal{Z}_i$ has been defined in Assumption 2.  By the definition of (3.3) and Lemma A.2, we have

\begin{eqnarray*} 
0 &\ge & \frac{1}{NT}Q_{\lambda}( \widehat{C}_\beta, \widehat{F}) - \frac{1}{NT}Q_{\lambda}( C_{\beta_0}, F_0)\nonumber \\
&=& \frac{1}{NT} \sum_{i=1}^N \left( \Delta\phi_i[\widehat{\beta}_m]+F_0\gamma_{0i} \right)'M_{\widehat{F}}\left(\Delta\phi_i[\widehat{\beta}_m]+F_0\gamma_{0i} \right) +\frac{1}{NT}\sum_{i=1}^N\mathcal{E}_i' M_{\widehat{F}} \mathcal{E}_i\nonumber \\
&& +\frac{2}{NT}\sum_{i=1}^N \left( \Delta\phi_i[\widehat{\beta}_m]+F_0\gamma(v_i) \right)'M_{\widehat{F}} \mathcal{E}_i +\sum_{j=1}^{p} \frac{\lambda_j}{NT} \|\widehat{C}_{\beta ,j} \| \nonumber \\
&&- \frac{1}{NT} \sum_{i=1}^N \left( \phi_i[\Delta_m]+\mathcal{E}_i\right)'M_{F_0}\left( \phi_i[\Delta_m]+\mathcal{E}_i \right) - \sum_{j=1}^{p^*} \frac{\lambda_j}{NT}\|C_{\beta_0 ,j} \|\nonumber \\
&=& \frac{1}{NT} \sum_{i=1}^N \left( \Delta\phi_i[\widehat{\beta}_m]+F_0\gamma_{0i} \right)'M_{\widehat{F}}\left(\Delta\phi_i[\widehat{\beta}_m]+F_0\gamma_{0i} \right)\nonumber\\
&&+\sum_{j=1}^{p} \frac{\lambda_j}{NT} \|\widehat{C}_{\beta ,j} \| - \sum_{j=1}^{p^*} \frac{\lambda_j}{NT}\|C_{\beta_0 ,j} \|+O_P\left(  \frac{1}{\sqrt[4]{\xi_{NT}}} + m^{-\frac{\mu}{2}}\right)\nonumber\\
&=& \vect(C_{\beta_0}-\widehat{C}_\beta )'\frac{1}{NT}\sum_{i=1}^{N}\mathcal{Z}_i'M_{\widehat{F}}\mathcal{Z}_i\vect(C_{\beta_0}-\widehat{C}_\beta )+\frac{1}{NT}\text{tr}\left( M_{\widehat{F}} F_0\Gamma _0'\Gamma _0F_0'M_{\widehat{F}}\right)\nonumber \\
&&+2\vect(C_{\beta_0}-\widehat{C}_\beta )'\frac{1}{NT}\sum_{i=1}^{N}\mathcal{Z}_i'M_{\widehat{F}}F_0\gamma_{0i} +\sum_{j=1}^{p} \frac{\lambda_j}{NT} \|\widehat{C}_{\beta ,j} \| - \sum_{j=1}^{p^*} \frac{\lambda_j}{NT}\|C_{\beta_0 ,j} \|\nonumber \\
&&+O_P\left(  \frac{1}{\sqrt[4]{\xi_{NT}}} + m^{-\frac{\mu}{2}}\right),\nonumber
\end{eqnarray*}
where the second equality follows from Lemma A.2. Thus, we can further write

\begin{eqnarray}\label{LDconsis0}
\sum_{j=1}^{p^*} \frac{\lambda_j}{NT}\|C_{\beta_0 ,j} \|&\ge &\vect(C_{\beta_0}-\widehat{C}_\beta )'\frac{1}{NT}\sum_{i=1}^{N}\mathcal{Z}_i'M_{\widehat{F}}\mathcal{Z}_i\vect(C_{\beta_0}-\widehat{C}_\beta )+\frac{1}{NT}\text{tr}\left( M_{\widehat{F}}F_0\Gamma _0'\Gamma _0F_0'M_{\widehat{F}}\right)\nonumber \\
&&+2\vect(C_{\beta_0}-\widehat{C}_\beta )'\frac{1}{NT}\sum_{i=1}^{N}\mathcal{Z}_i'M_{\widehat{F}}F_0\gamma_{0i} +O_P\left(  \frac{1}{\sqrt[4]{\xi_{NT}}} + m^{-\frac{\mu}{2}}\right) \nonumber \\
&\ge &\vect(C_{\beta_0}-\widehat{C}_\beta )'\left(A_{1\widehat{F}}-A_{3\widehat{F}}'A_{2}^{-1}A_{3\widehat{F}}\right)\vect(C_{\beta_0}-\widehat{C}_\beta ) \nonumber \\
&&+[\xi_{\widehat{F}}'+\vect(C_{\beta_0}-\widehat{C}_\beta )'A_{3\widehat{F}}'A_{2}^{-1}]A_{2}[\xi_{\widehat{F}}+A_{2}^{-1}A_{3\widehat{F}}\vect(C_{\beta_0}-\widehat{C}_\beta )]\nonumber \\
&&+O_P\left(  \frac{1}{\sqrt[4]{\xi_{NT}}} + m^{-\frac{\mu}{2}}\right) \nonumber\\
&\ge &O_P(1)  \|C_{\beta_0}-\widehat{C}_\beta \|^2+O_P\left(  \frac{1}{\sqrt[4]{\xi_{NT}}} + m^{-\frac{\mu}{2}}\right).
\end{eqnarray}
Till now, we can conclude that 

\begin{eqnarray}\label{LDconsis1}
\|C_{\beta_0}-\widehat{C}_\beta \|^2 =O_P\left( \frac{1}{\sqrt[4]{\xi_{NT}}} +m^{-\frac{\mu}{2}}+\frac{\lambda_{\text{max}}^* }{NT}\right)=o_P(1),
\end{eqnarray}
where the second equality follows from Assumption 3.

\medskip

(2). By (\ref{LDconsis0}) and (\ref{LDconsis1}), we can further obtain that

\begin{eqnarray*}
o_p(1)\ge \frac{1}{NT}\text{tr}\left[ \left( F_0'M_{\widehat{F}}F_0\right) \left( \Gamma_0'\Gamma _0\right) \right] +o_{P}\left( 1\right) ,\nonumber
\end{eqnarray*}
so $\frac{1}{NT}\text{tr}\left[ \left( F_0'M_{\widehat{F}}F_0\right) \left( \Gamma_0'\Gamma _0\right) \right] =o_{P}\left( 1\right) $. As in \citet[p. 1265]{Bai}, we can further conclude that $\frac{1}{T}\textrm{tr}\left( F_0'M_{\widehat{F}}F_0\right)=o_{P}\left( 1\right) $, $\left\Vert P_{\widehat{F}}-P_{F_0}\right\Vert =o_{P}\left( 1\right) $, and $\frac{1}{T}\widehat{F}'F_0$ is invertible with probability approaching one. Thus, the second result of this lemma follows. 

\medskip

(3). For the results result of this lemma, note that minimizing (3.2) with respect to $F$ is equivalent to minimizing $\sum_{i=1}^N \left( Y_i -\phi_i[\beta_m]  \right)'M_F\left( Y_i -\phi_i[\beta_m]  \right)$ without involving the penalty term. Thus, following the same arguments as in \citet[p. 1236]{Bai}, the estimate $\widehat{F}$ of (3.3) is obtained by

\begin{eqnarray}\label{EstF}
\frac{1}{NT}\sum_{i=1}^N\left( Y_i - \phi_i[\widehat{\beta}_m] \right)\left( Y_i - \phi_i[\widehat{\beta}_m] \right)'\widehat{F} = \widehat{F}V_{NT},
\end{eqnarray}
where $V_{NT}$ is a diagonal matrix with the diagonal being the $r$ largest eigenvalues of

\begin{eqnarray*}
\frac{1}{NT}\sum_{i=1}^N\left( Y_i - \phi_i[\widehat{\beta}_m] \right)\left( Y_i - \phi_i[\widehat{\beta}_m] \right)'\nonumber
\end{eqnarray*}
arranged in descending order.  

We now consider $V_{NT}$ and write

\begin{eqnarray*}
&&\widehat{F} V_{NT}=\left[\frac{1}{NT}\sum_{i=1}^N\left( Y_i - \phi_i[\widehat{\beta}_m  ] \right)\left( Y_i - \phi_i[\widehat{\beta}_m  ] \right)'\right]\widehat{F}  \nonumber\\
&=&\left[\frac{1}{NT}\sum_{i=1}^N\left( \phi_i [\beta_0 ] +F_0\gamma_{0i}+\mathcal{E}_i-  \phi_i[\widehat{\beta}_m  ]\right)\left( \phi_i [\beta_0 ] +F_0\gamma_{0i}+\mathcal{E}_i-  \phi_i[\widehat{\beta}_m  ]\right)'\right]\widehat{F} \nonumber \\
&=& \frac{1}{NT}\sum_{i=1}^N\left(\phi_i [\beta_0 ] - \phi_i [\widehat{\beta}_m  ]\right)\left( \phi_i [\beta_0 ] - \phi_i [\widehat{\beta}_m  ]\right)' \widehat{F}\nonumber\\
&&+ \frac{1}{NT}\sum_{i=1}^N\left( \phi_i [\beta_0 ] - \phi_i [\widehat{\beta}_m  ]\right)\left(F_0\gamma_{0i}\right)'\widehat{F} + \frac{1}{NT}\sum_{i=1}^N\left(F_0\gamma_{0i}\right)\left( \phi_i
[\beta_0 ] - \phi_i [\widehat{\beta}_m  ]\right)^{\prime }\widehat{F} \nonumber \\
&&+ \frac{1}{NT}\sum_{i=1}^N\left(\phi_i [\beta_0 ] - \phi_i [\widehat{\beta}_m  ]\right)\mathcal{E}_i' \widehat{F} + \frac{1}{NT}\sum_{i=1}^N\mathcal{E}_i\left( \phi_i [\beta_0 ] - \phi_i [\widehat{\beta}_m   ]\right)' \widehat{F}  \nonumber \\
&&+ \frac{1}{NT}\sum_{i=1}^N \mathcal{E}_i \mathcal{E}_i'\widehat{F}   + \frac{1}{NT}\sum_{i=1}^NF_0\gamma_{0i} \mathcal{E}_i' \widehat{F} + \frac{1}{NT}\sum_{i=1}^N \mathcal{E}_i\gamma_{0i}' F_0' \widehat{F} + \frac{1}{NT}\sum_{i=1}^NF_0\gamma_{0i}\gamma_{0i}' F_0' \widehat{F} \nonumber\\
&:=&I_{1NT}(\widehat{\beta}_m   ,\widehat{F}  )+\cdots +I_{5NT} (\widehat{\beta}_m   , \widehat{F}) +I_{6NT}(\widehat{F} )+\cdots +I_{9NT}(\widehat{F} ),\nonumber
\end{eqnarray*}
where the definitions of $I_{1NT}(\beta,F)$ to $I_{5NT}(\beta,F)$ and $I_{6NT}(F)$ to $I_{9NT}(F)$ should be obvious.

Note that $I_{9NT} (\widehat{F}  )=F_0(\Gamma_0'\Gamma_0/N)(F_0'\widehat{F}  /T)$. Thus, we can write

\begin{eqnarray}  \label{EqA15}
&&\widehat{F}  V_{NT}-F_0(\Gamma_0'\Gamma_0/N)(F_0'\widehat{F}  /T)\nonumber\\
&=&I_{1NT}(\widehat{\beta}_m   ,\widehat{F}  )+\cdots +I_{5NT} (\widehat{\beta}_m   , \widehat{F}  ) +I_{6NT}(\widehat{F}  )+\cdots +I_{8NT}(\widehat{F}  ).
\end{eqnarray}
Right multiplying each side of (\ref{EqA15}) by $(F_0'\widehat{F}  /T)^{-1}(\Gamma_0'\Gamma_0/N)^{-1}$, we obtain

\begin{eqnarray}  \label{EqA16}
&&\widehat{F}V_{NT}(F_0'\widehat{F}/T)^{-1}(\Gamma_0'\Gamma_0/N)^{-1}-F_0  \nonumber \\
&=&\left[I_{1NT}(\widehat{\beta}_m  ,\widehat{F} )+\cdots +I_{8NT}(\widehat{F} )\right](F_0'\widehat{F} /T)^{-1}(\Gamma_0' \Gamma_0/N)^{-1}.
\end{eqnarray}

Below, we examine each term on the right hand side of (\ref{EqA16}) and show that $V_{NT}$ is non-singular. Write

\begin{eqnarray}  \label{EqA17}
&&\frac{1}{\sqrt{T}}\left\|\widehat{F}  V_{NT}(F_0' \widehat{F}  /T)^{-1}(\Gamma_0'\Gamma_0/N)^{-1}-F_0 \right\|   \nonumber \\
&\le &\frac{1}{\sqrt{T}}\left[ \| I_{1NT}(\widehat{\beta}_m  ,\widehat{F} )\|+\cdots +\|I_{8NT}(\widehat{F} )\|\right] \cdot \|(F_0'\widehat{F}  /T)^{-1}(\Gamma_0'\Gamma_0/N)^{-1}\|.
\end{eqnarray}
We already know $(F_0'\widehat{F}  /T)^{-1}=O_P(1)$ by the proofs of the second result of this lemma and $(\Gamma_0' \Gamma_0/N)^{-1}=O_P(1)$, so focus on $\frac{1}{\sqrt{T}}\| I_{jNT}(\widehat{\beta}_m   ,\widehat{F} )\|$ with $j=1,2,\ldots, 5$ and $\frac{1}{\sqrt{T}}\| I_{jNT}(\widehat{F} )\|$ with $j=6, 7, 8$.

For $I_{1NT}(\widehat{\beta}_m  , \widehat{F} )$, we have

\begin{eqnarray*}
&&\frac{1}{\sqrt{T}} \left\| I_{1NT}(\widehat{\beta}_m  , \widehat{F} )\right\|\le \frac{\sqrt{r}}{NT}\sum_{i=1}^N \left\| \phi_i [\beta_0 ] - \phi_i [\widehat{\beta}_m  ]\right\|^2 \nonumber\\
&\le & \frac{\sqrt{r}}{NT}\sum_{i=1}^N \left\| \phi_i [\beta_{0,m} ] - \phi_i [\widehat{\beta}_m  ]\right\|^2 + \frac{\sqrt{r}}{NT}\sum_{i=1}^N \left\|\Delta_{m} \right\|^2\nonumber\\
&= &\vect(C_{\beta_0} -\widehat{C}_\beta   )'\frac{\sqrt{r}}{NT}\sum_{i=1}^{N} \mathcal{Z}_i'\mathcal{Z}_i\vect(C_{\beta_0} -\widehat{C}_\beta   )+ O_P( m^{-\mu} )\nonumber\\
&=& O_P(\| C_{\beta_0}  -\widehat{C}_\beta   \|^2) +O_P( m^{-\mu} )=O_P(\|\widehat{\beta}_m   -\beta_0 \|_{L^2}^2),\nonumber
\end{eqnarray*}
where the first and second equalities follow from Assumption 2.1.

For $I_{2NT}(\widehat{\beta}_m   , \widehat{F}  )$, write

\begin{eqnarray}\label{EqA18}
&&\frac{1}{\sqrt{T}}\left\| I_{2NT}(\widehat{\beta}_m   ,\widehat{F}  )\right\| \le \frac{\sqrt{r}}{NT}\sum_{i=1}^N \left\| \left( \phi_i [\beta_0 ] - \phi_i [\widehat{\beta}_m   ]\right)\left(F_0\gamma_{0i}\right)'\right\| \nonumber \\
&\le & \sqrt{r} \Big\{ \frac{1}{NT}\sum_{i=1}^N\left\| \phi_i [\beta_0 ] - \phi_i [\widehat{\beta}_m   ] \right\|^2\Big\}^{1/2} \Big\{\frac{1}{NT}\sum_{i=1}^N \|F_0\gamma_{0i} \|^2\Big\}^{1/2} \nonumber \\
&=&O_P(\|\widehat{\beta}_m    -\beta_0 \|_{L^2}),
\end{eqnarray}
where the second inequality follows from Cauchy-Schwarz inequality; and the last line follows from the same arguments given for $I_{1NT}(\widehat{\beta}_m   ,\widehat{F}  )$ and the fact that $\frac{1}{NT}\sum_{i=1}^N \|F_0\gamma_{0i} \|^2=O_P(1)$.

Similar to (\ref{EqA18}), we have $\frac{1}{\sqrt{T}}\left\| I_{jNT}(\widehat{\beta}_m   ,\widehat{F}  )\right\|= O_P(\|\widehat{\beta}_m -\beta_0 \|_{L^2})$ for $j=3,4,5$. By (1) of Lemma A.2 and $\frac{1}{\sqrt{T}}\|\widehat{F}  \|=O(1)$, we also obtain $\frac{1}{\sqrt{T}} \|I_{6NT}(\widehat{F}  )\|=O_P\left(\frac{1}{\sqrt{N}}\right)+O_P\left(\frac{1}{\sqrt{T}}\right)$.

For $I_{7NT}(\widehat{F}  )$ and $I_{8NT}(\widehat{F}  )$, write

\begin{eqnarray*}
&&E\left\|\frac{1}{NT}\sum_{i=1}^N F_0 \gamma_{0i} \mathcal{E}_i'\right\|^2= \sum_{t=1}^T \sum_{s=1}^T \frac{1}{N^2 T^2}\sum_{i=1}^N \sum_{j=1}^N E[f_{0t}' \gamma_{0i} \varepsilon_{is}f_{0t}' \gamma_{0j}\varepsilon_{js}] \nonumber\\
&=& \sum_{t=1}^T\sum_{s=1}^T \frac{1}{N^2 T^2}\sum_{i=1}^N \sum_{j=1}^N E[f_{0t}' \gamma_{0i}f_{0t}'\gamma_{0j}]E[\varepsilon_{is} \varepsilon_{js}] \nonumber\\
&\le &O(1) \sum_{t=1}^T\sum_{s=1}^T \frac{1}{N^2 T^2}\sum_{i=1}^N\sum_{j=1}^N\left\{E\| f_{0t}\|^4 E\|\gamma_{0i} \|^4 E\| f_{0t}\|^4 E\| \gamma_{0j}\|^4 \right\}^{1/4} |\sigma_{ij}|\nonumber \\
&\le &O(1) \frac{1}{N^2} \sum_{i=1}^N \sum_{j=1}^N
 |\sigma_{ij}| =O\left(\frac{1}{N}\right),\nonumber
\end{eqnarray*}
where the first inequality follows from Cauchy-Schwarz inequality and Assumption 1. We then can conclude that $\frac{1}{\sqrt{T}} \|I_{7NT}(\widehat{F}  )\|=\frac{1}{\sqrt{T}} \|I_{8NT}(\widehat{F}  )\|=O_P\left(\frac{1}{\sqrt{N}}\right).$

Based on the above analysis and by left multiplying (\ref{EqA15}) by $\widehat{F}  '/T$, we obtain

\begin{eqnarray*}
V_{NT}-(\widehat{F}  ' F_0/T)(\Gamma_0'\Gamma_0/N)(F_0'\widehat{F}   /T)=\frac{1}{T}\widehat{F}  '\left[I_{1NT}(\widehat{\beta}_m   ,\widehat{F}  )+\cdots +I_{8NT}(\widehat{F}  )\right]=o_P(1). \nonumber
\end{eqnarray*}
Thus,  $V_{NT} = (\widehat{F}  ' F_0/T)(\Gamma_0'\Gamma_0/N)(F_0'\widehat{F}   /T)+ o_P(1).$ When proving the second result of this lemma, we have shown that $F_0'\widehat{F}   /T $ is non-singular with probability approaching one, which implies that $V_{NT}$ is invertible with probability approaching one. We now left multiply (\ref{EqA15}) by $F_0'/T$ to obtain

\begin{eqnarray*}
(F_0'\widehat{F}  /T)V_{NT} = (F_0'F_0/T)(\Gamma_0'\Gamma_0/N)(F_0'\widehat{F}  /T) + o_P(1)\nonumber
\end{eqnarray*}
based on the above analysis. It shows that the columns of $F_0'\widehat{F}/T$ are the (non-normalized) eigenvectors of the matrix $(F_0'F_0/T)(\Gamma_0'\Gamma_0/N)$, and $V_{NT}$ consists of the eigenvalues of the same matrix (in the limit). Thus, the first result of this lemma follows.

\medskip

(4). According to the above analysis, (\ref{EqA16}) can be summarized by

\begin{eqnarray*}
\frac{1}{\sqrt{T}}\|\widehat{F}  \Pi_{NT}^{-1} - F_0 \| =O_P(\|\widehat{\beta}_m    -\beta_0 \|_{L^2})+O_P\left( \frac{1}{\sqrt{N}}\right) +O_P\left( \frac{1}{\sqrt{T}}\right) .\nonumber
\end{eqnarray*}

\medskip

(5). According to (\ref{EqA16}),

\begin{eqnarray*}
\frac{1}{T}F_0'(\widehat{F}   -F_0 \Pi_{NT}) &=& \frac{1}{T}F_0'\left[I_{1NT}(\widehat{\beta}_m   ,\widehat{F}  )+\cdots +I_{8NT}(\widehat{F}  )\right]V_{NT}^{-1}.\nonumber
\end{eqnarray*}
Note that $V_{NT}^{-1}=O_P(1)$, so we focus on $\frac{1}{T}F_0'\left[I_{1NT}(\widehat{\beta}_m   ,\widehat{F}  )+\cdots +I_{8NT}(\widehat{F}  )\right]$ below. By the proof given for the first result of this lemma, it is easy to show that

\begin{eqnarray*}
\left\|\frac{1}{T}F_0' \left[I_{1NT}(\widehat{\beta}_m   ,\widehat{F}  )+\cdots +I_{5NT}(\widehat{\beta}_m   ,\widehat{F}  )\right]\right\|= O_P(\|\widehat{\beta}_m    -\beta_0 \|_{L^2}).\nonumber
\end{eqnarray*}

We now consider $\left\|\frac{1}{T}F_0' I_{6NT}(\widehat{F}  ) \right\|$. Firstly, note that it is easy to show $\frac{1}{NT}\sum_{i=1}^{N}\| F_0'\mathcal{E}_i\|^2 =O_P(1)$. Secondly, for $\frac{1}{NT}\sum_{i=1}^{N}\left\| \mathcal{E} _{i}'\widehat{F}\right\| ^{2},$ we have

\begin{eqnarray*}
\frac{1}{NT}\sum_{i=1}^{N}\left\| \mathcal{E}_{i}'\widehat{F}  \right\| ^{2} &\le &\frac{2}{NT}\sum_{i=1}^{N}\left\| \mathcal{E}_{i}'F_0\Pi_{NT}\right\Vert ^{2}+\frac{2}{NT}\sum_{i=1}^{N}\left\|\mathcal{E}_{i}'\left( \widehat{F}  -F_0\Pi_{NT}\right) \right\Vert ^{2} \nonumber\\
&= &\frac{2}{NT}\sum_{i=1}^{N}\left\| \mathcal{E}_{i}'F_0\Pi_{NT}\right\Vert ^{2}+\frac{2}{NT}\sum_{i=1}^{N}\textrm{tr}\left\{\mathcal{E}_{i}'\left( \widehat{F}  -F_0\Pi_{NT}\right)\left( \widehat{F}  -F_0\Pi_{NT}\right)' \mathcal{E}_{i} \right\}\nonumber\\
&= &\frac{2}{NT}\sum_{i=1}^{N}\left\| \mathcal{E}_{i}'F_0\Pi_{NT}\right\Vert ^{2}+\frac{2}{NT} \textrm{tr}\left\{\left( \widehat{F}  -F_0\Pi_{NT}\right)\left( \widehat{F}  -F_0\Pi_{NT}\right)'\mathcal{E}'\mathcal{E} \right\}\nonumber\\
&\leq &O_{P}\left( 1\right)  +O(1)\frac{1}{N}\left\| \mathcal{E}'\mathcal{E} \right\| \frac{1}{T}\left\| \widehat{F}  -F_0\Pi_{NT}\right\|^{2},\nonumber
\end{eqnarray*}
where $\mathcal{E}$ has been defined in Lemma A.2. In connection with (1) of Lemma A.2 and (2) of this lemma, it gives that

\begin{eqnarray*}
&&\left\Vert \frac{1}{T}F_0' I_{6NT}(\widehat{F}  )\right\Vert \le \frac{1}{T}\left(\frac{1}{NT}\sum_{i=1}^{N}\| F_0'\mathcal{E}_i\|^2 \right)^{1/2}\left(\frac{1}{NT}\sum_{i=1}^{N} \left\| \mathcal{E}_{i}'\widehat{F}  \right\|^2 \right)^{1/2}\nonumber\\
&=&O_P(1)\frac{1}{T} + O_P\left(\frac{1}{\sqrt{T}}\right)\left\{\frac{1}{NT}\left\| \mathcal{E}'\mathcal{E} \right\| \frac{1}{T}\left\| \widehat{F}  -F_0\Pi_{NT}\right\|^{2}\right\}^{1/2}\nonumber\\
&=&O_P(1)\frac{1}{T} + O_P\left(\frac{1}{\sqrt{T}}\right)O_P\left(\frac{1}{\sqrt[4]{N}}+\frac{1}{\sqrt[4]{T}}\right) O_P\left(\|\widehat{\beta}_m    -\beta_0 \|_{L^2} +\frac{1}{\sqrt{N}}+\frac{1}{\sqrt{T}}\right)\nonumber\\
&=&O_P(1)\left\{\frac{1}{T}+\frac{ \|\widehat{\beta}_m    -\beta_0 \|_{L^2}}{\sqrt{T} \sqrt[4]{N}}  +\frac{1}{\sqrt{T}\sqrt[4]{N^3}}\right\}\nonumber\\
&\le &o_P(1) \|\widehat{\beta}_m    -\beta_0 \|_{L^2}+O_P(1)\frac{1}{T} +O_P(1)\frac{1}{\sqrt{T}\sqrt[4]{N^3}} ,\nonumber
\end{eqnarray*}
where the second equality follows from (1) of Lemma A.2 and the second result of this lemma.

For $\left\|\frac{1}{T}F_0' I_{7NT} (\widehat{F}  )\right\|$, we have

\begin{eqnarray*}
\left\|\frac{1}{T}F_0'I_{7NT} (\widehat{F}  )\right\| &\le & \left\|\frac{1}{T}F_0' F_0 \right\|\cdot \left\|\frac{1}{NT}\sum_{i=1}^N\gamma_{0i}\mathcal{E}_i'(\widehat{F}  - F_0 \Pi_{NT})\right\|\nonumber\\
&&+ \left\|\frac{1}{T}F_0'F_0 \right\|\cdot \left\| \frac{1}{NT}\sum_{i=1}^N \gamma_{0i}\mathcal{E}_i' F_0\Pi_{NT}\right\| \nonumber\\
&\le & \left\|\frac{1}{T}F_0' F_0 \right\|\cdot \left\|\frac{1}{N\sqrt{T}}\sum_{i=1}^N\gamma_{0i} \mathcal{E}_i'\right\|\cdot \frac{1}{\sqrt{T}}\| \widehat{F}  - F_0\Pi_{NT}\|\nonumber\\
&&+ \left\|\frac{1}{T}F_0'F_0\right\|\cdot \left\| \frac{1}{NT}\sum_{i=1}^N\gamma_{0i} \mathcal{E}_i' F_0\right\|\cdot \|\Pi_{NT} \|.\nonumber
\end{eqnarray*}
By Assumption 1.2, $\left\|\frac{1}{T}F_0' F_0\right\|=O_P(1)$. By the first two results of this lemma, we have $\|\Pi_{NT}\|=O_P(1)$ and $\frac{1}{\sqrt{T}}\| \widehat{F}  - F_0\Pi_{NT}\|=O_P(\|\widehat{\beta}_m    -\beta_0 \|_{L^2})+O_P\left(\frac{1}{\sqrt{N}}\right)+O_P\left(\frac{1}{\sqrt{T}}\right)$. Therefore, we focus on $\left\|\frac{1}{N\sqrt{T}}\sum_{i=1}^N\gamma_{0i} \mathcal{E}_i'\right\|$ and $\left\| \frac{1}{NT}\sum_{i=1}^N\gamma_{0i}\mathcal{E}_i' F_0\right\|$ below. Write

\begin{eqnarray}\label{gammaE1}
&&E\left\|\frac{1}{N\sqrt{T}}\sum_{i=1}^N\gamma_{0i} \mathcal{E}_i'\right\|^2=
\frac{1}{N^2T}\sum_{i=1}^N\sum_{j=1}^N\sum_{t=1}^T E[\gamma_{0i}'\gamma_{0j}]E[\varepsilon_{it}\varepsilon_{jt}] \nonumber \\
&\le &O(1)\frac{1}{N^2T}\sum_{i=1}^N\sum_{j=1}^N\sum_{t=1}^T |E[\varepsilon_{it}\varepsilon_{jt}]|=O(1)\frac{1}{N^2}\sum_{i=1}^N\sum_{j=1}^N |\sigma_{ij}|=O\left(\frac{1}{N}\right)
\end{eqnarray}
and using Assumption 1, it is easy to show that

\begin{eqnarray}\label{gammaE2}
E\left\| \frac{1}{NT}\sum_{i=1}^N\gamma_{0i} \mathcal{E}_i' F_0\right\|^2=O\left(\frac{1}{NT}\right),
\end{eqnarray}
which immediately yields

\begin{eqnarray*}
\left\|\frac{1}{T}F_0' I_{7NT}(\widehat{F}  ) \right\| &=&O_P\left( \|\widehat{\beta}_m   -\beta_0 \|_{L^2}\cdot \frac{1}{\sqrt{N}}\right)+O_P\left(\frac{1}{N}\right)  +O_P\left( \frac{1}{\sqrt{NT}}\right)\nonumber\\
&\le &O_P\left(\|\widehat{\beta}_m    -\beta_0 \|_{L^2}^2 \right)+O_P\left(\frac{1}{N}\right)  +O_P\left(\frac{1}{T}\right).\nonumber
\end{eqnarray*}
Similarly, $\left\|\frac{1}{T}F_0' I_{8NT}(\widehat{F}  ) \right\| =O_P\left(\|\widehat{\beta}_m    -\beta_0 \|_{L^2}^2 \right)+O_P\left(\frac{1}{N}\right)  +O_P\left(\frac{1}{T}\right)$.

Based on the above analysis, we have
\begin{eqnarray}\label{EqA19}
\left\|\frac{1}{T}F_0' (\widehat{F}  -F_0\Pi_{NT}) \right\| =O_P( \|\widehat{\beta}_m    -\beta_0 \|_{L^2}) +O_P\left( \frac{1}{N}\right)+O_P\left( \frac{1}{T}\right),
\end{eqnarray}
which further indicates 

\begin{eqnarray}\label{EqA20}
\left\|\frac{1}{T}\widehat{F}  '(\widehat{F}  -F_0\Pi_{NT})\right\|&=& \left\|\frac{1}{T}(\widehat{F}  -F_0\Pi_{NT} +F_0\Pi_{NT})' (\widehat{F}  -F_0\Pi_{NT})\right\|  \nonumber \\
&\le & \left\|\frac{1}{T}(\widehat{F}  -F_0\Pi_{NT} )'(\widehat{F}  -F_0\Pi_{NT})\right\| +\|\Pi_{NT}\|\cdot \left\|\frac{1}{T} F_0' (\widehat{F}  -F_0\Pi_{NT})\right\| \nonumber\\
&=& O_P(\|\widehat{\beta}_m    -\beta_0 \|_{L^2}) +O_P\left( \frac{1}{N}\right)+O_P\left( \frac{1}{T}\right).
\end{eqnarray}

\medskip

(6). Note (\ref{EqA19}) and (\ref{EqA20}) can be respectively expressed as

\begin{eqnarray*}
\frac{1}{T}F_0'\widehat{F}  -\frac{1}{T}F_0'F_0\Pi_{NT} = O_P( \|\widehat{\beta}_m    -\beta_0 \|_{L^2}) +O_P\left( \frac{1}{N}\right)+O_P\left( \frac{1}{T}\right)\nonumber
\end{eqnarray*}
and

\begin{eqnarray*}
I_{r}-\frac{1}{T}\widehat{F}  'F_0\Pi_{NT} =O_P(\|\widehat{\beta}_m    -\beta_0\|_{L^2}) +O_P\left( \frac{1}{N}\right)+O_P\left( \frac{1}{T}\right),\nonumber
\end{eqnarray*}
which further give

\begin{eqnarray*}
\frac{1}{T}\Pi_{NT}'F_0'\widehat{F}  -\frac{1}{T}\Pi_{NT}'F_0'F_0\Pi_{NT} = O_P( \|\widehat{\beta}_m    -\beta_0 \|_{L^2}) +O_P\left( \frac{1}{N}\right)+O_P\left( \frac{1}{T}\right)\nonumber
\end{eqnarray*}
and

\begin{eqnarray*}
I_{r} - \frac{1}{T}\Pi_{NT}'F_0'\widehat{F}   =O_P( \|\widehat{\beta}_m    -\beta_0 \|_{L^2}) +O_P\left( \frac{1}{N}\right)+O_P\left( \frac{1}{T}\right) .\nonumber
\end{eqnarray*}

Summing up the above two equations yields

\begin{eqnarray}  \label{EqA21}
I_{r} - \frac{1}{T}\Pi_{NT}'F_0'F_0\Pi_{NT} = O_P(\|\widehat{\beta}_m    -\beta_0 \|_{L^2}) +O_P\left( \frac{1}{N}\right)+O_P\left( \frac{1}{T}\right).
\end{eqnarray}
Note that it is easy to show that

\begin{eqnarray*}
\left\| P_{\widehat{F}} -P_{F_0}\right\|^2 &=& \text{tr}\left[ (P_{\widehat{F}  }-P_{F_0})^2\right] = \text{tr}\left[ P_{\widehat{F}  } - P_{\widehat{F}  } P_{F_0}- P_{F_0}P_{\widehat{F}  }+P_{F_0}\right]\nonumber\\
&=& \text{tr}\left[ I_{r}\right] - 2\text{tr}\left[P_{\widehat{F}  } P_{F_0}\right]+\text{tr}\left[I_{r}\right]=2 \text{tr}\left[ I_{r}-\widehat{F}  'P_{F_0} \widehat{F}  /T\right]\nonumber
\end{eqnarray*}
and, when proving this lemma, we have shown that

\begin{eqnarray*}
\frac{F_0'\widehat{F}  }{T} = \frac{F_0'F_0}{T}\Pi_{NT} +O_P( \|\widehat{\beta}_m    -\beta_0 \|_{L^2}) +O_P\left( \frac{1}{N}\right)+O_P\left( \frac{1}{T}\right).\nonumber
\end{eqnarray*}

Therefore, we can write

\begin{eqnarray*}
\widehat{F}  ' P_{F_0} \widehat{F}   /T = \Pi_{NT}'\left(\frac{F_0'F_0}{T}\right)\Pi_{NT}+O_P( \|\widehat{\beta}_m    -\beta_0 \|_{L^2}) +O_P\left( \frac{1}{N}\right)+O_P\left( \frac{1}{T}\right).\nonumber
\end{eqnarray*}
In connection with (\ref{EqA21}), we then obtain that

\begin{eqnarray}\label{APF}
\widehat{F}  'P_{F_0} \widehat{F}  /T = I_{r}+O_P(\|\widehat{\beta}_m    -\beta_0 \|_{L^2}) +O_P\left( \frac{1}{N}\right)+O_P\left( \frac{1}{T}\right).
\end{eqnarray}
Then the proof of the last result of this lemma is completed. \hspace*{\fill}{$\blacksquare$}

\bigskip

\noindent \noindent \textbf{Proof of Lemma A.4:}

For simplicity, we show that $\Pr (\|\widehat{C}_{\beta,p} \| = 0 ) \to 1$ only. The proofs for $\|\widehat{C}_{\beta,j} \| $ with $j=p^*+1,\ldots,p-1$ are the same. By (\ref{LDconsis1}) and Assumption 3, we can conclude that 

\begin{eqnarray}\label{slowrate}
 \|C_{\beta_0}-\widehat{C}_\beta \| =O_P\left(  \frac{1}{\sqrt[8]{\xi_{NT}}}\right).
\end{eqnarray} 

If $\|\widehat{C}_{\beta,p} \|  \ne 0$,  the following equation must hold:

\begin{eqnarray}\label{Lemma2}
0= \frac{\partial}{\partial C_{\beta,p}} Q_{\lambda} (C_\beta,F)\big|_{(C,F)=(\widehat{C}_\beta, \widehat{F})}  = -2B_1 + B_2,
\end{eqnarray}
where $B_1 = \sum_{i=1}^N  \mathsf{Z}_{ip}' M_{\widehat{F}}(Y_i-\phi_i[\widehat{\beta}_m])$, $\mathsf{Z}_{ip}  = (x_{i1,p} H_m(z_{i1}),\ldots, x_{iT,p} H_m(z_{iT}))'$ and $B_2 = \frac{\lambda_p }{\|\widehat{C}_{\beta,p} \|} \widehat{C}_{\beta,p}'$. For $B_1$, write

\begin{eqnarray*}
\frac{1}{NT} B_1 &=& \frac{1}{NT} \sum_{i=1}^N  \mathsf{Z}_{ip}' M_{\widehat{F}}(\phi_i[\beta_0] -\phi_i[\widehat{\beta}_m] +F_0\gamma_i +\mathcal{E}_i) \nonumber
\end{eqnarray*}
In view of (\ref{slowrate}) and the development of Lemma A.3, it is easy to know $\frac{\sqrt[8]{\xi_{NT}}}{\sqrt{m}}\cdot\frac{1}{NT}B_1=O_P(1)$. On the other hand, $\left\| \frac{\sqrt[8]{\xi_{NT}}}{\sqrt{m}}\cdot\frac{1}{NT}B_2 \right\|  \ge \frac{\sqrt[8]{\xi_{NT}} \lambda_{\text{min}}^\dagger }{\sqrt{m}NT}  \ge \kappa_1 $ by Assumption 3. Therefore, $\Pr(\|B_1 \|<\|B_2\|)\to 1$, which implies that, with a probability tending to 1, (\ref{Lemma2}) does not hold. The above analysis implies that $\widehat{C}_{\beta,p}$ must be located at a place where the objective function $Q_{\lambda} (C_\beta,F)$ is not differentiable with respect to $C_{\beta,p}$. Since  $Q_{\lambda} (C_\beta,F)$ is not differentiable with respect to $C_{\beta,p}$ only at the origin, we immediately obtain that $\Pr(\| \widehat{C}_{\beta,p} \| = 0) \to 1$. Similarly, we can show $\Pr (\| \widehat{C}_{\beta,j}\|=0 )\to 1$ with $j=p^*+1,\ldots,p-1$. The proof is complete. \hspace*{\fill}{$\blacksquare$}

\bigskip

\noindent \textbf{Proof of Lemma A.5:}

Note that (\ref{slowrate}) only gives a slow rate. Below, we aim to improve this rate. Having proved $\Pr (\| \widehat{C}_{\beta}^\dagger\|=0 )\to 1$, we delete the corresponding rows of $\widehat{C}_\beta$ and $x_{it,j}$ for $j=p^*+1,\ldots,p$ from the objective function. Thus, following the same arguments as in \citet[p. 1236]{Bai}, the estimator $\widehat{C}_\beta^*$ given by (3.3) can be written as

\begin{eqnarray*}
\vect( \widehat{C}_\beta^*)= \left(\sum_{i=1}^N {\mathcal{Z}_i^*}' M_{\widehat{F}}{\mathcal{Z}_i^*} + \frac{D_{m,p^*}}{2}\right)^{-1} \sum_{i=1}^N  {\mathcal{Z}_i^*}' M_{\widehat{F}} Y_i,\nonumber
\end{eqnarray*}
where $\mathcal{Z}_i^*=  (\mathcal{Z}_{i1}^*,\ldots, \mathcal{Z}_{iT}^*)'$ and $D_{m,p^*} = I_m\otimes \text{diag}\left\{\frac{\lambda_1}{ \|  \widehat{C}_{\beta,1}\| } ,\ldots, \frac{\lambda_{p^*}}{ \| \widehat{C}_{\beta,p^*}\| }\right\}$. Correspondingly, we denote that 

\begin{eqnarray*}
\vect(\widehat{C}_{\beta}^\sharp) = \left(\sum_{i=1}^N {\mathcal{Z}_i^*}' M_{\widehat{F}}{\mathcal{Z}_i^*} \right)^{-1} \sum_{i=1}^N {\mathcal{Z}_i^*}' M_{\widehat{F}} Y_i.\nonumber
\end{eqnarray*}
Thus, we can write

\begin{eqnarray}\label{startsharp}
 \widehat{C}_\beta^* -  C_{\beta_0}^*=(\widehat{C}_\beta^* - \widehat{C}_\beta^\sharp)+ (\widehat{C}_\beta^\sharp - C_{\beta_0}^*).
\end{eqnarray}
Below, we investigate each term on the right hand side of (\ref{startsharp}).

Firstly, consider $\widehat{C}_\beta^* - \widehat{C}_\beta^\sharp$, and write

\begin{eqnarray*}
\vect( \widehat{C}_\beta^*)-\vect(\widehat{C}_{\beta}^\sharp) &=&\left\{  \left(\sum_{i=1}^N {\mathcal{Z}_i^*}' M_{\widehat{F}}{\mathcal{Z}_i^*} + \frac{D_{m,p^*}}{2}\right)^{-1} -  \left(\sum_{i=1}^N {\mathcal{Z}_i^*}' M_{\widehat{F}}{\mathcal{Z}_i^*} \right)^{-1}\right\}\sum_{i=1}^N  {\mathcal{Z}_i^*}' M_{\widehat{F}} Y_i.\nonumber
\end{eqnarray*}
By Lemma A.1 and Assumption 2, we just need to consider the next term in order to get the difference between $\widehat{C}_\beta^*$ and $\widehat{C}_\beta^\sharp$.

\begin{eqnarray}\label{Deno}
&&\left\| \frac{1}{NT}\sum_{i=1}^N {\mathcal{Z}_i^*}' M_{\widehat{F}}{\mathcal{Z}_i^*} + \frac{D_{m,p^*}}{2NT}-\frac{1}{NT}\sum_{i=1}^N {\mathcal{Z}_i^*}' M_{\widehat{F}}{\mathcal{Z}_i^*} \right\|\nonumber \\
&=&  \left\| \frac{D_{m,p^*}}{2NT}\right\| =O\left(\frac{\sqrt{m}\lambda_{\text{max}}^*}{NT} \right).
\end{eqnarray}
Moreover, it is easy to know $\left\|\frac{1}{NT}\sum_{i=1}^N  {\mathcal{Z}_i^*}' M_{\widehat{F}} Y_i\right\|=O_P(\sqrt{m})$, which in connection with (\ref{Deno}) indicates $\| \widehat{C}_\beta^*- \widehat{C}_{\beta}^\sharp\| =O_P\left(\frac{m \lambda_{\text{max}}^* }{NT} \right)$.

We now focus on $\widehat{C}_{\beta}^\sharp   - C_{\beta_0}^* $, and write

\begin{eqnarray*}
\vect(\widehat{C}_{\beta}^\sharp   )-\vect(C_{\beta_0}^* )&=& \left[\sum_{i=1}^N  {\mathcal{Z}_i^*}' M_{\widehat{F}  }{\mathcal{Z}_i^*}\right]^{-1}\sum_{i=1}^N  {\mathcal{Z}_i^*}' M_{\widehat{F}  } \mathcal{E}_{i} \nonumber\\
&&+  \left[\sum_{i=1}^N  {\mathcal{Z}_i^*}' M_{\widehat{F}  }{\mathcal{Z}_i^*}\right]^{-1}\sum_{i=1}^N  {\mathcal{Z}_i^*}' M_{\widehat{F}  } F_0\gamma_{0i}\nonumber\\
&&+  \left[\sum_{i=1}^N  {\mathcal{Z}_i^*}' M_{\widehat{F}  }{\mathcal{Z}_i^*}\right]^{-1}\sum_{i=1}^N  {\mathcal{Z}_i^*}' M_{\widehat{F}  }\phi_i^*[\Delta_{m}^*]\nonumber\\
&:=&\Lambda_1 + \Lambda_2 + \Lambda_3,\nonumber
\end{eqnarray*}
where the definitions of $\Lambda_1$-$\Lambda_3$ should be obvious. Note

\begin{eqnarray*}
\frac{1}{NT}\sum_{i=1}^N  {\mathcal{Z}_i^*}' M_{\widehat{F}  }{\mathcal{Z}_i^*}= \frac{1}{NT}\sum_{i=1}^N {\mathcal{Z}_i^*}' M_{F_0}{\mathcal{Z}_i^*}\cdot(1  +o_P(1)) =\Sigma_{\mathcal{Z},f}^*\cdot(1  +o_P(1)) ,\nonumber
\end{eqnarray*}
where    $\Sigma_{\mathcal{Z},f}^*=E[\mathcal{Z}_{11}^* {\mathcal{Z}_{11}^*}'] -E[\mathcal{Z}_{11}^* f_{01}' ]\Sigma_f^{-1} E[f_{01}{\mathcal{Z}_{11}^* }' ]$. Similar to (A.5) of \cite{SuJin}, we obtain $\|\Lambda_3  \|=O_P\left(m^{-\frac{\mu}{2}}\right)$. In the following, we focus on studying $\Lambda_2$ at first, and then turn to $\Lambda_1$.

In the rest proofs of this lemma, we always let $\Xi_{NT}=(F_0'\widehat{F} /T)^{-1}(\Gamma_0' \Gamma_0/N)^{-1}$ for simplicity, and we have shown $\| \Xi_{NT}\|=O_P(1)$ in the proof of Lemma A.3. Recall that we have denoted $\Pi_{NT}$ and $V_{NT}$ in Lemma A.3, so $\Pi_{NT}^{-1} = V_{NT} \Xi_{NT}$. Then we start our investigation on $\Lambda_2$, and write

\begin{eqnarray*}
\frac{1}{NT} \sum_{i=1}^N  {\mathcal{Z}_i^*}'M_{\widehat{F} } F_0\gamma_{0i} &=& \frac{1}{NT } \sum_{i=1}^N  {\mathcal{Z}_i^*}'M_{\widehat{F} } \left(\widehat{F} \Pi_{NT}^{-1} - F_0\right)\gamma_{0i} \nonumber\\
&=& \frac{1}{NT} \sum_{i=1}^N  {\mathcal{Z}_i^*}'M_{\widehat{F} } \left[I_{1NT}(\widehat{\beta}_m^*,\widehat{F} )+\cdots +I_{8NT}(\widehat{F} )\right]\Xi_{NT} \gamma_{0i} \nonumber\\
&:=&J_{1NT} +\cdots + J_{8NT},\nonumber
\end{eqnarray*}
where the second equality follows from (\ref{EqA16}); $I_{1NT}(\beta,F)$ to $I_{8NT}(F)$ have been defined in the proof of Lemma A.3 but excluding $x_{it,j}$ for $j=p^*+1,\ldots,p$; and the definitions of $J_{1NT}$ to $ J_{8NT}$ should be obvious. In view of the decomposition of $J_{2NT}$ below, it is easy to know that $\| J_{1NT}\| = o_P(\| \widehat{C}_\beta^*  -C_{\beta_0}^*\|)$. Thus, we start from $J_{2NT}$ and write

\begin{eqnarray*}
J_{2NT} &=&\frac{1}{NT} \sum_{i=1}^N {\mathcal{Z}_i^*}'M_{\widehat{F} } I_{2NT}(\widehat{\beta}_m^* ,\widehat{F} )\Xi_{NT} \gamma_{0i}  \nonumber\\
&=&\frac{1}{NT} \sum_{i=1}^N {\mathcal{Z}_i^*}'M_{\widehat{F} }  \frac{1}{NT}\sum_{j=1}^N\left( \phi_j^* [\beta_{0,m}^*] - \phi_j^* [\widehat{\beta}_m^* ]\right)\left(F_0\gamma_{0j} \right)'\widehat{F} \Xi_{NT}\gamma_{0i} \nonumber \\
&&+\frac{1}{NT} \sum_{i=1}^N {\mathcal{Z}_i^*}'M_{\widehat{F} }  \frac{1}{NT}\sum_{j=1}^N \phi_j^*[\Delta_{m}^*] \left(F_0\gamma_{0j}\right)'\widehat{F} \Xi_{NT}\gamma_{0i}\nonumber\\
&=&\frac{1}{N^2T} \sum_{i=1}^N\sum_{j=1}^N {\mathcal{Z}_i^*}'M_{\widehat{F} } {\mathcal{Z}_j^*} \gamma_{0j}'\Big(\frac{F_0'\widehat{F} }{T}\Big)\Big(\frac{F_0'\widehat{F} }{T}
\Big)^{-1}\Big(\frac{\Gamma_0' \Gamma_0}{N}\Big)^{-1} \gamma_{0i} \left[ \vect(\widehat{C}_{\beta}^*  )-\vect(C_{\beta_0}^* )\right]\nonumber\\
&&+\frac{1}{NT} \sum_{i=1}^N {\mathcal{Z}_i^*}'M_{\widehat{F} }  \frac{1}{NT}\sum_{j=1}^N \phi_j^*[\Delta_{m}^*] \left(F_0\gamma_{0j}\right)'\widehat{F} \Xi_{NT}\gamma_{0i}\nonumber\\
&=&\frac{1}{N^2T} \sum_{i=1}^N\sum_{j=1}^N {\mathcal{Z}_i^*}'M_{\widehat{F} } {\mathcal{Z}_j^*} \gamma_{0j}' \Big(\frac{\Gamma_0' \Gamma_0}{N}\Big)^{-1} \gamma_{0i}  \left[ \vect(\widehat{C}_{\beta}^*  )-\vect(C_{\beta_0}^* )\right]\nonumber\\
&&+\frac{1}{NT} \sum_{i=1}^N {\mathcal{Z}_i^*}'M_{\widehat{F} }  \frac{1}{NT}\sum_{j=1}^N \phi_j^*[\Delta_{m}^*] \left(F_0\gamma_{0j}\right)'\widehat{F} \Xi_{NT}\gamma_{0i}\nonumber\\
&:=& J_{2NT,1} + J_{2NT,2}.\nonumber
\end{eqnarray*}

By a derivation similar (A.5) of \cite{SuJin}, we know $\left\| \left[ \sum_{i=1}^N  {\mathcal{Z}_i^*}' M_{\widehat{F} }{\mathcal{Z}_i^*}\right]^{-1}NTJ_{2NT,2}\right\|=O_P\left(m^{-\frac{\mu}{2}}\right)$, so negligible. We will further study $J_{2NT,1} $ later.

For $J_{3NT}$, write

\begin{eqnarray*}
 J_{3NT} &=&  \frac{1}{NT} \sum_{i=1}^N {\mathcal{Z}_i^*}'M_{\widehat{F} } I_{3NT}(\widehat{\beta}_m^* ,\widehat{F} )\Xi_{NT} \gamma_{0i} \nonumber \\
&=& \frac{1}{NT} \sum_{i=1}^N {\mathcal{Z}_i^*}'M_{\widehat{F} } \frac{1}{NT}\sum_{j=1}^N F_0\gamma_{0j} \left( \phi_j^* [\beta_0^*] - \phi_j^* [\widehat{\beta}_m^* ]\right)^{\prime }\widehat{F}  \Xi_{NT} \gamma_{0i} \nonumber\\
&= & \frac{1}{N T} \sum_{i=1}^N  {\mathcal{Z}_i^*}'M_{\widehat{F} }( \widehat{F}  \Pi_{NT}^{-1}- F_0)\frac{1}{NT}\sum_{j=1}^N\gamma_{0j} \left( \phi_j^* [\beta_0^*] - \phi_j^* [\widehat{\beta}_m] \right)'\widehat{F} \Xi_{NT} \gamma_{0i}\nonumber\\
&:=&\frac{1}{N T} \sum_{i=1}^N  {\mathcal{Z}_i^*}'M_{\widehat{F} } J_{3NT,i},\nonumber
\end{eqnarray*}
where $J_{3NT,i} =( \widehat{F}  \Pi_{NT}^{-1}- F_0)\frac{1}{NT}\sum_{j=1}^N\gamma_{0j} ( \phi_j^* [\beta_0^*] - \phi_j^* [\widehat{\beta}_m^* ] )'\widehat{F} \Xi_{NT} \gamma_{0i}.$ Below, we are going to show that 

\begin{eqnarray}\label{J3NTi}
 \left\| \left[ \sum_{i=1}^N  {\mathcal{Z}_i^*}' M_{\widehat{F} }{\mathcal{Z}_i^*}\right]^{-1}NTJ_{3NT}\right\| =o_P(\| \widehat{C}_{\beta} -C_{\beta_0}\|)
\end{eqnarray}
By the procedure similar to (A.5) of \cite{SuJin}, we just need to focus on $\frac{1}{NT}\sum_{i=1}^N \|  J_{3NT,i}\|^2$.

\begin{eqnarray*}
\frac{1}{NT}\sum_{i=1}^N \|  J_{3NT,i}\|^2&\le& \frac{1}{NT}\sum_{i=1}^N\|\widehat{F}  \Pi_{NT}^{-1}- F_0 \|^2 \Big\| \frac{1}{NT}\sum_{j=1}^N\gamma_{0j} ( \phi_j^* [\beta_0^*] - \phi_j^* [\widehat{\beta}_m^* ] )' \Big\|^2  \| \widehat{F} \Xi_{NT} \gamma_{0i}\|^2\nonumber\\
&\le &O_P(1)\frac{1}{T}\|\widehat{F}  \Pi_{NT}^{-1}- F_0 \|^2 \left(\frac{1}{N \sqrt{T} } \sum_{j=1}^N\| \phi_j^* [\beta_0^*] - \phi_j^* [\widehat{\beta}_m^* ] \|  \right)^2\nonumber\\
&= &O_P(1)\frac{1}{T}\|\widehat{F}  \Pi_{NT}^{-1}- F_0 \|^2 \left(\frac{1}{N } \sum_{j =1}^N\left\{\frac{1}{T}\| \phi_j^* [\beta_0^*] - \phi_j^* [\widehat{\beta}_m^* ] \|^2\right\}^{1/2}\right)^2 \nonumber\\
&=&o_P(\| \widehat{C}_{\beta} -C_{\beta_0}\|^2),\nonumber
\end{eqnarray*}
where the second inequality follows from $\Xi_{NT}=O_P(1)$ and $\frac{1}{\sqrt{T}}\| \widehat{F} \|=O(1)$; and the last equality follows from $\frac{1}{\sqrt{T}}\|\widehat{F}  \Pi_{NT}^{-1}- F_0 \| =o_P(1)$. Thus, we can conclude that (\ref{J3NTi}) holds.

For $ J_{4NT}$, write

\begin{eqnarray*}
 J_{4NT} & =&   \frac{1}{NT} \sum_{i=1}^N {\mathcal{Z}_i^*}'M_{\widehat{F} } I_{4NT}(\widehat{\beta}_m^* ,\widehat{F} )\Xi_{NT}  \gamma_{0i} \nonumber\\
&\le &   \frac{1}{N^2T^2} \sum_{i=1}^N \sum_{j=1}^N{\mathcal{Z}_i^*}'M_{\widehat{F} }  \left( \phi_j^* [\beta_{0,m}^*] - \phi_j^* [\widehat{\beta}_m^* ]\right) \mathcal{E}_j' F_{0}  \Pi_{NT}\Xi_{NT}  \gamma_{0i} \nonumber\\
&&+ \frac{1}{N^2T^2} \sum_{i=1}^N \sum_{j=1}^N{\mathcal{Z}_i^*}'M_{\widehat{F} } \phi_j^* [\Delta_{m}^*]  \mathcal{E}_j' F_{0} \Pi_{NT}\Xi_{NT}  \gamma_{0i} \nonumber\\
&& +\frac{1}{N^2T^2} \sum_{i=1}^N \sum_{j=1}^N{\mathcal{Z}_i^*}'M_{\widehat{F} }  \left( \phi_j^* [\beta_0^*] - \phi_j^* [\widehat{\beta}_m^* ]\right)\left[ \mathcal{E}_j'(\widehat{F} -F_{0}\Pi_{NT}) \right]\Xi_{NT}  \gamma_{0i}\nonumber\\
&:=& J_{4NT,1} +J_{4NT,2}+J_{4NT,3}.\nonumber
\end{eqnarray*}

For $J_{4NT, 1}$, write

\begin{eqnarray*}
\|J_{4NT,1}\| &=&\left\|\frac{1}{N^2T^2} \sum_{i=1}^N \sum_{j=1}^N{\mathcal{Z}_i^*}'M_{\widehat{F} }  \left( \phi_j^* [\beta_{0,m}^*] - \phi_j^* [\widehat{\beta}_m^* ]\right) \mathcal{E}_j' F_{0}\Pi_{NT} \Xi_{NT}  \gamma_{0i}\right\|\nonumber\\
&=&\left\|\frac{1}{N^2T^2} \sum_{i=1}^N \sum_{j=1}^N{\mathcal{Z}_i^*}'M_{\widehat{F} } \mathcal{Z}_{j}^* \vect(\widehat{C}_\beta^* -C_{\beta_0}^*) \mathcal{E}_j' F_{0}\Pi_{NT} \Xi_{NT}  \gamma_{0i}\right\|\nonumber\\
&\le &\left\|\frac{1}{N^2T^2} \sum_{i=1}^N \sum_{j=1}^N{\mathcal{Z}_i^*}'M_{\widehat{F} } \mathcal{Z}_{j}^*  \mathcal{E}_j' F_{0}\Pi_{NT}  \Xi_{NT}  \gamma_{0i}\right\| \cdot \|\vect(\widehat{C}_\beta^*  -C_{\beta_0}^*)\|\nonumber\\
&\le &O_P(1) \frac{1}{N T } \sum_{i=1}^N \left\| {\mathcal{Z}_i^*}' M_{\widehat{F} }\right\| \|\gamma_{0i} \| \cdot \frac{1}{N}\sum_{j=1}^N\left\| \mathcal{Z}_j^*\right\| \frac{1}{T}\|\mathcal{E}_j' F_{0}\| \cdot \|\vect(\widehat{C}_\beta^*  -C_{\beta_0}^*)\|\nonumber\\
&\le &\frac{1}{T} O_P(\sqrt{mT} ) \cdot O_P(\sqrt{mT} ) \cdot O_P(T^{-1/2}) \cdot \|\vect(\widehat{C}_\beta -C_{\beta_0})\|\nonumber\\
&=&o_P(\|\widehat{C}_\beta^*  -C_{\beta_0}^* \|),\nonumber
\end{eqnarray*}
where the last line follows from $\frac{m^2}{T}\to 0$. Thus, $\|J_{4NT,1}\| $ is negligible. Similarly, we can show both $\|J_{4NT,2}\| $ and $\|J_{4NT,3}\| $ are negligible by taking $\frac{1}{T}\|\phi_j^* [\Delta_{m}^*]  \|^2 =O(m^{-\mu})$ and $\frac{1}{\sqrt{T}}\|\widehat{F}  \Pi_{NT}^{-1}- F_0 \| =o_P(1)$ into account, respectively. Analogous to the derivations of $J_{3NT}$ and $J_{4NT}$, we can obtain that $\| J_{5NT}\| $ is negligible.

Below, we take a careful look at $J_{6NT}$. According to Assumption 1, let $\Omega_e = E[\mathcal{E}_i \mathcal{E}_i']$, which is a deterministic matrix uniformly in $i$. Thus, write

\begin{eqnarray*}
J_{6NT}  &=&\frac{1}{NT} \sum_{i=1}^N  {\mathcal{Z}_i^*}'M_{\widehat{F} }  \frac{1}{NT} \sum_{j=1}^N \mathcal{E}_j \mathcal{E}_j'   \widehat{F}  \Xi_{NT}  \gamma_{0i}\nonumber\\
&=& \frac{1}{NT^2} \sum_{i=1}^N  {\mathcal{Z}_i^*}'M_{\widehat{F} } \Omega_e \widehat{F}  \Xi_{NT}  \gamma_{0i} \nonumber\\
&&+\frac{1}{NT} \sum_{i=1}^N  {\mathcal{Z}_i^*}'M_{\widehat{F} }  \frac{1}{NT} \sum_{j=1}^N \left(\mathcal{E}_j \mathcal{E}_j' -\Omega_e\right)  \widehat{F}  \Xi_{NT}  \gamma_{0i} \nonumber\\
&:=&J_{6NT,1} + J_{6NT,2}.\nonumber
\end{eqnarray*}
We focus on $J_{6NT,2}$ at first.

\begin{eqnarray*}
 J_{6NT,2} &=& \frac{1}{N^2T^2} \sum_{i=1}^N    \sum_{j=1}^N {\mathcal{Z}_i^*}' \left(\mathcal{E}_j \mathcal{E}_j' -\Omega_e\right)  \widehat{F}  \Xi_{NT}  \gamma_{0i}\nonumber\\
& &+ \frac{1}{N^2T^2} \sum_{i=1}^N    \sum_{j=1}^N {\mathcal{Z}_i^*}' P_{\widehat{F} }   \left(\mathcal{E}_j \mathcal{E}_j' -\Omega_e\right)  \widehat{F}  \Xi_{NT} \gamma_{0i}\nonumber\\
&:=& J_{6NT,21} + J_{6NT,22}.\nonumber
\end{eqnarray*}
Further decompose $J_{6NT,21}$ as

\begin{eqnarray*}
J_{6NT,21}&=& \frac{1}{N^2T^2} \sum_{i=1}^N    \sum_{j=1}^N {\mathcal{Z}_i^*}' \left(\mathcal{E}_j \mathcal{E}_j' -\Omega_e\right)  F_0 \Pi_{NT}\Xi_{NT}  \gamma_{0i} \nonumber\\
&&+  \frac{1}{N^2T^2} \sum_{i=1}^N    \sum_{j=1}^N {\mathcal{Z}_i^*}' \left(\mathcal{E}_j \mathcal{E}_j' -\Omega_e\right)  (\widehat{F} -F_0\Pi_{NT}) \Xi_{NT}  \gamma_{0i} \nonumber\\
&:=&J_{6NT,211} + J_{6NT,212}.\nonumber
\end{eqnarray*}
Then by a development similar to \citet[pp. 30-31]{JYGH}, we obtain that $\|J_{6NT,21}\|=o_P\left(\sqrt{\frac{m}{NT}}\right)$. Similarly,  $\|J_{6NT,22} \|  = o_P\left(\sqrt{\frac{m}{NT}}\right).$ Therefore, we obtain $\|J_{6NT,2} \| =o_P\left(\sqrt{\frac{m}{NT}}\right)$. 

We will consider $J_{6NT,1} $ together with $J_{2NT,1}$ and $J_{8NT}$ later on. Then we only have one term $J_{7NT}$ left to consider.

\begin{eqnarray*}
J_{7NT} &=& \frac{1}{NT} \sum_{i=1}^N {\mathcal{Z}_i^*}'M_{\widehat{F} }\frac{1}{NT}\sum_{j=1}^NF_0\gamma_{0j}  \mathcal{E}_j' \widehat{F} \Xi_{NT}  \gamma_{0i} \nonumber\\
&=& \frac{1}{NT} \sum_{i=1}^N {\mathcal{Z}_i^*}'M_{\widehat{F} }(F_0 -\widehat{F} \Pi_{NT}^{-1})\frac{1}{NT}\sum_{j=1}^N\gamma_{0j} \mathcal{E}_j' \widehat{F} \Xi_{NT}  \gamma_{0i} .\nonumber
\end{eqnarray*}
Notice that

\begin{eqnarray*}
\frac{1}{NT}\sum_{j=1}^N\gamma_{0j}  \mathcal{E}_j' \widehat{F} &=& \frac{1}{NT}\sum_{j=1}^N\gamma_{0j} \mathcal{E}_j' F_0 +\frac{1}{NT}\sum_{j=1}^N\gamma_{0j} \mathcal{E}_j' (F_0 -\widehat{F} \Pi_{NT}^{-1})\nonumber\\
&=& O_P\left(\frac{1}{\sqrt{NT}}\right) +\left\| \frac{1}{N\sqrt{T}}\sum_{j=1}^N\gamma_{0j} \mathcal{E}_j'\right\| \frac{1}{\sqrt{T}} \| F_0 -\widehat{F} \Pi_{NT}^{-1}\|\nonumber\\
&=& O_P\left(\frac{1}{\sqrt{NT}}\right) +O_P\left( \frac{1}{\sqrt{N}}\right) \frac{1}{\sqrt{T}} \| F_0 -\widehat{F} \Pi_{NT}^{-1}\|,\nonumber
\end{eqnarray*}
where the second equality follows from (\ref{gammaE2}); and the third equality follows from (\ref{gammaE1}). Following the arguments given for $J6$ of \citet[pp. 1271-1272]{Bai}, it is easy to show that $\|J_{7NT}\|=o_P\left(\sqrt{\frac{m}{NT}}\right) +o_P(\|\widehat{C}_\beta -C_{\beta_0} \|)$.

Based on the above analyses and Assumption 4, we have

\begin{eqnarray*}
&&\vect(\widehat{C}_{\beta}^\sharp )-\vect(C_{\beta_0}^* ) - {\Sigma_{\mathcal{Z},f}^{*\, -1}} J_{2NT,1}\cdot (1+o_P(1))\nonumber \\
&=&{\Sigma_{\mathcal{Z},f}^{*\, -1}} \left\{\frac{1}{NT}\sum_{i=1}^N  {\mathcal{Z}_i^*}'M_{\widehat{F} } \mathcal{E}_{i}+ J_{6NT,1}+J_{8NT}\right\}\cdot(1+o_P(1))\nonumber\\
&=&{\Sigma_{\mathcal{Z},f}^{*\, -1}} \cdot\frac{1}{NT}\sum_{i=1}^N  \left\{ {\mathcal{Z}_i^*}'M_{\widehat{F} }+ \frac{1}{N}  \sum_{j=1}^N{\mathcal{Z}_{j}^*} 'M_{\widehat{F} } \gamma_{0j}'  (\Gamma_0' \Gamma_0/N)^{-1} \gamma_{0i}\right\}\mathcal{E}_{i}\cdot(1+o_P(1))\nonumber\\
&&+{\Sigma_{\mathcal{Z},f}^{*\, -1}}  \cdot J_{6NT,1}\cdot(1+o_P(1)).\nonumber
\end{eqnarray*}
Further organise the above equation, we have

\begin{eqnarray*}
\vect(\widehat{C}_{\beta}^\sharp )-\vect(C_{\beta_0}^* )&=&A_{1NT}^{-1}{\Sigma_{\mathcal{Z},f}^{*\, -1}} \cdot\frac{1}{NT }\sum_{i=1}^N  \left\{ {\mathcal{Z}_i^*}' M_{\widehat{F} }+A_{3,i}\right\}\mathcal{E}_{i}\cdot(1+o_P(1))\nonumber\\
&& +A_{1NT}^{-1} {\Sigma_{\mathcal{Z},f}^{*\, -1}}  \cdot J_{6NT,1}\cdot(1+o_P(1)),\nonumber
\end{eqnarray*}
where

\begin{eqnarray}\label{ANTS}
A_{1NT}&=&I_{m p^*} - {\Sigma_{\mathcal{Z},f}^{*\, -1}}  A_{2NT}\cdot (1+o_P(1)),\nonumber \\
A_{2NT}&=&\frac{1}{N^2T} \sum_{i=1}^N\sum_{j=1}^N {\mathcal{Z}_i^*}' M_{\widehat{F} }  \mathcal{Z}_j^* \gamma_{0j}' \Big(\frac{\Gamma_0' \Gamma_0}{N}\Big)^{-1} \gamma_{0i} , \nonumber \\
A_{3,i}&=&\frac{1}{N}  \sum_{j=1}^N {\mathcal{Z}_j^*}'M_{\widehat{F} } \gamma_{0j} '  (\Gamma_0' \Gamma_0/N)^{-1} \gamma_{0i}.
\end{eqnarray}

Note that

\begin{eqnarray}\label{EqA22}
\sqrt{\frac{NT}{m}} J_{6NT,1}&=&\frac{ 1}{(mN)^{\frac{1}{2}}T^{\frac{3}{2}}} \sum_{i=1}^N  {\mathcal{Z}_i^*}'M_{\widehat{F} } \Omega_e \widehat{F}  \Xi_{NT}  \gamma_{0i} \nonumber \\
 &=&\frac{\sqrt{N}}{\sqrt{mT}} \cdot \frac{ 1}{N T } \sum_{i=1}^N  {\mathcal{Z}_i^*}'M_{\widehat{F} } \Omega_e \widehat{F}  \Xi_{NT}  \gamma_{0}(v_{i})=O_P\left(\sqrt{\frac{N}{T}} \right)=O_P(1).
\end{eqnarray}
where the last equality follows from Assumption 3. Thus, we obtain $\|J_{6NT,1} \| =O_P\left(\sqrt{\frac{m}{NT}}\right)$. Moreover, it is easy to show $\frac{1}{NT }\sum_{i=1}^N  \left\{ {\mathcal{Z}_i^*}'M_{\widehat{F} }+A_{3,i}\right\}\mathcal{E}_{i} =O_P\left(\sqrt{\frac{m}{NT}}\right)$.  Based on the above development,  the proof is complete.\hspace*{\fill}{$\blacksquare$}

\bigskip

Note that under the HD setting, the elements of $\beta_m(z)$ belonging to $ L^2 (V_z)$ indicates that $\|C_\beta\|\le a_0 \sqrt{p}$ with $a_0$ being a large constant. We will be repeatedly using this fact below.

\bigskip

\noindent \textbf{Proof of Lemma A.6:}

(1). Write

\begin{eqnarray*}
&&\frac{1}{NT}\sum_{i=1}^N \left(\phi_i[\beta_{0,m}]-\phi_i[\beta_m] \right)' M_F \mathcal{E}_i \nonumber\\
&=&\frac{1}{NT}\sum_{i=1}^N \left(\phi_i[\beta_{0,m}]-\phi_i[\beta_m] \right)'\mathcal{E}_i+\frac{1}{NT}\sum_{i=1}^N \left(\phi_i[\beta_{0,m}]-\phi_i[\beta_m] \right)' P_F \mathcal{E}_i\nonumber\\
&:=&\Lambda_1 +\Lambda_2.\nonumber
\end{eqnarray*}

For $\Lambda_1$, write 

\begin{eqnarray*}
&&\sup_{\|  C_\beta \|\le a_0\sqrt{p} }\left|\frac{1}{NT}\sum_{i=1}^N \left(\phi_i[\beta_{0,m}]-\phi_i[\beta_m] \right)'\mathcal{E}_i\right|\le \left\| \frac{1}{NT}\sum_{i=1}^N\mathcal{Z}_i' \mathcal{E}_i\right\|\cdot  \sup_{\|  C_\beta \|\le a_0\sqrt{p}}\left\|\vect(C_{\beta_0} - C_\beta)\right\|\nonumber\\
&=& O\left(\sqrt{\frac{mp}{NT}}\right) \cdot\sup_{\|  C_\beta \|\le a_0\sqrt{p} }\left\|C_{\beta_0} - C_\beta\right\|= O_P\left( \sqrt{\frac{mp^2}{NT}} \right)\nonumber
\end{eqnarray*}
where the first equality follows from some standard analysis on the term $\frac{1}{NT}\sum_{i=1}^N\mathcal{Z}_i' \mathcal{E}_i$ using Assumption 1.1; and the last equality follows from $\|C_\beta\|\le a_0\sqrt{p}$.

In order to consider $\Lambda_2$, let $\Delta b= (\phi_1[\beta_{0,m}]-\phi_1[\beta_m],\ldots ,\phi_N[ \beta_{0,m}]-\phi_N[ \beta_m])$, and note that

\begin{eqnarray}\label{EqLPA10}
&&\sup_{\|  C_\beta \|\le a_0\sqrt{p}}\frac{1}{NT} \left\|\Delta b\right\|^2 = \sup_{\|  C_\beta \|\le a_0\sqrt{p}}\frac{1}{N}\sum_{i=1}^N  (C_\beta -C_{\beta_0})'\mathcal{Z}_i'  \mathcal{Z}_i (C_\beta -C_{\beta_0})\nonumber \\
&\le &O_P(1)\sup_{\|  C_\beta \|\le a_0\sqrt{p}} \| C_\beta -C_{\beta_0} \|^2 =O_P(p),
\end{eqnarray}
where the inequality follows from Assumption 2.1.

Then we are able to write

\begin{eqnarray*}
&&\sup_{\|  C_\beta \|\le a_0\sqrt{p}, \, F\in \mathsf{D}_F}\left| \frac{1}{NT}\sum_{i=1}^N \left(\phi_i[\beta_{0,m}]-\phi_i[\beta_m] \right)' P_F \mathcal{E}_i\right| \nonumber\\
&=&\sup_{\|  C_\beta \|\le a_0\sqrt{p}, \, F\in \mathsf{D}_F}   \left| \frac{1}{NT}  \textrm{tr}\left( P_F  \mathcal{E}' \Delta b'\right)\right| \le  \frac{r}{NT} \sup_{\|  C_\beta \|\le a_0\sqrt{p}, \, F\in \mathsf{D}_F} \|P_F  \mathcal{E}' \Delta b' \|_{\textrm{sp}}\nonumber\\
&\le&    \sup_{\|  C_\beta \|\le a_0\sqrt{p}, \, F\in \mathsf{D}_F}\frac{r}{NT}  \|P_F\|_{\textrm{sp}} \| \mathcal{E}\|_{\textrm{sp}}\|  \Delta b \|_{\textrm{sp}}=O_P\left(\sqrt{\frac{p\,\xi_{NT}}{NT} }  \right) ,\nonumber
\end{eqnarray*}
where the last equality follows from Assumption 5.1 and (\ref{EqLPA10}).

Based on the above development, the result follows.

\medskip

(2). Write

\begin{eqnarray}\label{EqLPRes}
&&\sup_{F\in  \mathsf{D}_F} \left\vert \frac{1}{NT}\sum_{i=1}^{N}\phi _{i}\left[  \Delta_{ m}\right]' M_{F}\phi _{i}\left[ \Delta_{ m}\right]\right\vert \le\left\vert \frac{1}{NT}\sum_{i=1}^{N}\phi _{i}\left[ \Delta_{ m}\right]'\phi _{i}\left[ \Delta_{ m}\right] \right\vert \nonumber \\
&=&\frac{1}{NT}\sum_{i=1}^N\sum_{t=1}^T \| x_{it}^*\|^2 \|\Delta_{m}^*(z_{it}) \|^2 =O_{P}\left(p^* m^{-\mu}\right),
\end{eqnarray}
where the last equality follows from Assumption 2.1 and $E\| x_{it}^*\|^2=O(p^*)$.

\medskip

(3). Write
\begin{eqnarray*}
&&\sup_{F\in  \mathsf{D}_F} \left\vert \frac{1}{NT}\sum_{i=1}^{N}\phi _{i}\left[  \Delta_{ m}\right] ^{\prime }M_{F}F_0\gamma _{0i}\right\vert =\sup_{F\in  \mathsf{D}_F} \left\vert \frac{1}{NT}\text{tr}\left( M_{F}F_0\Gamma _0^{\prime }\Delta^{\prime }\right) \right\vert \nonumber\\
&\leq &\frac{r}{NT}\sup_{F\in  \mathsf{D}_F} \left\Vert M_{F}\right\Vert _{\text{sp}}\left\Vert
F_0\right\Vert _{\text{sp}}\left\Vert \Gamma _0\right\Vert _{\text{sp}}\left\Vert \Delta \right\Vert _{\text{sp}} =O_P(\sqrt{p^*} m^{-\frac{\mu}{2}}), \nonumber
\end{eqnarray*}
where $\Delta = (\phi_1(\Delta_m),\ldots, \phi_N(\Delta_m))$; and the second equality follows from that $\frac{1}{NT}\left\Vert \Delta  \right\Vert ^2=O_{P}\left(p^* m^{-\mu}\right)$ as in (2) of this lemma.

\medskip

(4). Similar to the proof for (3) of this lemma, the result follows.

\medskip

(5). Write

\begin{eqnarray*}
&&\sup_{\|  C_\beta \|\le a_0\sqrt{p}, \, F\in \mathsf{D}_F}\left\vert \frac{1}{NT}\sum_{i=1}^{N}\phi _{i}\left[  \Delta_{m}\right] ^{\prime }M_{F}\left\{\phi _{i} \left[\beta_m\right] -\phi _{i} \left[\beta_{0,m}\right] \right\} \right\vert \nonumber\\
&\leq & \left\{ \frac{1}{NT}\sum_{i=1}^{N}\|\phi _{i}\left[  \Delta_{m}\right] \|^2\right\} ^{1/2}  \cdot\sup_{\|  C_\beta \|\le a_0\sqrt{p}}\left\{ \frac{1}{NT}\sum_{i=1}^{N}\|\phi _{i} \left[\beta_m\right] -\phi _{i} \left[\beta_{0,m}\right] \|^2 \right\}^{1/2}\nonumber \\
&=& O_P(\sqrt{p^*}m^{-\frac{\mu}{2}})\cdot O_P(\sqrt{p}) ,\nonumber
\end{eqnarray*}
where the equality follows from (\ref{EqLPA10}) and (\ref{EqLPRes}). Then the proof is complete. \hspace*{\fill}{$\blacksquare$}

\bigskip

\noindent \textbf{Proof of Lemma A.7:}

(1). Still, let $\Delta\phi_i[\beta_m] = \phi_i[\beta_{0,m}] - \phi_i[\beta_{m}]$, $\xi_F =\text{vec}\left( M_F F_0\right) ,$ $A_{1F}=\frac{1}{NT}\sum_{i=1}^{N}\mathcal{Z}_i'M_{F}\mathcal{Z}_i$, $A_{2}=\frac{1}{NT}\left( \Gamma_0'\Gamma_0\right) \otimes I_{T},$ and $A_{3F}=\frac{1}{NT}\sum_{i=1}^{N}\gamma_{0i} \otimes (M_F\mathcal{Z}_i)$. By the definition of (3.3) and Lemma A.6, we have

\begin{eqnarray*} 
0 &\ge & \frac{1}{NT}Q_{\lambda}( \widehat{C}_\beta, \widehat{F}) - \frac{1}{NT}Q_{\lambda}( C_{\beta_0}, F_0) \nonumber\\
&=& \frac{1}{NT} \sum_{i=1}^N \left( \Delta\phi_i[\widehat{\beta}_m]+F_0\gamma_{0i} \right)'M_{\widehat{F}}\left(\Delta\phi_i[\widehat{\beta}_m]+F_0\gamma_{0i} \right) +\frac{1}{NT}\sum_{i=1}^N\mathcal{E}_i' M_{\widehat{F}} \mathcal{E}_i\nonumber\\
&& +\frac{2}{NT}\sum_{i=1}^N \left( \Delta\phi_i[\widehat{\beta}_m]+F_0\gamma(v_i) \right)'M_{\widehat{F}} \mathcal{E}_i +\sum_{j=1}^{p} \frac{\lambda_j}{NT} \|\widehat{C}_{\beta ,j} \| \nonumber \\
&&- \frac{1}{NT} \sum_{i=1}^N \left( \phi_i[\Delta_m]+\mathcal{E}_i\right)'M_{F_0}\left( \phi_i[\Delta_m]+\mathcal{E}_i \right) - \sum_{j=1}^{p^*} \frac{\lambda_j}{NT}\|C_{\beta_0 ,j} \|\nonumber\\
&=& \frac{1}{NT} \sum_{i=1}^N \left( \Delta\phi_i[\widehat{\beta}_m]+F_0\gamma_{0i} \right)'M_{\widehat{F}}\left(\Delta\phi_i[\widehat{\beta}_m]+F_0\gamma_{0i} \right)\nonumber\\
&&+\sum_{j=1}^{p} \frac{\lambda_j}{NT} \|\widehat{C}_{\beta ,j} \| - \sum_{j=1}^{p^*} \frac{\lambda_j}{NT}\|C_{\beta_0 ,j} \|+O_P\left(\frac{1}{\sqrt[4]{\xi_{NT}}} +  \sqrt{\frac{p(\xi_{NT}+mp)}{NT}} + \sqrt{p\, p^*} m^{-\frac{\mu}{2}}\right)\nonumber\\
&=& \vect(C_{\beta_0}-\widehat{C}_\beta )'\frac{1}{NT}\sum_{i=1}^{N}\mathcal{Z}_i'M_{\widehat{F}}\mathcal{Z}_i\vect(C_{\beta_0}-\widehat{C}_\beta )+\frac{1}{NT}\text{tr}\left( M_{\widehat{F}} F_0\Gamma _0'\Gamma _0F_0'M_{\widehat{F}}\right)\nonumber \\
&&+2\vect(C_{\beta_0}-\widehat{C}_\beta )'\frac{1}{NT}\sum_{i=1}^{N}\mathcal{Z}_i'M_{\widehat{F}}F_0\gamma_{0i} +\sum_{j=1}^{p} \frac{\lambda_j}{NT} \|\widehat{C}_{\beta ,j} \| - \sum_{j=1}^{p^*} \frac{\lambda_j}{NT}\|C_{\beta_0 ,j} \| \nonumber\\
&&+O_P\left(\frac{1}{\sqrt[4]{\xi_{NT}}} +  \sqrt{\frac{p(\xi_{NT}+mp)}{NT}} + \sqrt{p\, p^*} m^{-\frac{\mu}{2}}\right),\nonumber
\end{eqnarray*}
where the second equality follows from (1) of Lemma A.2, and Lemma A.6. Thus, we can further write

\begin{eqnarray}\label{HDconsis0}
\sum_{j=1}^{p^*} \frac{\lambda_j}{NT}\|C_{\beta_0 ,j} \|&\ge &\vect(C_{\beta_0}-\widehat{C}_\beta )'\frac{1}{NT}\sum_{i=1}^{N}\mathcal{Z}_i'M_{\widehat{F}}\mathcal{Z}_i\vect(C_{\beta_0}-\widehat{C}_\beta )+\frac{1}{NT}\text{tr}\left( M_{\widehat{F}}F_0\Gamma _0'\Gamma _0F_0'M_{\widehat{F}}\right)\nonumber \\
&&+2\vect(C_{\beta_0}-\widehat{C}_\beta )'\frac{1}{NT}\sum_{i=1}^{N}\mathcal{Z}_i'M_{\widehat{F}}F_0\gamma_{0i} \nonumber \\
&&+O_P\left(\frac{1}{\sqrt[4]{\xi_{NT}}} +  \sqrt{\frac{p(\xi_{NT}+mp)}{NT}} + \sqrt{p\, p^*} m^{-\frac{\mu}{2}}\right) \nonumber \\
&\ge &\vect(C_{\beta_0}-\widehat{C}_\beta )'\left(A_{1\widehat{F}}-A_{3\widehat{F}}'A_{2}^{-1}A_{3\widehat{F}}\right)\vect(C_{\beta_0}-\widehat{C}_\beta ) \nonumber \\
&&+[\xi_{\widehat{F}}'+\vect(C_{\beta_0}-\widehat{C}_\beta )'A_{3\widehat{F}}'A_{2}^{-1}]A_{2}[\xi_{\widehat{F}}+A_{2}^{-1}A_{3\widehat{F}}\vect(C_{\beta_0}-\widehat{C}_\beta )]\nonumber \\
&&+O_P\left(\frac{1}{\sqrt[4]{\xi_{NT}}} +  \sqrt{\frac{p(\xi_{NT}+mp)}{NT}} + \sqrt{p\, p^*} m^{-\frac{\mu}{2}}\right)\nonumber\\
&\ge &O_P(1)  \|C_{\beta_0}-\widehat{C}_\beta \|^2+O_P\left(\frac{1}{\sqrt[4]{\xi_{NT}}} +  \sqrt{\frac{p(\xi_{NT}+mp)}{NT}} + \sqrt{p\, p^*} m^{-\frac{\mu}{2}}\right).
\end{eqnarray}
Till now, we can conclude that 

\begin{eqnarray}\label{HDconsis1}
\|C_{\beta_0}-\widehat{C}_\beta \|^2 =O_P\left(\frac{1}{\sqrt[4]{\xi_{NT}}} +  \sqrt{\frac{p(\xi_{NT}+mp)}{NT}} + \sqrt{p\, p^*} m^{-\frac{\mu}{2}} +\frac{p^*\lambda_{\text{max}}^* }{NT}\right)=o_P(1),
\end{eqnarray}
where the second equality follows from Assumption 5.2.

\medskip

(2). By (\ref{HDconsis0}) and (\ref{HDconsis1}), we can further write that

\begin{eqnarray*}
o_p(1)\ge \frac{1}{NT}\text{tr}\left[ \left( F_0'M_{\widehat{F}}F_0\right) \left( \Gamma_0'\Gamma _0\right) \right] +o_{P}\left( 1\right) ,\nonumber
\end{eqnarray*}
so $\frac{1}{NT}\text{tr}\left[ \left( F_0'M_{\widehat{F}}F_0\right) \left( \Gamma_0'\Gamma _0\right) \right] =o_{P}\left( 1\right) $. As in \citet[p. 1265]{Bai}, we can further conclude that $\frac{1}{T}\textrm{tr}\left( F_0'M_{\widehat{F}}F_0\right)=o_{P}\left( 1\right) $, $\left\Vert P_{\widehat{F}}-P_{F_0}\right\Vert =o_{P}\left( 1\right) $, and $\frac{1}{T}\widehat{F}'F_0$ is invertible with probability approaching one. Thus, the second result of this theorem follows. \hspace*{\fill}{$\blacksquare$}

\end{document}